\documentclass[twocolumn]{aastex631}

\usepackage{lipsum}
\usepackage{mathtools}
\usepackage{xcolor}
\usepackage{soul}

\newcommand{\vhelio}{V_{\mathrm{HELIO}}}
\newcommand{\teff}{T_{\mathrm{eff}}}
\newcommand{\logg}{\log g}
\newcommand{\age}{\mathrm{Age}}
\newcommand{\feh}{[\mathrm{Fe}/\mathrm{H}]}
\newcommand{\alphafe}{[\alpha/\mathrm{Fe}]}

\newcommand{\npossdibs}{133\@}

\newcommand{\nfinaldibs}{84\@}
\newcommand{\nfinaldibsPROB}{73.3\@}
\newcommand{\nfinaldibsSIG}{25\@}

\newcommand{\nfinalRANDdibs}{0\@}
\newcommand{\nfinalSTRONGdibs}{55\@}
\newcommand{\nfinalFeHdibs}{25\@}
\newcommand{\nfinalAKdibs}{29\@}
\newcommand{\nfinalSNRdibs}{24\@}

\newcommand{\nfinaldibsEMISSION}{8\@}
\newcommand{\nfinaldibsEMISSIONtextupper}{Eight\@}

\newcommand{\nfinaldibsABSORPTION}{76\@}

\newcommand{\nfinaldibsABSORPTIONtextlower}{seventy-six\@}

\shorttitle{DIBs in APOGEE}
\shortauthors{McKinnon et al.}

\received{July 11, 2023}
\revised{February 6, 2024}
\accepted{February 6, 2024}
\submitjournal{The Astrophysical Journal}

\begin{document}

\title{Data-driven Discovery of Diffuse Interstellar Bands with APOGEE Spectra}

\correspondingauthor{Kevin McKinnon}
\email{kevin.mckinnon@ucsc.edu}

\author[0000-0001-7494-5910]{Kevin A. McKinnon}
\affiliation{Department of Astronomy \& Astrophysics, University of California, Santa Cruz, 1156 High Street, Santa Cruz, CA 95064, USA}

\author[0000-0001-5082-6693]{Melissa K. Ness}
\affiliation{Department of Astronomy, Columbia University, Pupin Physics Laboratories, New York, NY 10027, USA}
\affiliation{Center for Computational Astrophysics, Flatiron Institute, 162 Fifth Avenue, New York, NY 10010, USA}

\author[0000-0002-6667-7028]{Constance M. Rockosi}
\affiliation{Department of Astronomy \& Astrophysics, University of California, Santa Cruz, 1156 High Street, Santa Cruz, CA 95064, USA}

\author[0000-0001-8867-4234]{Puragra Guhathakurta}
\affiliation{Department of Astronomy \& Astrophysics, University of California, Santa Cruz, 1156 High Street, Santa Cruz, CA 95064, USA}

\begin{abstract}
Data-driven models of stellar spectra are useful tools to study non-stellar information, such as the Diffuse Interstellar Bands (DIBs) caused by intervening interstellar material. Using $\sim 55000$ spectra of $\sim 17000$ red clump stars from the APOGEE DR16 dataset, we create 2nd order polynomial models of the continuum-normalized flux as a function of stellar parameters ($\teff$, $\logg$, $\feh$, $\alphafe$, and $\age$). The model and data show good agreement within uncertainties across the APOGEE wavelength range, although many regions reveal residuals that are not in the stellar rest-frame. We show that many of these residual features -- having average extrema at the level of $\sim3\%$ in stellar flux on average -- can be attributed to incompletely-removed spectral lines from the Earth's atmosphere and DIBs from the interstellar medium (ISM). After removing most of the remaining contamination from the Earth's sky, we identify \nfinaldibs\@ absorption features not seen in unreddened sightlights that have $<50\%$ probability of being noise artifacts -- with \nfinaldibsSIG\@ of these features having $<5\%$ probability of being noise artifacts -- including all 10 previously-known DIBs in the APOGEE wavelength range. Because many of these features occur in the wavelength windows that APOGEE uses to measure chemical abundances, characterization and removal of this non-stellar contamination is an important step in reaching the precision required for chemical tagging experiments. Proper characterization of these features will benefit Galactic ISM science and the currently-ongoing Milky Way Mapper program of SDSS-V, which relies on the APOGEE spectrograph. 
\end{abstract}


\section{Introduction} \label{sec:intro}
Stellar spectra capture the parameters of a star's evolutionary state and record the chemical composition of the material in which it formed. Small samples of high resolution stellar spectra have been used to describe the individual element abundance distributions of the Milky Way (MW) in the local neighbourhood \citep[e.g.][]{Edvardsson_1993, Feltzing_1998, Prochaska_2000, Bensby_2003}. With the advent of large surveys -- such as RAVE \citep{Steinmetz_2006}, SEGUE \citep{Yanny_2009,Rockosi_2022}, APOGEE \citep{Wilson_2012,Majewski_2016,Majewski_2017}, Gaia-ESO \citep{Gilmore_2012}, GALAH \citep{DeSilva_2015,Martell_2017},  LAMOST \citep{Zhao_2012}, and H3 \citep{Conroy_2019a} -- has come the ability to map abundances across the disk, bulge, and halo of our Galaxy \citep[e.g.][]{Bergemann_2014,RojasArriagada_2014,Nidever_2014,Hayden_2015,Buder_2019,Buder_2022,Wylie_2021,Eilers_2022}. These large data ensembles have also enabled new, statistically-motivated questions to be tackled about topics such as the underlying dimensionality of individual abundance distributions and the information content of stellar spectra \citep[e.g.][]{Mitschang_2014,Ting_2015,Price-Jones_2018,Ness_2019,Ness_2022,Ting_2019,Feeney_2021, Weinberg_2022, Griffith_2022}. The answers to these questions are key to understanding the origin of individual elements and the utility of those elements to reconstruct the assembly history of the MW.

Chemical tagging -- the ability to distinguish co-natal stars based on chemical abundances derived from spectra -- is one of the foundational ideas of stellar surveys. Understanding the conditions that create particular populations of stars informs our stellar physics models and puts constraints on models of galaxy formation and evolution. In theory, stars that are born together were formed from the same gas cloud and thus share a chemical signature in their atmospheres. In practice, the level of precision required for chemical tagging is not currently feasible \citep[$< 0.02$~dex;][]{Ness_2019}.

The difficulties around chemical tagging become even more severe if there are unknown or unmodeled features in a spectrum, especially if those features impact wavelength regions used for measuring chemical abundances. In the visible and infrared (IR) regimes, the largest and most obvious source of non-stellar signal comes from the Earth's atmosphere, including sky line emission from OH and telluric absorption from CH$_4$, CO$_2$, and H$_2$O. Because of detailed measurements of the night sky's effects as well as knowing the rest-frame that spectral features are produced in, astronomers are able to account for and remove the bulk of Earth's atmosphere's signature. However, many spectra suffer from imperfect sky line and telluric removal, which leaves residual features capable of confusing spectral analysis pipelines. 

Another (often ignored) source of contamination comes from intervening dust and gas along the line-of-sight (LOS) to a star. Due to the velocity offset between gas/dust clouds and stars, spectral features from the Interstellar Medium (ISM) can appear at different wavelength locations in a set of observations at different LOS in the Galaxy. This issue is complicated further when the identification or central wavelength of an ISM-based feature is unknown or poorly constrained. Without a complete and accurate model of a star's light, it is often difficult to know a priori whether a particular residual feature is caused by non-stellar sources or is simply unknown physics/missing chemical species in the model. 

One common detection and characterization method for diffuse interstellar bands (DIBs) is to measure a feature's presence in multiple spectra of different stars and then to show correlations between ISM properties (e.g. extinction from dust) and that feature's strength. Typically, O- and B-type stars have been used as the preferred background ``lighthouses'' to study intervening structure because their spectra exhibit weak or rotationally-broadened features that make DIB identification much easier. However, accurate spectral modelling of the more-numerous later stellar types can reveal previously-unknown DIBs that were hidden by stellar features while also probing DIB properties along many more lines of sight \citep[e.g.][]{Ebenbichler_2022}.

Efforts to detect, characterize, and map these DIBs have historically been focused on the optical regime \citep[e.g.][]{Herbig_1995}, though a growing number of studies have been exploring the near-IR \citep[e.g.][]{Joblin_1990,Geballe_2011,Cox_2014,Zasowski_2015,Elyajouri_2016,Elyajouri_2017,Tchernyshyov_2017,Tchernyshyov_2018,Ebenbichler_2022,Smoker_2023}. For instance, the ten currently-known DIBs that fall in the near-IR $H$-band ($1.51-1.7~\mu$m) wavelengths seen by the APOGEE spectrograph are summarized in Table \ref{tab:previous_DIBs}, which is in stark contrast to the hundreds of known optical DIBs \citep[e.g. 559 DIBs in $5000-9000$~\AA\, as measured by][]{Fan_2019}. It is particularly important to understand sources of IR features as this regime is able to peer through the dusty regions of our Galaxy's disk. 

\begin{deluxetable}{cc}
\tablecaption{Most precise measurements of rest-frame wavelengths for currently-known DIBs that fall inside of the wavelength regions covered by the APOGEE spectrograph\tablenotemark{a}. DIBs that fall between the wavelength coverage of the three APOGEE detectors have been omitted.
\label{tab:previous_DIBs}}
\tablehead{
\colhead{$\lambda_0$ (\AA)} & \colhead{Reference}} 
\startdata
$15225 \pm 10$ & \citet{Geballe_2011} \\
$15272.42 \pm 0.04$ & \citet{Zasowski_2015} \\
$15616.13 \pm 0.07$ & \citet{Elyajouri_2017}\\
$15651.38 \pm 0.07$ & \citet{Elyajouri_2017} \\
$15671.82 \pm 0.03$ & \citet{Elyajouri_2017} \\
$15990 \pm 10$ & \citet{Geballe_2011} \\
$16231.1 \pm 0.5$ & \citet{Cox_2014} \\
$16571.5 \pm 0.5$ & \citet{Cox_2014} \\
$16582.5 \pm 0.5$ & \citet{Cox_2014} \\
$16592.5 \pm 0.5$ & \citet{Cox_2014} \\
\enddata
\tablenotetext{a}{The \citet{Cox_2014} values have been converted from their reported Air wavelengths to Vacuum.}
\end{deluxetable}

Astronomy's burgeoning ``Big Data Era'' has facilitated the development of novel data-driven approaches to understand stellar spectra that are less reliant on underlying physical models. A few successful techniques to characterize stellar light include using deep learning \citep{Leung_2019}, polynomial models of stellar labels \citep[e.g. The Cannon;][]{Ness_2015}, and non-Gaussian Processes \citep[e.g.][]{Feeney_2021}. One significant benefit of data-motivated models is that they can describe stellar features -- and correlations between features -- in spectra that are currently unknown to physics-based models. Additionally, the data models do not rely on many of the simplifying assumptions that are common in synthetic models (e.g. local thermal equilibrium, 1D radial stellar models and atmospheres). Finally, data-driven models are especially useful when the physics is not well constrained, such as in the low-density environments of the ISM that are currently impossible to recreate on Earth. As a star's light passes through intervening gas and dust on its way to our telescopes, it is imprinted with ISM signatures from many chemical species whose identities and properties are generally unknown. Detailed characterization of all the signatures in a spectrum are therefore important in disentangling the origin of various spectral features -- which furthers the science goals of both abundance measurements and ISM studies. 

Currently, the Milky Way Mapper (MWM) program of SDSS-V is using the APOGEE spectrograph to collect millions of stellar across all regions of the MW to understand its formation history and the physics of its stars \citep{Kollmeier_2017}; any improvements in the APOGEE analysis pipeline will therefore have compounding effects on the MWM science goals. Finally, better constraints on near-IR DIBs can be studied with tomography techniques \citep[e.g.][]{Tchernyshyov_2017,Tchernyshyov_2018} to develop a more complete picture of our Galaxy's ISM. 

In this paper, we describe the APOGEE spectra and stellar parameters used in our analysis in Section \ref{sec:data} and then present a data-driven model of those spectra in Section \ref{sec:modelling}. In Section \ref{sec:stucture_in_residuals}, we study the spectral residuals to show that DIBs, tellurics, and sky lines are responsible for many of the relatively-large remaining features. We remove the Earth-based residuals to detect and characterize the remaining DIB features in Section \ref{sec:finding_DIBs}. Finally, we summarize our results in Section \ref{sec:summary}. 

\section{Data} \label{sec:data}

\begin{figure}[h]
\begin{center}
\begin{minipage}[c]{\linewidth}
\includegraphics[width=\linewidth]{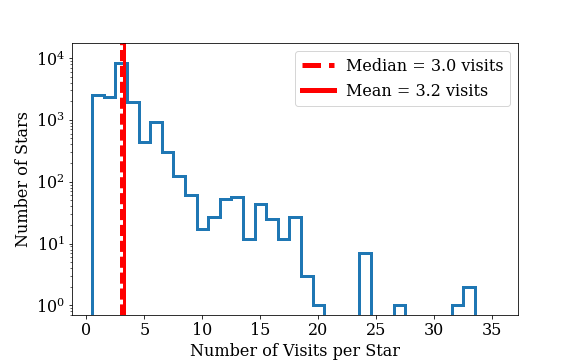}
\end{minipage}\\
\begin{minipage}[c]{\linewidth}
\includegraphics[width=\linewidth]{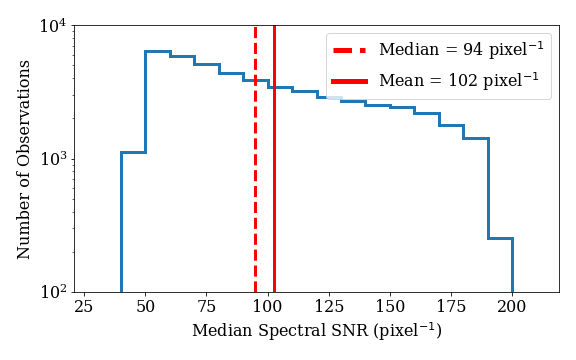}
\end{minipage}
\caption{\textbf{Top:} Histograms of the number of spectroscopic visits per star. The bin width has been set to 1. The median and mean number of visits per star are shown with vertical red lines. There are a total of 55028 individual visit spectra for the 17104 RC stars in our sample. \textbf{Bottom:} Histogram of the median spectral SNR for each visit used in our analysis, with mean and median shown with vertical red lines.}
\label{fig:nvisits_and_SNR_hists}
\end{center}
\end{figure}

This work makes extensive use of stellar spectra, abundances, and parameters of Red Clump (RC) stars in the MW as measured by the APOGEE spectrograph \citep{Wilson_2012,Majewski_2016,Majewski_2017} on the Sloan Telescope at the Apache Point Observatory as a component of the Sloan Digital Sky Survey \citep[SDSS;][]{York_2000,Eisenstein_2011,Blanton_2017}. 
The RC sample -- defined by \citet{Bovy_2014} using stellar parameters and simulated stellar evolution -- boasts high spectral signal-to-noise ratios (SNR) as well as precise stellar parameters and abundances. These properties make the RC sample an ideal population for data-driven modelling and for studying non-stellar residuals. 

The APOGEE spectra cover the $H$-band ($\sim 15000 - 17000~\mathrm{\AA}$) with high spectral resolving power ($R\sim 22500$) and fine pixel spacing ($\Delta \lambda \sim 0.2~\AA\cdot\mathrm{pixel}^{-1}$). The publicly-available spectral data were reduced using the APOGEE data reduction pipeline as described in \citet{Nidever_2015} and are given in the rest-frame of each star. Our analysis uses the individual visit spectra instead of the combined visit spectra to account for changes in the LOS velocity -- and, therefore, the location of non-stellar features -- of each observation. Distributions of the number of visits per star and the median spectral SNR of the individual visit spectra are given in Figure \ref{fig:nvisits_and_SNR_hists}.

\begin{figure*}[t]
\begin{center}
\begin{minipage}[c]{\linewidth}
\includegraphics[width=\linewidth]{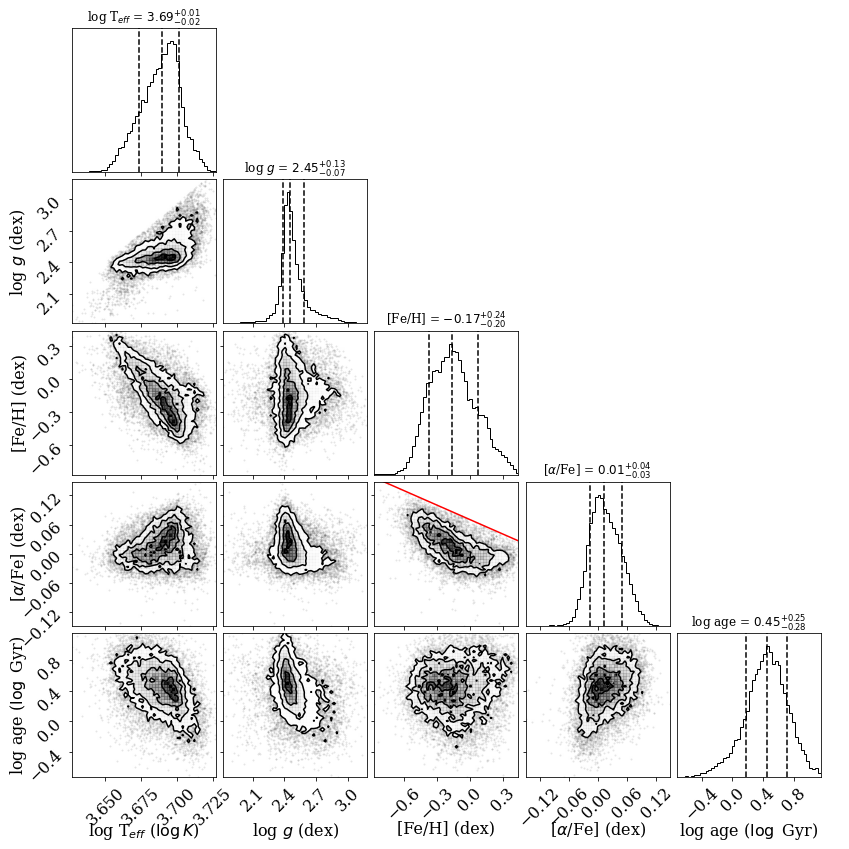}
\end{minipage}
\caption{Distribution of the stellar parameters for the 17104 RC stars in our sample. The titles above each histogram on the diagonal summarize the median and 68\% region of each distribution, which are displayed graphically with the vertical dashed black lines. The red line in the $\feh$ versus $\alphafe$ plot near the center of the figure shows where we mask out approximately 2600 stars that lie above this relationship; we do this to keep the 2D abundance distribution singly-peaked.}
\label{fig:RC_stellar_params}
\end{center}
\end{figure*}

The spectra were analysed by the the APOGEE Stellar Parameter and Chemical Abundances \citep[ASPCAP][]{Garcia-Perez_2016} pipeline. We use the ASPCAP $\teff$, $\logg$, $\feh$, and $\alphafe$ measurements and uncertainties from Data Release 16 \citep[DR16;][]{Jonsson_2020}, while stellar ages come from the catalogue of \citet{Sit_2020}. As exemplified in Figure \ref{fig:RC_stellar_params}, the stars in our sample occupy a relatively narrow range in stellar parameters and abundances. After noticing a minor secondary peak near $(\feh,\alphafe)=(-0.6,+0.2)~\mathrm{dex}$ in the $\alphafe$ versus $\feh$ panel, we removed stars with abundances above the red line to ensure that our modelling only focuses on a single chemical population. 

For the individual visit spectra in our sample, we remove pixels that have $\mathrm{SNR}< 50~\mathrm{pixel}^{-1}$. We also set the maximum pixel SNR to $200~\mathrm{pixel}^{-1}$ as recommended by ASPCAP, which suggests that the ``uncertainty floor floor is at the level of 0.5\%''\footnote{see the Uncertainty Arrays section at \newline\url{https://www.sdss4.org/dr16/irspec/spectra/}}. Because of known superpersistence issues in the blue detector, we mask out spectral observations where the fiber number is $\leq 100$; this selection removes approximately 6500 RC stars that do not have a single observation with a fiber number greater than 100. Finally, following the approach of \citet{Price-Jones_2018}, we remove data from pixels that have any of the the bitmask flags listed in Table \ref{tab:bitmasking}.

\begin{table}[t]
\caption{Pixel bitmasking of spectra\footnote{This is a subset of Table 1 in \citet{Price-Jones_2018}, which masks use of the bitmask definitions of \citet{Holtzman_2015}}.
\label{tab:bitmasking}}
\centering
\begin{tabular}{cc}
\hline \hline
Bitmask                               & Name\\ \hline
0 & Pixel marked as BAD in bad pixel mask \\
1 & Pixel marked as cosmic ray in ap3d \\
2 & Pixel marked as saturated in ap3d \\
3 & Pixel marked as unfixable in ap3d \\
4 & Pixel marked as bad as determined                 \\
  &                                   from dark frame \\
5 & Pixel marked as bad as determined                 \\
  &                                   from flat frame \\
6 & Pixel set to have very high error            \\
  &                                   (not used) \\
7 & No sky available for this pixel                 \\
  &                                 from sky fibers \\
12 & Pixel falls near sky line \\ \hline
\end{tabular}
\end{table}

These APOGEE individual visit spectra in the stellar rest-frame, along with the \citet{Sit_2020} ages and ASPCAP parameters and abundances, are used in combination to build data-driven models in Section \ref{sec:modelling}, which are the basis of our analysis.

\section{Modelling RC Spectra} \label{sec:modelling}

\subsection{Preprocessing Spectra} \label{ssec:preprocessing_spectra}

First, we continuum normalize the individual visit spectra using iterative B-spline fitting. At each iteration, the B-spline is defined by 50~\AA-spaced knots. For the first iteration, all flux measurements in a spectrum are used to define the initial spline. For subsequent iterations, the new spline is measured using only flux values that are within $3\sigma$ (in flux uncertainty) for fluxes below the spline or $5\sigma$ for fluxes above the spline. This masking of fluxes is done to ensure that strong absorption features do not overly impact the continuum measurement. 

We iterate the spline fitting up to 100 times per spectrum, but stop iterating if the current iteration's spline is very similar to the previous one:
$$\sum_{i}\frac{| c'_{i,j,k}-c_{i,j,k}|}{c_{i,j,k}} < 1\times 10^{-5}$$
where $c_{i,j,k}$ is the continuum spline value at pixel $i$ from the previous iteration and $c'_{i,j,k}$ is the current iteration's continuum spline value at pixel $i$ for observation/visit number $k$ of star $j$. This condition is such that the summed absolute fractional change is smaller than 0.001\%, which tends to occur when the subset of fluxes being masked hasn't changed from one iteration to another. This process usually only takes a handful of iterations (i.e. $\leq 5$), and virtually all of the spectra converge on a spline well before the 100 maximum iterations. 

To capture any remaining continuum, we repeat this continuum-fitting process after the first model fit to define a continuum-adjustment B-spline. We divide each individual visit spectrum by the previously-defined continuum spline and the best-fit model to get an approximate noise spectrum that may still have some continuum trends in it. We then use the same iterative B-spline fitting process as above, but using 25~\AA-spaced knots and a $3\sigma$ threshold above and below for masking. The finer-spaced knots and the narrower threshold are because the residual spectrum ideally only consists of noise and any trends larger than $\sim 20~\mathrm{\AA}$\@ likely arise from an incomplete initial continuum removal. 

The continuum-normalized flux in pixel $i$ for spectral observation $k$ of star $j$ is then $y_{i,j,k} = f_{i,j,k}/c_{i,j,k}$ with corresponding uncertainty of $\sigma_{i,j,k} = \sigma_{f,i,j,k}/c_{i,j,k}$, where $f_{i,j,k}$ and $\sigma_{f,i,j,k}$ are the raw flux and uncertainty values. The continuum-normalized fluxes are then used as the inputs for the data-driven modelling. 

\subsection{Modelling Flux using Stellar Labels} \label{ssec:modelling_math}

\begin{figure*}[t]
\begin{center}
\includegraphics[width=\linewidth, trim={0cm 0 0cm 0},clip]{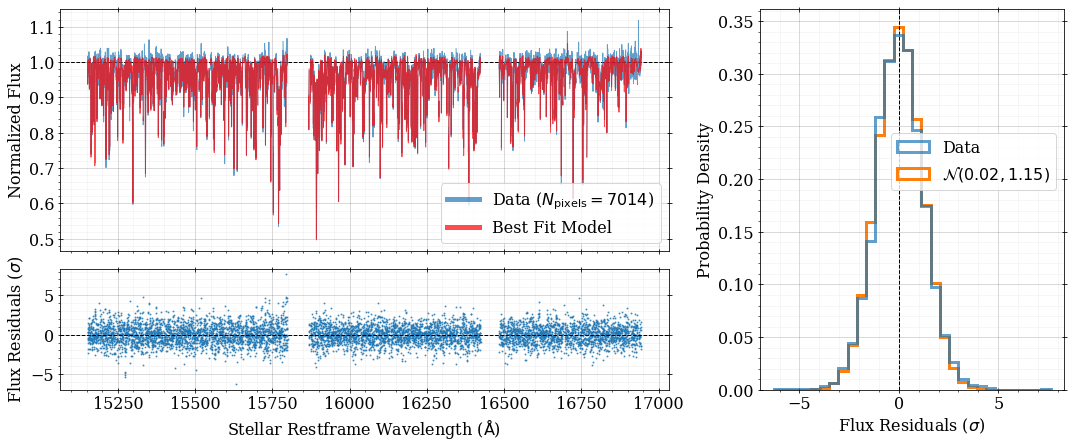}
\caption{Comparison of best fit model (red line) and normalized flux (blue line) for one RC star with median SNR of $105~\mathrm{pixel}^{-1}$. The residuals in the lower left panel and the right panel have been scaled by their corresponding uncertainties. The orange histogram in the right panel is the best fit normal distribution, which shows that this star has good agreement with the data (mean near 0, standard deviation near 1) and that the difference between the data and model are larger than expected by the flux uncertainties at the level of $\sim 15\%$\@. The gaps in the data around 15850~\AA\@ and 16450~\AA\@ correspond to the wavelength gaps between the three APOGEE CCDs.}
\label{fig:model_comparison_example}
\end{center}
\end{figure*}

Following the approach of \citet{Ness_2016}, we define the continuum-normalized flux in each pixel to be a 2nd-order polynomial of stellar labels \citep[see also][]{Price-Jones_2018}. 
We note that a key difference between these bodies of work is the stellar labels used in the fitting: \citet{Price-Jones_2018} uses $\teff$, $\logg$, and $\feh$, \citet{Ness_2016} uses the same, but also includes $\alphafe$ and mass, while our model uses $\teff$, $\logg$, $\feh$, $\alphafe$, and age. Though a detailed comparison of these models is outside the scope of this work, we expect that our model will produce extremely similar results to the \citet{Ness_2016} case because of the tight relationship between mass and age for RC stars \citep[e.g.][]{Martig_2015}.  The vector of stellar labels in our analysis for star $j$ is therefore given as
\begin{equation} \label{eq:stellar_label_vector}
    \vec x_{j} = \begin{pmatrix}
                    T_{\mathrm{eff},j}^2 \\
                    T_{\mathrm{eff},j} \times \logg_j \\
                    T_{\mathrm{eff},j} \times \feh_j \\
                    T_{\mathrm{eff},j} \times \alphafe_j \\
                    T_{\mathrm{eff},j} \times \age_j \\
                    T_{\mathrm{eff},j} \\
                    \logg^2_j \\
                    \vdots \\
                    \feh^2_j \\
                    \vdots \\
                    \alphafe^2_j \\
                    \vdots \\
                    \age^2_j \\ 
                    \vdots \\
                    1
                \end{pmatrix}
\end{equation}
such that the vector contains all the parameters to the first and second powers, cross terms, and a constant. Our model of the continuum-normalized fluxes is defined as
\begin{equation} \label{eq:model}
    \begin{split}
        y_{i,j,k} = \vec x_{j}^T \cdot \vec \theta_{i} +\varepsilon_{i,j,k}
    \end{split}
\end{equation}
where $\vec \theta_{i}$ are the coefficients for the stellar label terms in Equation \ref{eq:stellar_label_vector} for pixel $i$ and 
\begin{equation}
    \varepsilon_{i,j,k} \sim \mathcal{N}\left(0,\sigma_{i,j,k}\right)
\end{equation} 
describes the noise as a result of the uncertainty in a pixel's flux. 
This functional form implies that the data likelihood at pixel $i$ is
\begin{equation} \label{eq:likelihood}
    \mathrm{likelihood}_i = \prod_{j}^{n_{*}}\prod_{k}^{n_{\mathrm{obs},j}} \mathcal{N}\left(y_{i,j,k} | \vec x_{j}^T \cdot \vec \theta_{i}, \sigma_{i,j,k} \right).
\end{equation} 

We then see that the likelihood at a given pixel is maximized when 
\begin{equation}
    \sum_{j}^{n_{*}}\sum_{k}^{n_{\mathrm{obs},j}} \left(\frac{y_{i,j,k} - \vec x_j^T \cdot \vec \theta_i}{\sigma_{i,j,k}}\right)^2
\end{equation} 
is minimized, which occurs when
\begin{equation} \label{eq:best_fit_coeffs}
    \hat \theta_{i} = \left[ \sum_{j}^{n_{*}}\sum_{k}^{n_{\mathrm{obs},j}} \frac{1}{\sigma_{i,j,k}^2} \vec x_j \cdot \vec x_j^T \right]^{-1} \cdot \left[ \sum_{j}^{n_{*}}\sum_{k}^{n_{\mathrm{obs},j}}  \frac{y_{i,j,k}}{\sigma_{i,j,k}^2} \vec x_j \right],
\end{equation} 
which defines the best-fit coefficient vector at pixel $i$ for a set of normalized fluxes, uncertainties, and stellar labels. 

To propagate the uncertainties on the stellar parameters to the model coefficients, we repeatedly draw realizations of the stellar parameters for each star and remeasure the best-fit coefficients at each pixel. After 500 iterations, we take the median of the best-fit coefficients to be the coefficients of final model. The residual flux for pixel $i$ of observation $k$ of star $j$ is defined to be
\begin{equation}
    r_{i,j,k,l} = y_{i,j,k} - \vec x_{j,l}^T \cdot \hat \theta_{i,l},
\end{equation}
for realization $l$ of the the stellar parameters samples, $\vec x_{j,l}$, which yields a best-fit coefficient measurement of $\hat \theta_{i,l}$. To propagate the uncertainty on the stellar labels -- and therefore the uncertainty on the best-fit coefficients -- to the residuals, we also repeatedly measure the residual flux values for the 500 realizations, giving samples of $r_{i,j,k,l}$. The final residual flux, $\hat r_{i,j,k}$, is taken to be the median of these realizations, with an uncertainty that is given by
\begin{equation}
    \sigma_{r,i,j,k}^2 = \sigma_{i,j,k}^2 + \mathrm{var}\left( r_{i,j,k,1}, \dots,  r_{i,j,k,500}\right),
\end{equation}
where $\mathrm{var}\left( r_{i,j,k,1}, \dots,  r_{i,j,k,500}\right)$ is the variance of the 500 residual measurements at a given pixel for a given spectrum. In words, the residual uncertainty is a quadrature sum of the normalized flux uncertainty and the propagation of the best-fit coefficient uncertainty. 

\begin{figure}[ht]
\begin{center}
\includegraphics[width=\linewidth]{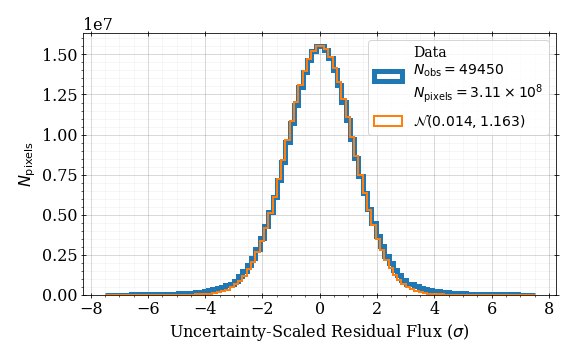}
\caption{Distribution of uncertainty-scaled residuals for $\sim 3\times 10^8$ pixels from all $\sim 5\times 10^5$ spectral observations of $\sim 17000$ stars in our sample (blue histogram). The orange histogram is the best fit Gaussian, which shows that the residuals are centered near zero and have a standard deviation near 1, as would be expected if the model and flux uncertainties perfectly describe the observed fluxes. Because the best fit Gaussian width is $1.164\pm0.004$, the model doesn't perfectly describe the data within their uncertainties, as is expected if there is non-modelled information remaining.}
\label{fig:residual_dist_all_stars}
\end{center}
\end{figure}

\begin{figure}[h]
\begin{center}
\includegraphics[width=\linewidth]{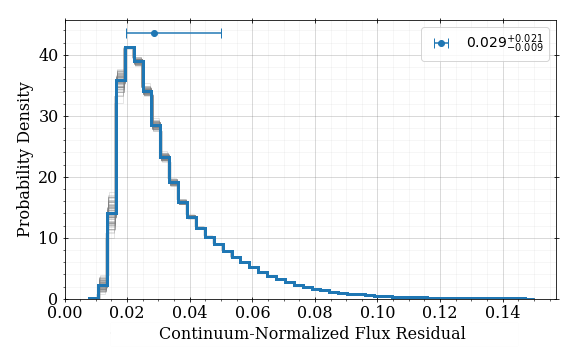}
\caption{Distribution of the size of residual fluxes that are in excess of the best-fit Gaussian in Figure \ref{fig:residual_dist_all_stars}. The grey histograms show different realizations of measuring the excess, while the blue histogram shows the median of these realizations. The blue point at the top of the figure summarizes the median and 68\% region of the distributions, revealing that the non-Gaussian excess of residuals is $\sim 3\%$ of the stellar flux on average.}
\label{fig:residual_dist_excess}
\end{center}
\end{figure}

\begin{figure*}[t]
\begin{center}
\begin{minipage}[c]{0.49\linewidth}
\includegraphics[width=\linewidth]{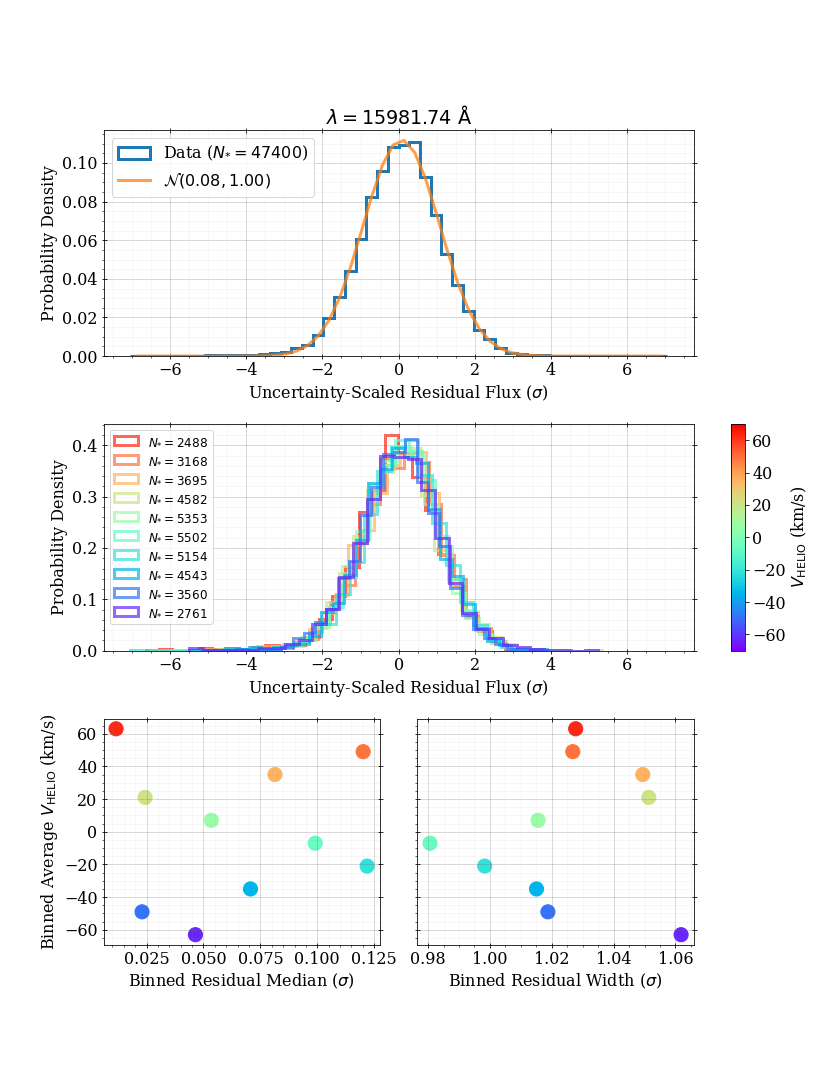}
\end{minipage} \hfill
\begin{minipage}[c]{0.49\linewidth}
\includegraphics[width=\linewidth]{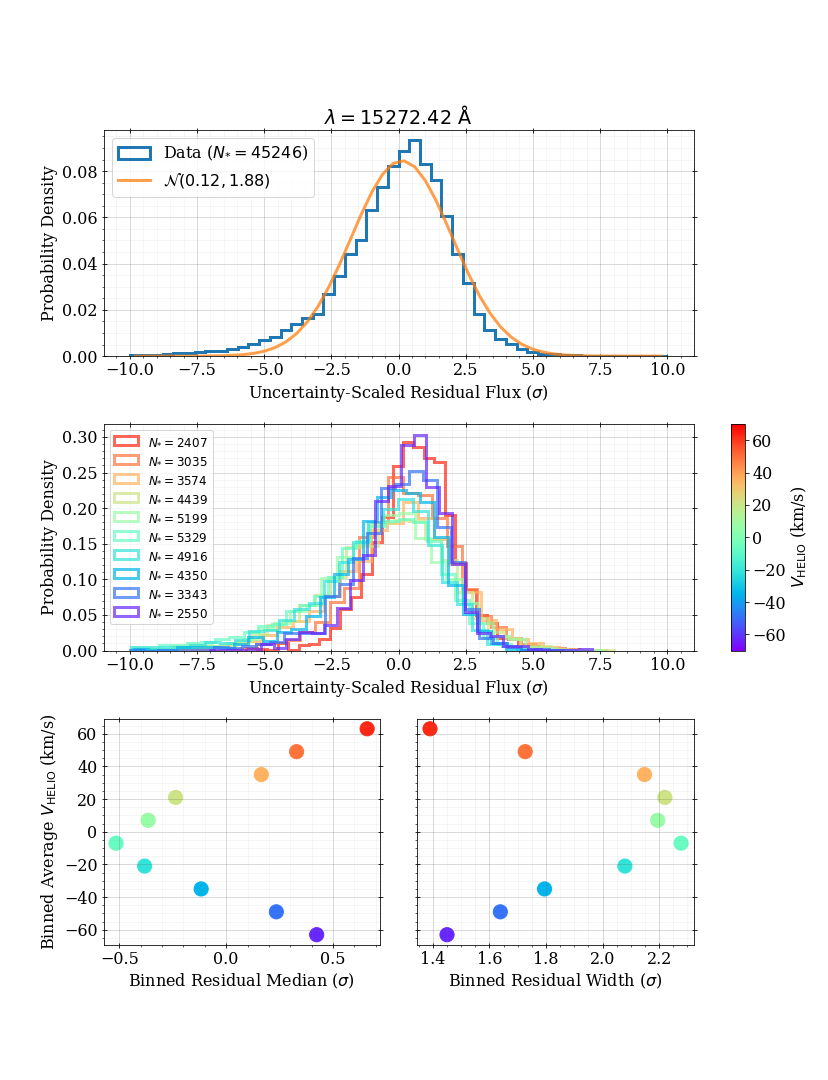}
\end{minipage}
\caption{Residual fluxes scaled by their uncertainty (i.e. [normalized flux$-$model]/uncertainty) for all the stars in two different APOGEE pixels. The wavelength of each pixel, in the stellar rest-frame, is listed above the top panel in columns of figures. By highlighting a single pixel, we focus on the residual information from all of the spectra at a particular stellar rest-frame wavelength. \textbf{Left:} A well-modelled pixel, where the uncertainty-scaled flux residuals (blue in top panel) follow the expected unit normal (orange in top panel). In the middle panel, the residuals are binned by heliocentric velocity (number of spectra in each bin given by $N_*$ in the legend), and we see no differences between the velocity bins. The bottom two panels show the median and standard deviations of the velocity-binned histograms, again showing no noticeable trend in the residuals with velocity. \textbf{Right:} A pixel where the residuals do not follow a unit Gaussian, or any Gaussian for that matter. After binning by heliocentric velocity in the middle panel, and plotting the binned medians and standard deviations in the bottom two panels, we see a noticeable trend in the median and width of the residual distribution with velocity.}
\label{fig:residual_distribution_comparison}
\end{center}
\end{figure*}

A comparison of the the best fit model and data for a single observation of one star in our sample is shown in Figure \ref{fig:model_comparison_example}. The uncertainty-scaled residual distribution for this star is close to the expected unit Gaussian distribution (mean of 0, standard deviation of 1), which shows that the model does a good job of capturing the information in the spectrum. If we assume that the fluxes and uncertainties reported by APOGEE accurately describe the data, then the best-fit Gaussian to the residual distribution (orange histogram) having a width of $\sim 1.15\pm0.1\sigma$ implies that there is information in this star's spectrum that the model does not capture at the level of $\sim 15\%$ of the flux uncertainties. 

If we instead look at the uncertainty-scaled residual distribution across all observations and all pixels, we get the distribution shown in Figure \ref{fig:residual_dist_all_stars}. As before, the distribution is quite similar to the unit Gaussian, which suggests that the model performs well across all the spectral observations in our sample. Again, that the width is greater than $1\sigma$ reveals that there may be more information in the spectra than the model is able to describe. 

Using the residual histogram and best-fit Gaussian distribution in Figure \ref{fig:residual_dist_all_stars}, we vertically align the distributions so that they have the same height at their peaks so that we can quantify their difference in the wings (i.e. excess in the data over the expectation). First, we find that the data excess in the wings corresponds to $\sim 3.3\%$ of the total pixels. Next, we draw many realizations of residual flux measurements from each bin in the blue data histogram where the data counts exceed the expected distribution. The results of this process are summarized in Figure \ref{fig:residual_dist_excess}, where the blue curve shows the median distribution of the size of flux residual in excess of the expectation. The blue point above the histogram shows the median and 68\% region of the distribution, and the grey histograms show individual realizations of residual flux samples. In summary, the residuals that are not explained by the model have a median size of $\sim 2.9\%$ of the stellar flux, with a 68\% region covering 2\% to 5\% of the stellar flux. 

\section{Structure in Residuals} \label{sec:stucture_in_residuals}

\begin{figure*}[ht]
\begin{center}
\includegraphics[width=\linewidth,trim={1cm 4cm 0cm 4cm},clip]{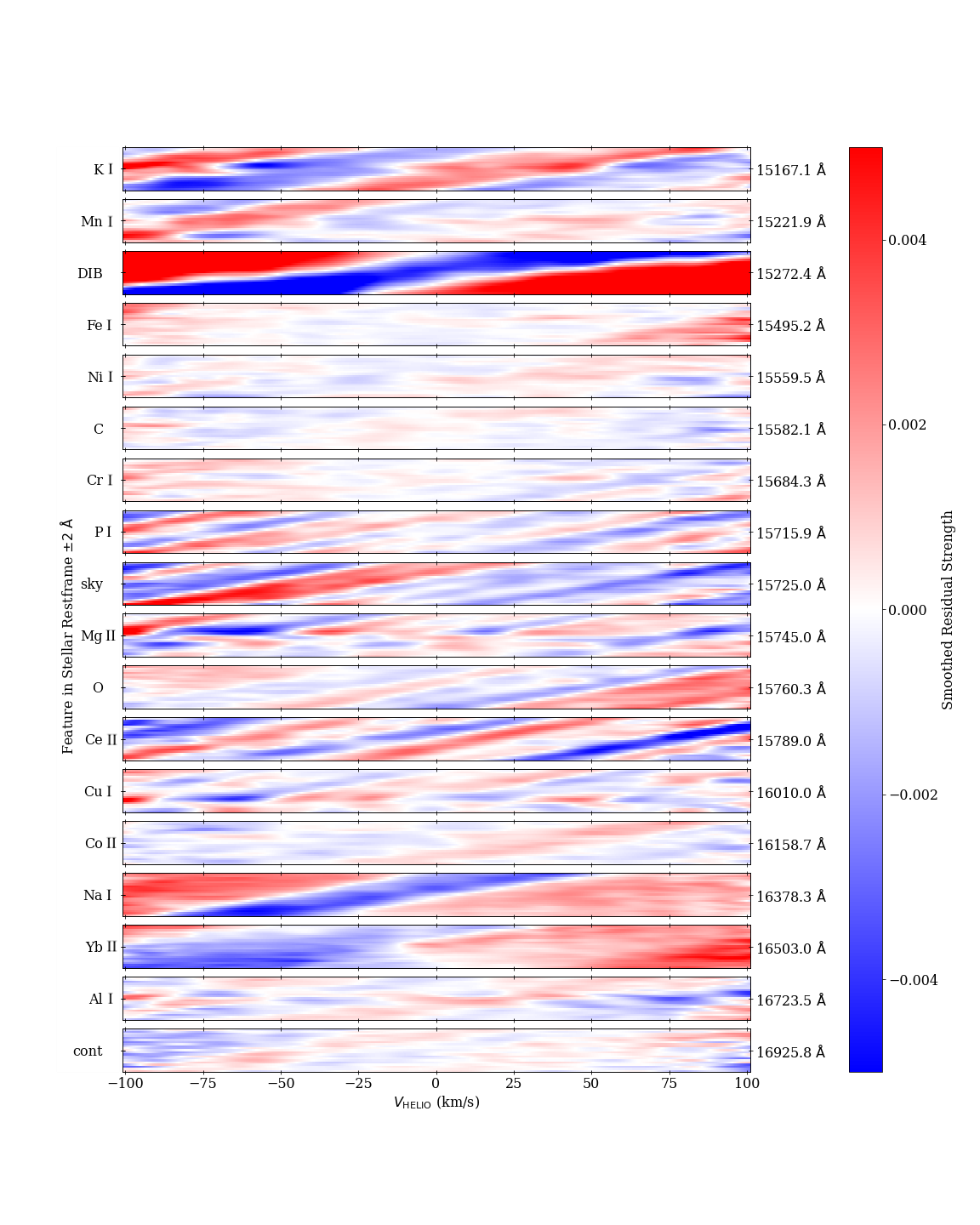}
\caption{Smoothed residual spectra of all RC stars with $|\vhelio|<100~\mathrm{km/s}$, sorted by the star's $\vhelio$. Each column corresponds to a residual spectrum and each row corresponds to a pixel/wavelength, with bluest wavelengths at the top and reddest wavelengths at the bottom. This image focuses on 18 different features (wavelengths listed on the right edge of the figure) with each panel consisting of 21 pixels; these cutouts include 15 elemental features, one region we've identified as being mostly continuum, one region of high telluric contamination, and one region near the strongest known DIB in APOGEE (15272~\AA\@ feature). The residual spectra have been Gaussian-smoothed in the horizontal direction with a kernel width of $5~$km/s. Where applicable, ionization levels are shown. For the C window, the dominant feature is from CO, and in the O window, the dominant feature is from OH. Some elemental regions (e.g. the DIB region, 3rd from the top) show a residual minimum that moves through the element window as a function of heliocentric velocity; this can be explained as an unmodelled residual feature that is not in the same reference frame as the star.}
\label{fig:smoothed_residuals_VHELIO_sorting}
\end{center}
\end{figure*}

To further explore the residuals, we begin by looking for trends in the flux residuals as a function of wavelength (in stellar rest-frame) and other explanatory variables. In many APOGEE pixels, we find that the model almost perfectly describes the data; for instance, the pixel summarized in the left panels of Figure \ref{fig:residual_distribution_comparison} shows a residual distribution (scaled by the residual uncertainties) that agrees very well with the unit Gaussian. When the stars are split up into different heliocentric velocity ($\vhelio$) bins -- as is shown in the left middle panel -- we see little difference between the residual distributions, and no obvious trend in those distributions' medians or standard deviations (left bottom panels). 

On the other hand, there are some pixels where the residual distributions are distinctly non-Gaussian and have widths that are much larger than $1\sigma$. An example of this is shown in the right panels of Figure \ref{fig:residual_distribution_comparison} for a pixel near the central wavelength of the strongest known DIB in the APOGEE wavelength range \citep[e.g. $\lambda_0 = 15272.42~\mathrm{\AA}$\@, from][]{Zasowski_2015}. Breaking the stars up into $\vhelio$ bins in the middle right panel, we now see significant differences between the residual distributions' shapes as well as their medians and widths (right bottom panels. In general, the extreme velocity bins have more positive residual medians and smaller standard deviations while the moderate velocity bins have negative residual medians and larger widths. 

We next look at the trends in the residuals with heliocentric velocity across neighbouring pixels at different wavelength cutouts. First, we sort the stars by their heliocentric velocity and then smooth the residuals (using a combination of inverse variance weighting and Gaussian smoothing based on $\vhelio$ at each pixel for each residual spectrum) to produce Figure \ref{fig:smoothed_residuals_VHELIO_sorting}. Each of the 18 panels show the smoothed residuals using 21 APOGEE pixels (i.e. width of $\sim4~\mathrm{\AA}$) centered on a particular feature, with the central wavelength of that feature listed on the right edge of the cutout. These regions correspond to 15 elemental features used by ASPCAP to measure abundances \citep[these are the same central wavelengths as used by][]{Feeney_2021}, one region we identify as having minimal visible features (i.e. continuum), one region where we notice a large amount of residual Earth-based contamination, and one region around the strongest DIB in the APOGEE wavelength range.

The y-axis of each panel in Figure \ref{fig:smoothed_residuals_VHELIO_sorting} corresponds to wavelength relative to the central wavelength label on the right of each plot, with blue wavelengths at the top and red wavelengths at the bottom. The x-axis is the sorted list of heliocentric velocities of the stars such that a vertical column in this figure corresponds to a single residual spectrum of a particular observation with that velocity.

By eye, some of the wavelength cutouts (e.g. Fe I, Ni I, C, Cr I, continuum) show small amplitude in smoothed residuals and not much correlation with heliocentric velocity. Other cutouts (e.g. K I, DIB, P I, sky, Na I, Yb II) show relatively strong trends in residuals with $\vhelio$. 

\begin{figure}[b]
\begin{center}
\includegraphics[width=\linewidth]{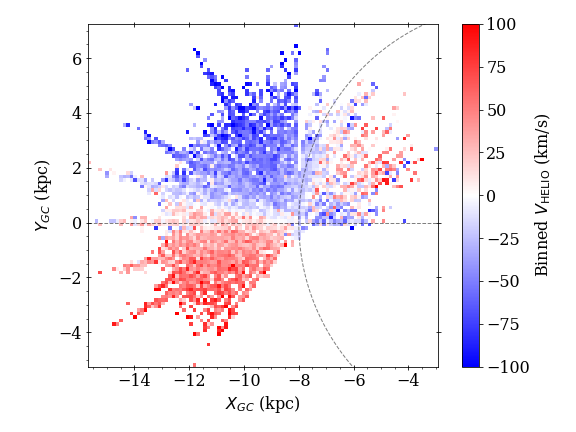}
\caption{Binned heliocentric velocity in the Galactic disk for the RC stars in the APOGEE  DR16 sample with $|Z|<0.75~\mathrm{kpc}$. The dashed circle of radius $8~\mathrm{kpc}$ and the horizontal dashed line intersect at the approximate location of the sun.}
\label{fig:stellar_vhelio}
\end{center}
\end{figure}

\begin{figure*}[ht]
\begin{center}
\includegraphics[width=\linewidth,trim={3cm 17cm 2cm 1cm},clip]{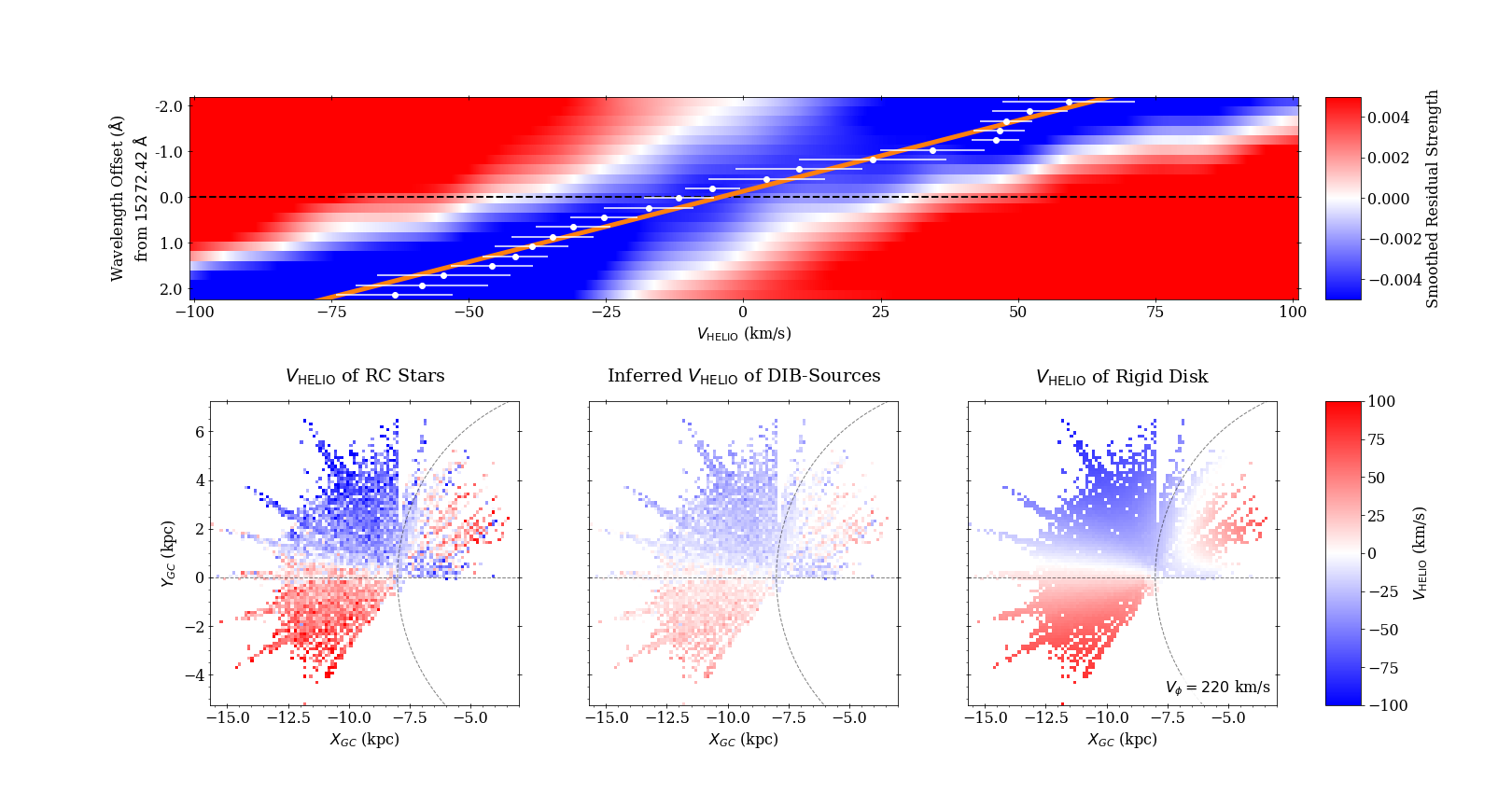}
\caption{Tracing the 15272~\AA\@ DIB feature location to estimate the velocity offset between DIB sources and RC stars. The smoothed residual data are the same as the 3rd-from-top panel of Figure \ref{fig:smoothed_residuals_VHELIO_sorting}. The white points showing the approximate location of the residual minimum at each pixel in stellar rest-frame, and the orange line shows the best fit line to these data. Using the known rest-frame wavelength of this DIB feature, the orange line is used to infer the $\vhelio$ of the DIB source as a function of stellar $\vhelio$.}
\label{fig:dib_minimum_location}
\end{center}
\end{figure*}

Focusing on the DIB panel in particular, we see a residual minimum (i.e. an absorption feature) move through the cutout as a function of heliocentric velocity. This is because the strong DIB feature, while present in many of the APOGEE spectra, appears at different wavelength locations in the stellar rest-frame spectrum as a result of the offset between the velocity of the star and the velocity of the DIB-producing source. The wavelength of a DIB feature in a star's rest-frame spectrum is given by:
\begin{equation} \label{eq:DIB_wave_loc}
    \lambda_{\mathrm{rest},*} = \lambda_{\mathrm{rest,\mathrm{DIB}}}\cdot \left(1+\frac{v_*+\Delta v}{c} \right)\cdot \left(1+\frac{v_*}{c} \right)^{-1}
\end{equation}
where $\lambda_{\mathrm{rest,\mathrm{DIB}}}$ is the wavelength of the DIB in its own rest-frame, $v_*$ is the star's heliocentric velocity, and $\Delta v$ is the LOS velocity offset between the DIB source and the star. Because DIB features will show up at different wavelength locations in the stellar rest-frame spectra, our model is not able to describe the feature and thus leaves DIB residuals behind. Similarly, features from the Earth's sky are also likely to show up as residuals because they too appear at different wavelengths in the stellar rest-frame spectra:
\begin{equation} \label{eq:Earth_wave_loc}
    \lambda_{\mathrm{rest},*} = \lambda_{\mathrm{rest,\mathrm{Earth}}}\cdot \left(1+\frac{v_*}{c} \right)^{-1}.
\end{equation}

For the Earth-based features, sorting by heliocentric velocity means that the features will occur at similar wavelength locations, which is why they appear strengthened in the velocity-sorted, smoothed residual panels. Comparing the features in the sky and the DIB panels, we also notice a difference in the horizontal width of the features; DIB features are known to be quite broad \citep[e.g. FWHMs of 3.5~\AA\@ to 6~\AA\@ for a handful of 15272~\AA\@ DIB detections in Table 1 of][]{Zasowski_2015}, so it agrees with expectations that the DIB panel residual has a visually larger width than, for example, the width of the local maximum diagonal stripe in the left half of the sky panel. Based on this difference, the narrow width of features in the P I and Ce II panels suggest these residuals are likely caused by the Earth instead of DIBs, while the large width of the residual minima stripes in the K I and Na I panels may originate from DIBs.

The particular pattern seen in the DIB panel of Figure \ref{fig:smoothed_residuals_VHELIO_sorting} is a result of the average RC heliocentric velocity having a correlation with the average ISM heliocentric velocity in the MW disk. Because the RC stars in our sample are generally quite old (average of $\sim2.8~\mathrm{Gyr}$, from Figure \ref{fig:RC_stellar_params}), they have had a longer time to kinematically decouple from the gaseous disk they were likely born in. In the Galactocentric cylindrical radius range of our stellar sample -- 95\% of the APOGEE RC sample is within $6.4<R<13.4~\mathrm{kpc}$ with a median at $10.1~\mathrm{kpc}$ -- the gaseous disk has been seen to have relatively flat rotation at $\sim220~\mathrm{km/s}$ \citep[e.g., from the HII measurements of][]{Brand_1993}. Using RC stars from APOGEE and GALAH, \citet{Khanna_2019} measure rotation velocities that are also relatively constant ($210<V_{\phi}<230~\mathrm{km/s}$) in the $6<R<11~\mathrm{kpc}$ region for $|Z|<0.75~\mathrm{kpc}$. In particular, their $R>9~\mathrm{kpc}$ measurements -- where approximately 75\% of our APOGEE RC stars fall -- shows remarkably stable $V_{\phi}\sim215~\mathrm{km/s}$\@. This suggests that RC stars can be thought of as belonging to a rigidly rotating disk that is spinning $\sim5~\mathrm{km/s}$ slower than the gaseous disk.

When we sort the DIB-based features by heliocentric velocity, we are largely sorting by azimuthal angle, as can be seen in Figure \ref{fig:stellar_vhelio}. If we assume both the RC stars and the ISM are described by rigid-rotating disks, then there is an average stellar $\vhelio$ and an average DIB $\vhelio$ for each bin in azimuthal angle; it is the relationship between the average stellar $\vhelio$ and average DIB $\vhelio$ that produces the pattern in the DIB panel. We can explore this relationship explicitly by tracing the location of the DIB minimum as a function of stellar rest-frame wavelength. This is done in Figure \ref{fig:dib_minimum_location}, where the orange best fit line to the DIB minimum location and the known rest wavelength of the DIB feature allow us to estimate the DIB source velocity as a function of stellar $\vhelio$. 

\begin{figure*}[ht]
\begin{center}
\includegraphics[width=\linewidth,trim={0cm 0cm 0cm 0cm},clip]{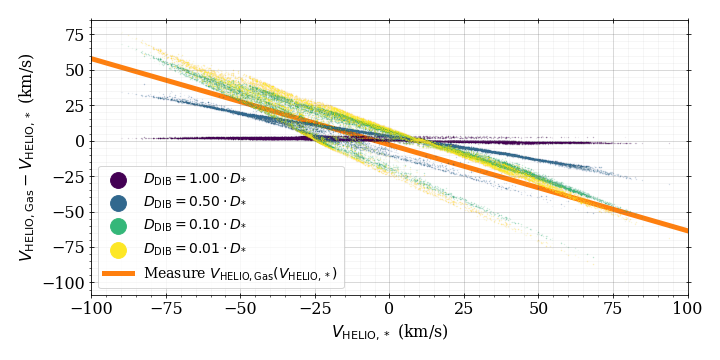}
\caption{Simulated comparison of the difference in heliocentric velocity between gas and stars when the intervening gas is at different fractions of the stellar LOS distance. The simulated stellar disk is rotating at $215~\mathrm{km/s}$ while the simulated gaseous disk is rotating at $220~\mathrm{km/s}$, and measurements are generated for each RC star in Figure \ref{fig:stellar_vhelio}. The orange line is the result of using the known DIB rest-frame wavelength with the the orange line in Figure \ref{fig:dib_minimum_location} to estimate DIB velocity as a function of stellar velocity. The agreement between the orange line and the results of placing the DIB sources between 1\% and 50\% of the distance to each star suggests that the DIB absorption feature is being produced by intervening material, as we would expect. The tight relationship between $\Delta \vhelio$ and $\vhelio$ explains why binning in stellar heliocentric velocity for the RC sample leads to increased DIB residual strength in Figure \ref{fig:smoothed_residuals_VHELIO_sorting}.}
\label{fig:dvhelio_sim_results}
\end{center}
\end{figure*}

\begin{figure*}[ht]
\begin{center}
\includegraphics[width=\linewidth]{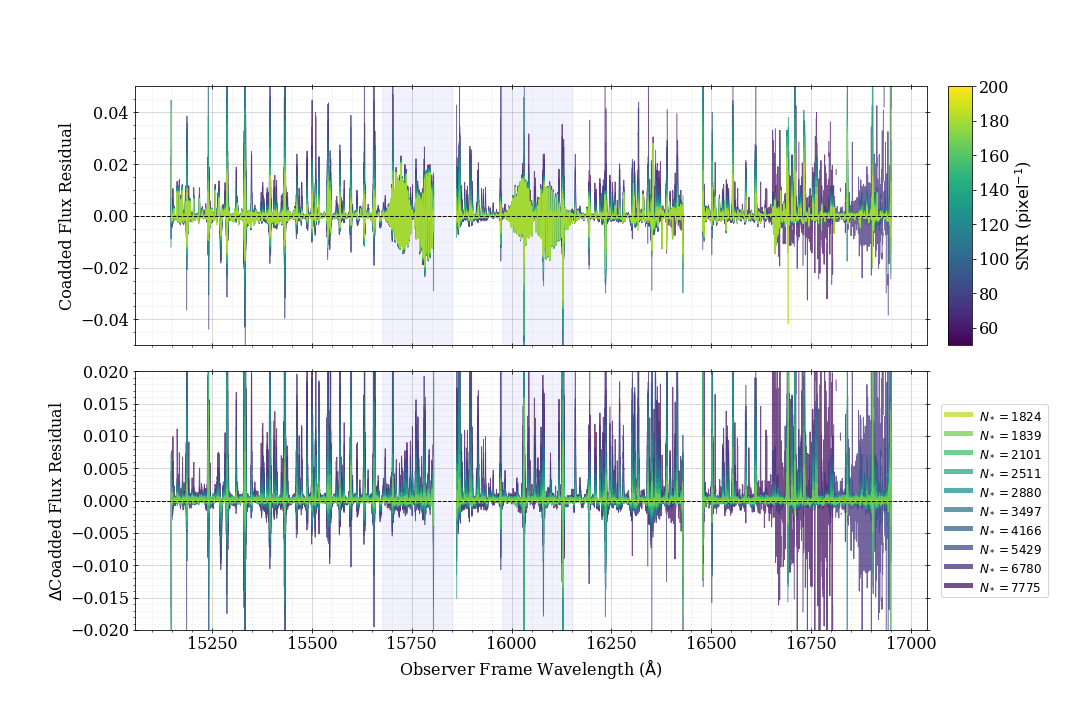}
\caption{Inverse-variance coadded residual spectra in the observer reference frame in different spectral SNR bins. The blue shaded regions highlight where the strongest telluric features (caused by CO$_2$) occur in APOGEE. The top panel shows the raw coadded residual spectra, and each combined spectrum has a similar median SNR of $\sim 550~\mathrm{pixel^{-1}}$; this panel demonstrates that the residual spikes of the lower SNR bins are larger than those of the high SNR bins. The bottom panel takes the coadded spectra from the top panel and subtracts off the combined spectrum of highest SNR bin (yellowest line in top panel) to better highlight the difference between the spectra of the SNR bins.}
\label{fig:coadded_residuals_SNR_bins}
\end{center}
\end{figure*}

We next compare the relationship we measure for gas $\vhelio$ as a function of stellar $\vhelio$ to our expectations. To that end, we assume the RC stars all belong to a rigidly rotating disk with $V_{\phi}\sim215~\mathrm{km/s}$. Similarly, we assume there is also a gaseous disk rotating at $V_{\phi}\sim220~\mathrm{km/s}$. For each RC star in Figure \ref{fig:stellar_vhelio}, we use the star's Galactic position to obtain the expected velocity vector from the rotating stellar disk, which is then transformed into a heliocentric velocity after accounting for the sun's position and motion\footnote{We use the same values for solar position and motion as \newline \citet{Khanna_2019}:  $$(X,Y,Z)_{\odot}=(-8,0,0)~\mathrm{kpc}$$ and $$(U,V,W)_{\odot}=(11.1,241.92,7.25)~\mathrm{km/s}.$$}. We follow a similar procedure for generating the simulated $\vhelio$ of the gas, but we now need to assume a distance from the sun to the DIB source along each LOS. To be agnositic of the exact DIB source distance, we choose a fraction of the total LOS distance to each star, and use those new implied Galactic coordinates and the gas disk to measure $\vhelio$ for the intervening gas. We repeat this process for distance fractions of 100\%, 50\%, 10\%, and 1\% of the total distance to each star. 

A comparison of the velocity difference between simulated gas and simulated stars is shown in Figure \ref{fig:dvhelio_sim_results}, where the data points are colored by the different fractional distances of the gas. The orange line is the result of using the best fit line in Figure \ref{fig:dib_minimum_location} with the known rest-frame wavelength of the DIB to measure heliocentric velocity of the DIB as a function of stellar $\vhelio$; to be clear, the orange line in Figure \ref{fig:dvhelio_sim_results} is not a fit to the simulated data. We see good agreement between the orange line and the simulated cases when the DIB sources are, on average, between 1\% and 50\% of the distance to the stars. This implies that the velocity offset function we measure from the location of the DIB minimum in the stellar rest-frame is caused by a source that is in the foreground of the stars, as we would expect for an ISM-based absorption feature. 

Additionally, we see that the gas velocity offset in the simulated data has a relatively tight correlation with stellar velocity. This is ultimately what causes the pattern we see in the DIB panel of the smoothed residual in Figure \ref{fig:smoothed_residuals_VHELIO_sorting}; for the RC sample, stellar velocity has an approximately linear relationship with the gas offset velocity, so the wavelength location of a DIB in the stellar rest-frame, Equation \ref{eq:DIB_wave_loc}, effectively becomes a function of stellar velocity alone. This explains why sorting by $\vhelio$ amplifies DIB signals and causes their location to move smoothly in wavelength. 

By comparing the apparent strength of residual spectral features with various parameters, we found a slight correlation in residual strength with spectral SNR. In particular, residual spectra with lower median SNR show higher levels of contamination, especially in wavelengths regions where tellurics and sky lines are known to occur. We shift the residual spectra to the observer reference frame using each observation's LOS velocity to allow for Earth-based features to align in wavelength. We then inverse-variance combine residual spectra in different spectral SNR bins, which yields the coadded spectra shown in Figure \ref{fig:coadded_residuals_SNR_bins}; the combined uncertainty in each of the SNR bins are approximately the same, with median combined SNRs of $\sim 550~\mathrm{pixel^{-1}}$. The top panel shows the coadded residual flux in 10 different SNR bins, and this panel reveals that the extrema of the low SNR bins are generally larger in magnitude than the extrema of the higher SNR bins. The bottom panel is a similar plot, but now the spectrum of the highest SNR bin has been subtracted from all the spectra of the other SNR bins to highlight the change in residuals as a function of SNR. The faint blue regions in both panels corresponds to the strongest telluric regions in APOGEE. These regions are dominated by CO$_2$ absorption, as shown in Section 6.4 and Figure 19 of \citet{Nidever_2015} using the LBLRTM model atmosphere code \citep{Clough_2005} with the US Standard 1976 atmosphere.

In even the highest SNR bins, the residual features are commonly on the order of $1\%-2\%$ of the stellar flux, and as high as $4\%-5\%$ in the extreme cases, which is similar to the peak absorption depths seen in 15272~\AA\@ DIB detections \citep[e.g. the bottom panel of Figure 3 in][]{Zasowski_2015}. The structure we observe in our residual spectra reveals that a complicated combination of Earth-based features and DIBs are common across the APOGEE wavelength range. These trends of residuals with spectral SNR and heliocentric velocity highlighted in this work are what prompted the authors of \citet{Ness_2022} to include SNR and $\vhelio$ as explanatory variables in their regression fitting to measure the relationship between abundances and stellar evolutionary state. 

We posit that the non-stellar residuals seen in Figure \ref{fig:smoothed_residuals_VHELIO_sorting} may be responsible for the larger-than-expected scatter in the some of abundance distributions measured by ASPCAP \citep[e.g.][]{Zasowski_2019,Jonsson_2020}; some elemental abundances (e.g. Na) are determined by flux measurements at a small number of pixels, so the presence of DIBs at those wavelengths may have a relatively large impact. If we attribute a large portion of the residual excess in Figure \ref{fig:residual_dist_excess} to non-stellar features, then it wouldn't be unexpected to see a large effect from features with local minima that are $\sim3\%$ of the stellar flux.

\section{Removing Sky lines and Tellurics to Identify Diffuse Interstellar Bands} \label{sec:finding_DIBs}  

\begin{figure}[b]
\begin{center}
\includegraphics[width=\linewidth]{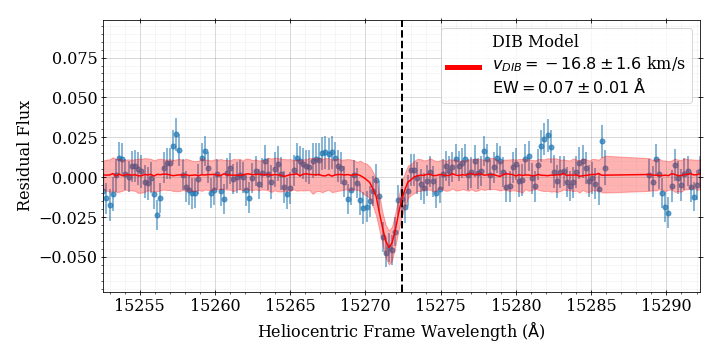}
\caption{Fitting the 15272~\AA\@ DIB feature with an inverted Gaussian to the residuals of the spectrum in Figure \ref{fig:model_comparison_example}. The blue data points show the residual data, the red line shows the best fit inverted Gaussian model with a 68\% error envelope, and the dashed vertical black line marks the rest-frame wavelength of the 15272~\AA\@ DIB feature.}
\label{fig:example_DIB_fit}
\end{center}
\end{figure}

To identify the DIBs present in the residual spectra, we first need to characterize and remove the more common contamination from Earth-based features. One advantage of isolating the Earth-based residuals versus the DIBs is that we are easily able to shift to the rest-frame of the sky line and telluric residuals. 

Because almost all the spectra have residual Earth-based features, but not all spectra have DIB contamination, it is useful to split the residual spectra up into two groups: those with and without the strongest DIB feature. Once we have these two groups, we can study the Earth-based contamination using the low-DIB-strength  group. Those results can then be used to remove the Earth-based contamination from the high-DIB-strength group. 

\begin{figure}[b]
\begin{center}
\includegraphics[width=\linewidth]{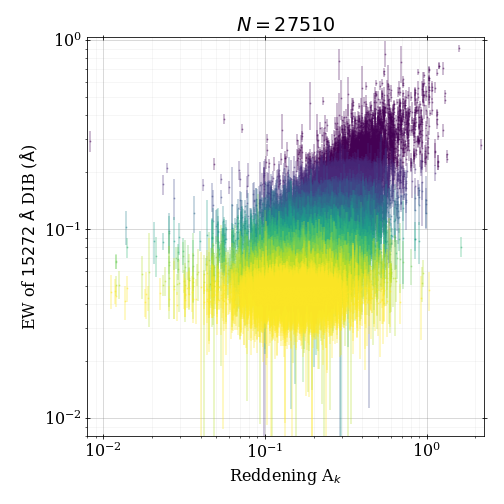}
\caption{Equivalent width strength of the 15272~\AA\@ DIB feature versus $K$-band reddening for the spectra in our sample with well-measured 15272~\AA\@ DIB features. The different colors correspond to the 10 different EW bins used later in combining the residuals. There are approximately 2100 residual spectra per bin.}
\label{fig:DIB_EW_versus_ext}
\end{center}
\end{figure}

To accomplish this, we need to rank the spectra by their DIB strength. We fit an inverted Gaussian to the strong DIB feature that is present around 15272~\AA\@ DIB in each residual spectrum, after shifting to the heliocentric frame using the $\vhelio$ for each observation; an example of this is show in Figure \ref{fig:example_DIB_fit}\footnote{A reader who is interested in exploring the APOGEE spectra in the 15272~\AA\@ DIB rest-frame can find the results of our inverted-Gaussian fitting process at \url{https://doi.org/10.5281/zenodo.10607694}}. This allows us to measure a LOS velocity and equivalent width for the strong DIB in each residual spectrum. We then use the EW measurements to define a low- and a high-DIB-strength group; residual spectra that failed to fit an absorption feature near 15272~\AA\@ DIB were automatically assigned to the low-DIB-strength group, and then a threshold of $\mathrm{EW}=1\times10^{-2}~\mathrm{\AA}$ was chosen to divide the spectra with DIB detections to ensure an approximately equal number of spectra ($\sim 21000$) in the low- and high-DIB-strength groups. For the residual spectra in the high-DIB-strength group, the 15272~\AA\@ DIB EW is compared to the $K$-band reddening\footnote{The $A_K$ are provided by the ASPCAP data tables; for the RC stars in our sample, approximately 80\% of the $A_K$ measurements are found using the Rayleigh-Jeans Color Excess method, while the remaining 20\% are estimated using the SFD dust maps of \citet{Schlegel_1998}.} in Figure \ref{fig:DIB_EW_versus_ext} with data points colored by the EW bins used in our later analysis. 

For the low-DIB-strength group, we fit a 2nd-order polynomial model to the observer-frame residual fluxes at each pixel using a process similar to what is described in Section \ref{ssec:modelling_math}. In this case, however, the stellar label vector in Equation \ref{eq:stellar_label_vector} is replaced by 
\begin{equation} 
    \vec x_{j} = \begin{pmatrix}
                    \mathrm{SNR}_{j}^2 \\
                    \mathrm{SNR}_{j} \\
                    1
                \end{pmatrix}
\end{equation}
and the data we are fitting are the residuals in the observer frame. This model enables us to quantify the predictive power of SNR alone to describe the spectral residuals that are not captured by our five label model from Section \ref{ssec:modelling_math}.

\begin{figure*}[ht]
\begin{center}
\includegraphics[width=\linewidth,height=9in,keepaspectratio,trim={0cm 5cm 3cm 4.5cm},clip]{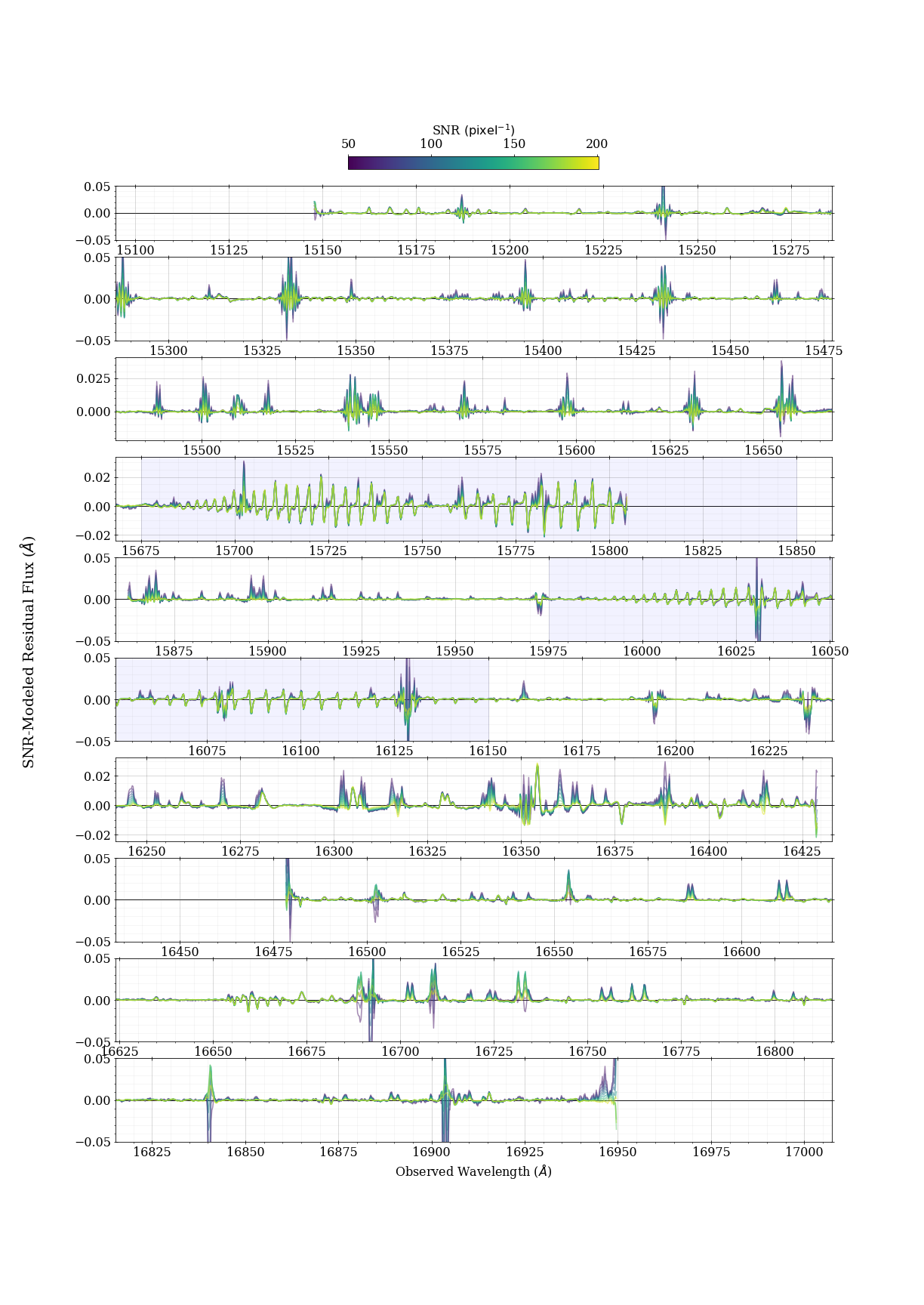}
\caption{SNR-modeled residual spectra in the observer rest-frame. The colors of the spectra in this figure correspond to the SNR values in the colorbar above the top panel, and not the EW bins in Figure \ref{fig:DIB_EW_versus_ext}. The blue shaded regions highlight where the strongest telluric features (caused by CO$_2$) occur in APOGEE.}
\label{fig:SNR_modeled_residuals}
\end{center}
\end{figure*}

Examples of the resulting best-fit model for different SNR bins are shown in Figure \ref{fig:SNR_modeled_residuals}. Overall, this model captures much of the behaviour we see in the coadded residuals of Figure \ref{fig:coadded_residuals_SNR_bins}. There are a few regions that stand out visually (e.g. in the blue shaded regions where CO$_2$ telluric features are particularly strong) that have a saw-toothed shape to the residuals; this is a shape that can be created by subtracting two Gaussians with the same width but a slight offset in mean. We suggest that these Earth-based residuals are caused by a wavelength mismatch between the raw spectra and the telluric models being subtracted. To be clear, these stacked residual spectra are produced by stacking a couple thousand spectra together, which means that any issues of random error in the individual wavelength solutions should be minimal. Instead, these stacked spectra are revealing systematic offsets in the wavelength solution from the telluric model.

As shown in \citet{Nidever_2015}, the APOGEE data reduction pipeline's telluric and sky removal performs very well in most cases. \citet{Holtzman_2018} further shows improvement in APOGEE's Earth-feature removal; however, they also point out that the reduction pipeline's approach is to flag and ignore pixels near particularly strong telluric and sky lines instead of a more computationally expensive approach that might achieve smaller residuals. Of particular interest, Figure 3 of \citet{Holtzman_2018} reveals the same saw-tooth residuals that we see in our Figure \ref{fig:SNR_modeled_residuals}, with their figure focusing on the $15680-15800~\mathrm{\AA}$ region that is sensitive to tellurics from CO$_2$. While the APOGEE reduction process works well for their science goals, our analysis has shown that detailed accounting of Earth's atmosphere's light can reveal previously-hidden information.  

By looking at the wavelengths and shapes of the Earth-based features in Figure \ref{fig:SNR_modeled_residuals}, we can see that some of the smoothed residual patterns in Figure \ref{fig:smoothed_residuals_VHELIO_sorting} are likely explained by the night sky. The Ce II panel of Figure \ref{fig:smoothed_residuals_VHELIO_sorting}, for example, shows a repeating alternation of residual fluxes from low to high, which is visually dissimilar to the DIB panel which shows a single feature moving through the cutout window. The Ce II panel's central wavelength of 15789.1~\AA\@ places it in a region of saw-toothed features in Figure \ref{fig:SNR_modeled_residuals}, so this mottled pattern that appears after sorting by heliocentric velocity is likely the result of these saw-tooth shapes constructively and destructively combining.  

While this SNR-dependent Earth-residual model performs well in describing the observer-frame residuals, we choose to not use this model in our subsequent DIB analysis without extensive further testing. Instead, we remove sky line and telluric features by pairing up residual spectra in the low- and high-DIB-strength groups based on spectral SNR. That is, for each high-DIB-strength residual spectrum, we identify a low-DIB-strength residual spectrum with median spectral SNR that is within $5~\mathrm{pixel}^{-1}$ of the high-DIB-strength spectrum's SNR. We ensure that each low-DIB-strength residual spectrum is used only once. 

Once each high-DIB-strength residual spectrum has a corresponding low-DIB-strength residual spectrum, we subtract each pair and add their uncertainties in quadrature in the observer frame. This has the effect of removing the Earth-based residuals from the high-DIB-strength residual spectra while leaving the possible DIB features relatively untouched. The subtracted spectra are then shifted from the observer frame to the 15272~\AA\@ DIB rest-frame using the velocity measured in the inverted-Gaussian fitting for the high-DIB-strength spectrum in the pair. Because each pair of spectra are likely not taken at the same exact time or with the same observing conditions, their subtraction may not perfectly remove Earth-based features in each resulting spectrum. However, averaging these subtracted spectra together does remove the vast majority of obvious telluric features. Because our subsequent analysis is based on combinations of spectra, we believe this approach is a simple and model-free method of dealing with telluric contamination. Furthermore, we find that our final results are insensitive to the particular pairings of low- and high-DIB-strength spectra, so long as the two spectra are similar enough in SNR.

The next step is to break the spectra up into different bins based on the strength of the 15272~\AA\@ DIB. We define 10 DIB-strength bins such that approximately 2100 spectra are in each bin; these are shown by the different colors in Figure \ref{fig:DIB_EW_versus_ext}. We then combine the residual spectra in these bins using population fitting to measure a population mean and uncertainty/standard deviation at each pixel. The precise statistics used for this combination are explained in Appendix \ref{sec:spectral_coaddition}, but the key takeaway is that the population fitting takes into account both the uncertainty in each residual measurement as well as the spread in those measurements to give back realistic means and uncertainties at each pixel. 

\begin{figure*}[ht]
\begin{center}
\includegraphics[width=\linewidth,height=8.5in,keepaspectratio]{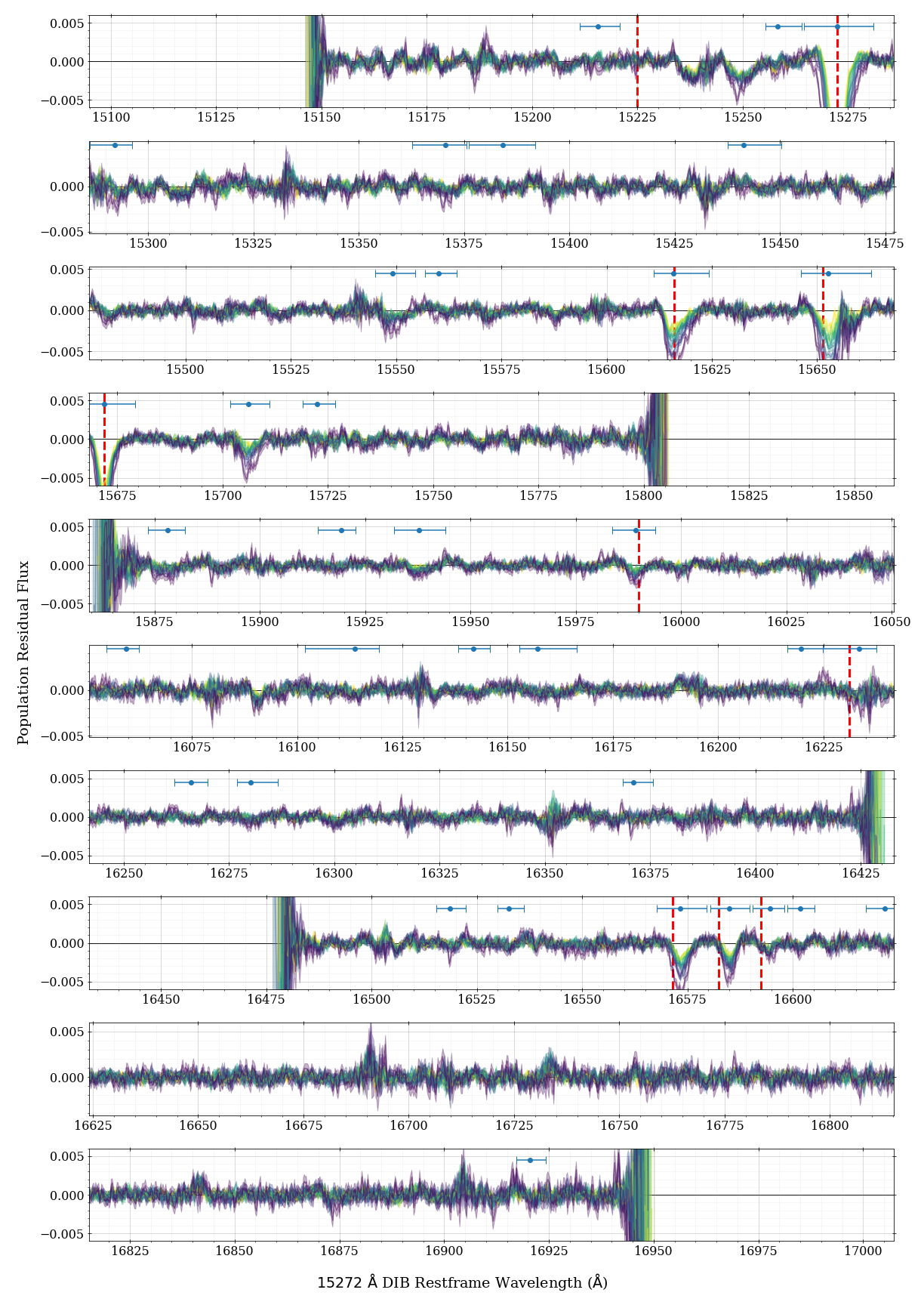}
\caption{Coadded residual spectra in the 15272~\AA\@ DIB rest-frame colored by the 15272~\AA\@ DIB strength as shown in Figure \ref{fig:DIB_EW_versus_ext}. The vertical red lines show the rest-frame wavelengths of the previously-known DIB features in Table \ref{tab:previous_DIBs}. There are many newly discovered features in emission and absorption that scale as a function of EW bin, making them new candidate DIB features. Thirty five of the possible DIBs we detect -- which we will later measure to have significant correlation with the 15272~\AA\@ DIB strength -- are shown by the blue points above the spectra with widths showing the region we use as a cutout when measuring equivalent widths.}
\label{fig:coadded_specrta_EW_binned}
\end{center}
\end{figure*}

The results of this population-fitting process are shown in Figure \ref{fig:coadded_specrta_EW_binned}, where the colors denote the same DIB-strength bins as in Figure \ref{fig:DIB_EW_versus_ext}. The vertical red lines show the locations of the previously-known DIBs in Table \ref{tab:previous_DIBs}; by eye, the increasing DIB-strength bins show increasing depth of features at most of these wavelengths. For the 15225~\AA\@ feature, the large literature uncertainty in central wavelength may suggest that the true central wavelength of that DIB should correspond to one of the features near 15250~\AA\@ instead. We also notice additional absorption features (e.g. slightly redward of 15700~\AA\@) that have shapes similar to the previously-known DIBs. 

We recognize that all the DIB features are likely not in the same rest-frame as the 15272~\AA\@ DIB, as they might be produced by different intervening clouds along the sight-line to each star and therefore have different relative LOS velocities. However, the velocity difference between DIB sources along a single LOS is not likely to be large enough to shift the wavelength by more than a few APOGEE pixels. For instance, the vast majority of the 15272~\AA\@ DIB velocities we measure are in the $\pm 30~\mathrm{km/s}$ range, so at $\sim 16000~\mathrm{\AA}$\@, a $10~\mathrm{km/s}$ velocity offset would manifest as a $\sim 0.5~$\AA\@ offset in wavelength. When we combine spectra in each DIB-strength bin, we may slightly reduce the signal of DIB features from other sources because of a combination of wavelength offsets and their strength not correlating with the 15272~\AA\@ strength, but we still expect to detect their presence. Our decision to use the 15272~\AA\@ DIB rest-frame will, however, preserve the strength of features that are being produced by the 15272~\AA\@ DIB source, so this technique is particularly useful for identifying DIBs that correlate with the strongest DIB in the APOGEE wavelength range. For future work where extremely precise central wavelengths are needed or the goal is to find every possible DIB in the APOGEE wavelength range, one approach would be to cross-correlate the residual spectra using different wavelength cutouts to find the velocities that amplify the signals of all DIBs. 

To identify possible locations of DIBs, we smooth the highest DIB-strength residual spectrum (i.e. the darkest purple line in Figure \ref{fig:coadded_specrta_EW_binned}) with a Gaussian kernel width of 5 pixels; after some visual vetting, this yields \npossdibs\@ local maxima and minima as possible DIBs in emission and absorption. Of the possible DIBs, we highlight the locations of 35 features in Figure \ref{fig:coadded_specrta_EW_binned} that we will ultimately find are likely produced by the same source that produces the 15272~\AA\@ DIB; we choose to not show all \npossdibs\@ possible features for visual clarity. As expected, this step finds features nearby to all of the previously-known DIBs. Many of the local extrema we find are likely spurious detections, but we choose to err on the side of testing too many extrema versus applying a more complicated thresholding at the detection step. For each local extremum, we then step redward and blueward from that wavelength in the smoothed spectrum until the slope is near 0 to define a useful wavelength region around each feature. We then manually check these regions to confirm that they cover the entirety of an apparent feature, tweaking the boundaries where necessary. This defines the wavelength cutouts we will later use for measuring the EW strength of the possible DIBs and for quantifying DIB detection probabilities. The list of possible features agree well with the features we find when we smooth a population-combined spectrum using all $\sim 21000$ residual spectral pairs; we choose to use the feature locations from the highest DIB-strength spectrum instead of the total population because the signal from true DIBs will be strongest in the high DIB-strength bin. 

As mentioned in the last few paragraphs, the central wavelengths we measure in the 15272~\AA\@ DIB rest-frame for DIBs that are produced by a different source may be offset from their true value by a few pixels, but this is still an improvement on the wavelength location for a few of the previously-known DIBs in Table \ref{tab:previous_DIBs}. Additionally, because the spectra in each DIB-strength bin are a combination of $\sim2100$ individual residual spectra, a systematic velocity offset between different DIB sources would be required to produce a significant shift in the central wavelength location. This implies that our central wavelength measurements are likely within a handful of pixels of the true rest-frame wavelength of each DIB feature. To be conservative, we estimate the uncertainty on the central wavelengths of the possible DIBs to be $\sim1~\mathrm{\AA}$\@. 

\begin{figure*}[ht]
\begin{center}
\includegraphics[width=\linewidth]{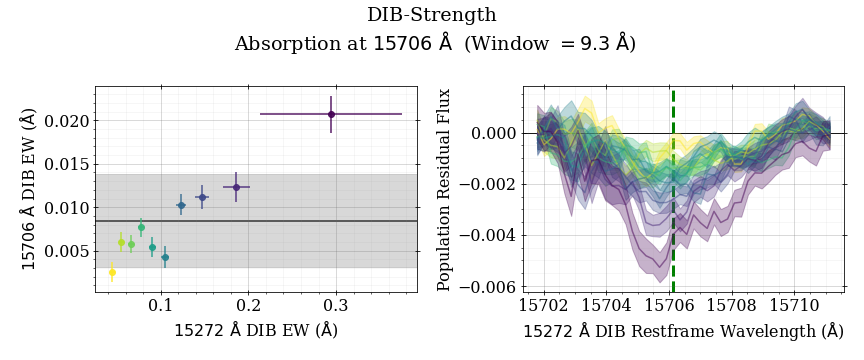}
\caption{Equivalent width strength of a previously-unknown DIB feature near 15706~\AA\@ for the different EW bins shown in Figure \ref{fig:DIB_EW_versus_ext}. The ``Window'' in the title defines the width of the wavelength cutout. The dashed vertical green line in the right panel shows the location of the local minimum identified in the smoothed highest DIB-strength residual spectrum, which is slightly redward of the feature minimum due to the feature's assymetric shape. The black horizontal line in the left panel and the corresponding grey shaded region are the population mean and width of the current DIB EW measurements (i.e. population fit to the y-axis values). This particular feature's EW shows a fairly strong correlation with the strength of the 15272~\AA\@ DIB.}
\label{fig:example_EW_binned_residual_EW_measures}
\end{center}
\end{figure*}

We next measure the equivalent width of the local extrema in each of the DIB-strength bins of Figure \ref{fig:coadded_specrta_EW_binned}. Using the wavelength cutouts defined above, we empirically integrate the flux. Specifically, we use the 5 pixels at the edges of each cutout (3 inside, 2 outside of the cutout) to define an average flux and wavelength on the blue and red edge; these blue and red average fluxes define a line for each DIB-strength bin that we use at the local continuum measurement. After subtracting off the local continuum line for each bin, we perform a trapezoid numerical integration to measure the area of the feature. To propagate uncertainties in the flux, we repeated sample flux measurements -- including the fluxes used to measure the local continuum line -- to get a distribution on the empirical EW for each bin. The EW measurements we report are the median and standard deviation of those realizations. An example of the resulting EW measurements is shown in the left panel of Figure \ref{fig:example_EW_binned_residual_EW_measures} for the possible DIB near 15706~\AA\@, with a cutout of the feature shown in the right panel for the different DIB-strength bins. We note that the nearest Earth-based feature is a sky emission line at 15720.6~\AA\@, which is sufficiently far away from the new 15706~\AA\@ DIB to have a negligible impact on our results.

To characterize how much the EW measurements change between the bins, we perform a population fit to the EW measurements of the possible DIB feature (e.g. the y-axis values of the left panel); this population fit again follows the approach detailed in Appendix \ref{sec:spectral_coaddition}, and an example of the resulting medians on the population mean and width are shown with a black line and grey region in the left panel of Figure \ref{fig:example_EW_binned_residual_EW_measures}. We are particularly interested in the population width because a large width indicates that the EW measurements between the bins are quite different. 

Of course, each possible DIB feature may only be a spurious detection; that is, a feature may only show up in the highest DIB-strength bin by random chance alone. To determine the significance of a detection, we repeat the binning, combining, and EW-measuring process, but this time, we bin randomly. To be clear, instead of using the strong-DIB-strength ranking to bin our spectra, we assign spectra randomly into 10 different bins, with the same $\sim2100$ spectra per bin. Using random binning, we expect that the features in each bin will be quite similar to each other, as will their EW measurements. 

\begin{figure*}[ht]
\begin{center}
\includegraphics[width=\linewidth]{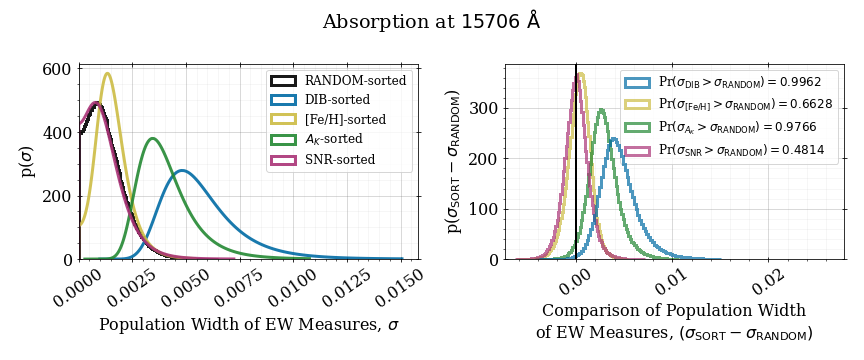}
\caption{ Comparison of the population width distributions of EW measurements around the possible DIB absorption feature shown in Figure \ref{fig:example_EW_binned_residual_EW_measures} when sorted and binned by different parameters. In each case, we measure the EW of the possible DIB feature from the combined residual spectra in different parameter bins, and then measure the population width of the EW measurements (e.g. the shaded grey region of Figure \ref{fig:example_EW_binned_residual_EW_measures}). \textbf{Left:} Distributions on the EW population widths for different parameter bins; the black histogram is the resulting average of 5 realizations of random sorting and binning. \textbf{Right:} Distribution on the difference between the population widths in the left panel and the random population width. In this case, the largest population width is achieved after binning by DIB-strength, suggesting that the absorption feature in Figure \ref{fig:example_EW_binned_residual_EW_measures} is best correlated with 15272~\AA\@ DIB-strength.}
\label{fig:example_population_width_distribution_comparison}
\end{center}
\end{figure*}

To quantify exactly how strong a detection is (i.e. its significance above randomness), we also measure the population width of the EW measurements from random binning. In particular, we are interested in how the population width distributions compare between the random binning and the binning by the strong-DIB-strength cases. To account for the variation in the random binning that might occur by chance alone, we repeat the random binning 5 times. The resulting population width distributions of the populations agree quite well between the realizations for each possible feature. This implies that our results aren't overly sensitive to the particular choice in random binning. To be careful, however, we average the population width distributions from the 5 random realizations together. Measurements from this point on that refer to ``random'' come from the ``averaged random'' results.

Recognizing that strong-DIB-strength might not be the only/best parameter to explain the change in a feature's strength between bins, we explore a few additional binning options: $\feh$, median spectral SNR, and $A_K$ reddening. Respectively, these can be thought of as testing for features that are the result of residual chemical information that the model did not capture, remaining Earth-based contamination or other SNR effects, and DIBs that correlate with the amount of LOS dust but don't originate from the 15272~\AA\@ DIB source. As before, we sort by a given parameter and then bin the spectra so that there are approximately 2100 spectra per bin. In every case, we measure EWs in each bin at each possible feature, and then use those EWs to measure a population width distribution, which are then compared to the results from random binning. An example of these population width distributions for the different sorting parameters is shown in the left panel of Figure \ref{fig:example_population_width_distribution_comparison}; in the right panel, samples from the random distribution have been subtracted from samples in the other distributions to get a distribution of $\sigma_{\mathrm{SORT}}- \sigma_{\mathrm{RANDOM}}$.

\begin{figure*}[ht]
\begin{center}
\includegraphics[width=\linewidth]{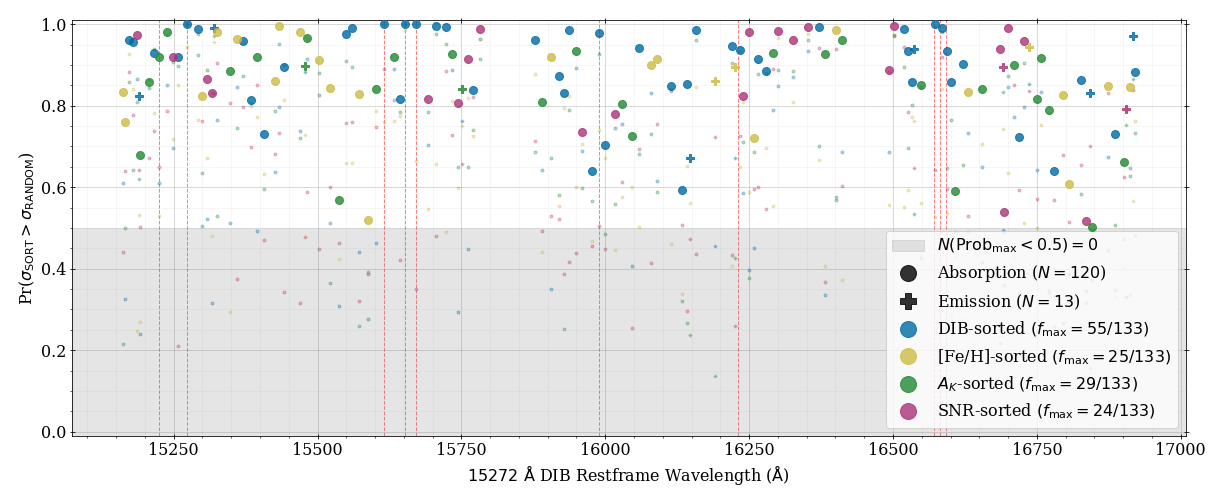}
\caption{Summary of the probabilities that a particular sorting parameter's population width of EW measurements is greater than random chance can explain (e.g. the legend values in the right panel of Figure \ref{fig:example_population_width_distribution_comparison}) for all possible DIB features. The possible DIB wavelengths are our best estimates of each feature's central wavelength; the exact locations of the features that are best described by $A_K$ or DIB-strength sorting are given provided in Table~\ref{tab:new_DIB_summary}. There are \nfinalRANDdibs\@ possible DIB features where the maximum probability of all sorting parameters is less than 50\%, meaning that none of the features are best explained by chance alone. At each possible DIB wavelength, the four sorting parameter's probabilities are plotted in a vertical line; the sorting parameter with the maximum median population width is enlarged compared to the others. Of the \npossdibs\@ possible DIB features, \nfinalSTRONGdibs\@ show the largest difference from random when binned by $15272~\mathrm{\AA}$\@ DIB strength, another \nfinalAKdibs\@ features are potentially DIBs produced by sources other than the $15272~\mathrm{\AA}$\@ DIB, \nfinalSNRdibs\@ features may originate from SNR effects, and \nfinalFeHdibs\@ features are best explained by $\feh$ binning which may be a result of the model imperfectly describing the data.}
\label{fig:possible_DIBs}
\end{center}
\end{figure*}

We can then directly integrate these population width difference distributions to measure the probability that the population width for a given sorting parameter is greater than the population width from random sorting. In cases where none of the probabilities are greater than 50\%, we decide that randomness alone likely produced the observed feature. We also use the medians of these distributions to determine which sorting parameter yields the greatest difference from random, implying that a particular feature is best explained by that parameter. Because the relationship between each feature's EW measurements and a given sorting parameter is potentially complicated (i.e. not a simple functional form), we argue that our approach is a cautious, statistically-motivated way of making detections without assuming a parameterized relationship. We are simply answering the question: ``Which of the sorting parameters produces the largest difference in EW measurements between the bins?'' If a feature has its best sorting from DIB-strength, it may suggest that feature is produced by the same source as the 15272~\AA\@ DIB. If the best sorting is $A_K$, it may be that feature is still truly a DIB, but that it has a different source than the 15272~\AA\@ DIB.

We summarize the probabilities from all the sorting parameters for all \npossdibs\@ possible features in Figure \ref{fig:possible_DIBs}. At our best estimate of the wavelength for each feature, a vertical line of 4 points are plotted to show the probability above random for each of the different sorting parameters; the point with the highest median population width is enlarged compared to the others, provided that the maximum probability above random is greater than 50\% (i.e. there is only one enlarged point per feature). These probabilities reveals that \nfinalRANDdibs\@ of the features are best explained by random chance alone (i.e. spurious detections), while the other \nfinalSTRONGdibs, \nfinalFeHdibs, \nfinalAKdibs, and \nfinalSNRdibs\@ features correlate best with the 15272~\AA\@ DIB strength, $\feh$, $A_K$, and spectral SNR respectively. Comparing to the previously-measured DIBs (vertical red lines), we recover $> 93\%$ detection probabilities in each case. Apart from the 15225~\AA\@ DIB -- which finds best sorting using $A_K$ -- all remaining previously-measured DIBs in Table \ref{tab:previous_DIBs} have best sorting from DIB-strength. These results may suggest that most of the previously-known DIBs in APOGEE are produced by the same source/species along each LOS while the remaining DIB has a different origin. 

While we may not have perfectly captured the central wavelength or cutout region around each of the possible features, we emphasize that our process is likely biased towards calling a true DIB spurious than the other way around. A more rigorously-defined cutout region and fitting of the local continuum would likely yield stronger detection signals than we report, suggesting there may be even more DIB features than we discover. Highly-detailed analyses in the future may reveal DIBs that are not included in our results, and possible DIBs we classify as having non-ISM origin may be found to have significant signal. 

\begin{table*}[t]
\caption{Number of the \npossdibs\@ possible DIBs with best sorting parameter detection probabilities above different thresholds.
\label{tab:number_of_DIBs}}
\centering
\begin{tabular}{cccccc}
\hline \hline
Probability & \multicolumn{4}{c}{Number of Detected DIBs by SORT Parameter} & Number below \\ 
\cline{2-5}
Threshold & DIB-strength & $\feh$ & $A_K$ & SNR & Threshold \\ \hline
0.50 & 55 & 25 & 29 & 24 & 0 \\
0.68 & 51 & 23 & 24 & 22 & 13 \\
0.95 & 22 & 5 & 3 & 9 & 94 \\
0.99 & 11 & 1 & 0 & 3 & 118 \\
\hline
\end{tabular}
\end{table*}

In general, the average probability above random for detected features is higher for the DIB-strength sorting compared to the other sorting parameters (e.g. mean $\mathrm{Pr}_{\mathrm{DIB}} = 0.89$ versus mean $\mathrm{Pr}_{\mathrm{A_K}} = 0.83$). This is likely a consequence of our choice to bin in the 15272~\AA\@ DIB rest-frame; as discussed previously, DIBs from other sources with slightly different rest-frames will experience a reduced signal, so their difference from random is not as strong. Table \ref{tab:number_of_DIBs} lists the number of features in each bin with detection probabilities (i.e. probability above random) above various thresholds. This is equivalent to asking how many of the enlarged points persist above changing heights of the shaded grey region in Figure \ref{fig:possible_DIBs}. We notice that the number of detected features in the $A_K$ sorting drops off faster than the number of features in the DIB-strength bin, and this again is likely caused by binning in the 15272~\AA\@ DIB rest-frame.

Linking the resulting DIBs back to the features seen in Figure \ref{fig:smoothed_residuals_VHELIO_sorting}, many of the element windows occur near newly-detected DIBs. The Na I panel, for instance, is near a detected feature at 16382.1~\AA\@ that is best explained by $A_K$ sorting with 93\% probability above random. This feature likely explains the diagonal streak in the smoothed residuals of the Na I panel. It may also be responsible for increasing the scatter in Na abundances that ASPCAP reports, emphasizing the need to account for DIBs in future stellar abundance pipelines. 

To summarize the patterns in the $\vhelio$-sorted, smoothed residuals, the Fe I, Ni I, C, and Cr I panels show minimal residual features suggesting these windows are free of significant non-stellar light. Like the DIB panel, the K I, Mn I, Co II, Na I, and Yb II panels have nearby detected DIB features that likely explain their broad diagonal stripes. The P I, Mg II, Ce II, Cu I, Al I, and even the continuum panels have narrower diagonal stripes like the sky panel and have no significant nearby detected DIBs, suggesting that these patterns are almost exclusively explained by Earth-based residuals. The O panel is unique in that it shows a mix of both DIB signal (i.e. broad emission near the right edge) and telluric residuals (i.e. narrower diagonals near the middle). 

As an initial step towards identifying the chemical species that producing each DIB, we compare the central wavelength locations to known hydrogen transitions in the APOGEE wavelength range. The newly-detected DIB in Figure \ref{fig:example_EW_binned_residual_EW_measures}, for instance, is within $1~\mathrm{\AA}$ of the Brackett series $n=15$ to $n=4$ transition ($\lambda_0=15705.0~\mathrm{\AA}$), suggesting that atomic hydrogen in the ISM is a probable source of this feature. That we are able to detect features at many known hydrogen wavelengths without a priori searching there bodes well for our general methods. 

A complete summary of the \nfinaldibs/\npossdibs\@ DIBs whose best sorting parameter is either DIB-strength or $A_K$ is listed in Table \ref{tab:new_DIB_summary}. \nfinaldibsEMISSIONtextupper\@ of the features are in emission and the remaining \nfinaldibsABSORPTIONtextlower\@ are in absorption. All wavelengths are given in the 15272~\AA\@ DIB rest-frame, and the Wavelength Range column gives the window we used to measure the EWs; the continuum wavelengths are defined by the 5 pixels nearest to the values in the Wavelength Range column. By summing the probabilities above random, we expect that \nfinaldibsPROB\@ of the \nfinaldibs\@ DIB detections are truly DIBs. For features that have central wavelengths within $2~\mathrm{\AA}$\@ of a known hydrogen recombination line, we give the name of the series and level. Many of these newly-discovered DIBs occur in the same wavelength regions that were obscured by incomplete sky line and telluric removal; this, combined with their sub-percent sizes, explains why our analysis has been able to reveal these features for the first time. 

\section{Summary} \label{sec:summary}

We have created data-driven models of RC stellar spectra using $\sim 5.5\times 10^5$ individual observations of $\sim 1.7\times 10^5$ stars from the APOGEE dataset. 
The modeling uses five parameters -- $\teff$, $\logg$, $\feh$, $\alphafe$, and age -- to predict the spectra, and the resulting models agree quite well with the data (residual Gaussian mean near 0 and width of $\sim 1.16 \sigma$) across the APOGEE wavelength range. 
This implies that these five labels are sufficient to explain the majority of information present in the spectra. Consequently, there is not a substantial amount of residual information that may be leveraged for pursuits such as chemical tagging. Though the residuals of the data-minus-model are relatively small ($\sim3\%$ of stellar flux on average), it is very important to understand and isolate their astrophysical origin. We discover that there are many pixels where the residuals suggest that a significant number of non-stellar features are also present in the stellar spectra. We identify which of these features are likely Earth-based and which are likely Diffuse Interstellar Bands. We find \nfinaldibs\@ possible DIBs in APOGEE spectra that have less than 50\% probability of appearing by chance alone, including all previously-discovered DIBs in this wavelength region. 

Our key results include:
\begin{enumerate}
    \item The residuals of our data compared to the model show correlations with heliocentric velocity, which we show is evidence that many of these residual features are not in the stellar rest-frame, such as sky lines, tellurics, and DIBs. These residual features appear at the level of 3\% of the stellar flux on average (Section \ref{ssec:modelling_math}, Figure \ref{fig:residual_dist_excess}; Section \ref{sec:stucture_in_residuals}, Figures \ref{fig:residual_distribution_comparison}, \ref{fig:smoothed_residuals_VHELIO_sorting}, and \ref{fig:coadded_residuals_SNR_bins});
    \item The size of Earth-based residuals appear anti-correlated with spectral SNR. The shape of these residuals suggest that they may be removed by correcting for a wavelength offset between the sky model and raw observations (Section \ref{sec:finding_DIBs}, Figure \ref{fig:SNR_modeled_residuals});
    \item After removing the Earth-based residuals, we combine residual spectra in the rest-frame of the strongest DIB in the APOGEE wavelength range ($\lambda_0 = 15272$~\AA). We detect \nfinaldibs\@ DIB features in absorption (including all of previously-measured DIBs) and emission that show highest correlation in strength with either $K$-band reddening ($A_K$) or EW strength of the 15272~\AA\@ DIB feature (Section \ref{sec:finding_DIBs}, Figures \ref{fig:coadded_specrta_EW_binned} and \ref{fig:possible_DIBs}, Tables \ref{tab:number_of_DIBs} and \ref{tab:new_DIB_summary}). 
\end{enumerate}

Future work based on our results will focus on measuring the impact that unaccounted-for DIB and Earth-based features have on ASPCAP-measured abundances. It would also be worthwhile to compare the DIB wavelengths to lines produced by chemical species (i.e. more than just atomic hydrogen) that are known components of the ISM, as well as to correlate the new DIB strengths with additional tracers of ISM density. Another useful undertaking that would advance this work is detailed joint modelling of the DIBs, sky lines, tellurics, and stellar spectra in APOGEE, such as employed by the \texttt{MADGICS} method of \citet{Saydjari_2023} using Gaia spectra. A complete understanding of all sources of spectral features is necessary for chemical tagging experiments, and this will have the added benefit of improving our understanding of the ISM. 

\begin{acknowledgments}

The authors thank the anonymous referee for comments that helped improve the clarity of this paper. KM, CMR, and PG were supported by NSF Grant AST-2206328. We thank Julianne Dalcanton and the Center for Computational Astrophysics Astrophysical Data Group for useful discussions. KM thanks his PhD advisors, CMR and PG, for enthusiastically supporting his decision to pursue a Designated Emphasis in Statistics, without which, this work would not have been possible.

\end{acknowledgments}

\vspace{5mm}
\facilities{APO, LCO, SDSS}

\software{Astropy \citep{astropy_2013,astropy_2018,astropy_2022}, Bovy's \texttt{APOGEE} Code \citep{Bovy_2014,bovy_2016}, corner \citep{corner_citation}, emcee \citep{emcee_citation}, IPython \citep{ipython_citation}, jupyter \citep{jupyter_citation}, matplotlib  \citep{matplotlib_citation}, numpy \citep{numpy_citation}, Price-Jones' \texttt{SPECTRAL SPACE PYTHON} Code (\url{https://github.com/npricejones/spectralspace}), scipy \citep{scipy_citation}}

\bibliography{refereed_arxiv}{}

\begin{thebibliography}{}
\expandafter\ifx\csname natexlab\endcsname\relax\def\natexlab#1{#1}\fi
\providecommand{\url}[1]{\href{#1}{#1}}
\providecommand{\dodoi}[1]{doi:~\href{http://doi.org/#1}{\nolinkurl{#1}}}
\providecommand{\doeprint}[1]{\href{http://ascl.net/#1}{\nolinkurl{http://ascl.net/#1}}}
\providecommand{\doarXiv}[1]{\href{https://arxiv.org/abs/#1}{\nolinkurl{https://arxiv.org/abs/#1}}}

\bibitem[{{Astropy Collaboration} {et~al.}(2013){Astropy Collaboration},
  {Robitaille}, {Tollerud}, {Greenfield}, {Droettboom}, {Bray}, {Aldcroft},
  {Davis}, {Ginsburg}, {Price-Whelan}, {Kerzendorf}, {Conley}, {Crighton},
  {Barbary}, {Muna}, {Ferguson}, {Grollier}, {Parikh}, {Nair}, {Unther},
  {Deil}, {Woillez}, {Conseil}, {Kramer}, {Turner}, {Singer}, {Fox}, {Weaver},
  {Zabalza}, {Edwards}, {Azalee Bostroem}, {Burke}, {Casey}, {Crawford},
  {Dencheva}, {Ely}, {Jenness}, {Labrie}, {Lim}, {Pierfederici}, {Pontzen},
  {Ptak}, {Refsdal}, {Servillat}, \& {Streicher}}]{astropy_2013}
{Astropy Collaboration}, {Robitaille}, T.~P., {Tollerud}, E.~J., {et~al.} 2013,
  \aap, 558, A33, \dodoi{10.1051/0004-6361/201322068}

\bibitem[{{Astropy Collaboration} {et~al.}(2018){Astropy Collaboration},
  {Price-Whelan}, {Sip{\H{o}}cz}, {G{\"u}nther}, {Lim}, {Crawford}, {Conseil},
  {Shupe}, {Craig}, {Dencheva}, {Ginsburg}, {Vand erPlas}, {Bradley},
  {P{\'e}rez-Su{\'a}rez}, {de Val-Borro}, {Aldcroft}, {Cruz}, {Robitaille},
  {Tollerud}, {Ardelean}, {Babej}, {Bach}, {Bachetti}, {Bakanov}, {Bamford},
  {Barentsen}, {Barmby}, {Baumbach}, {Berry}, {Biscani}, {Boquien}, {Bostroem},
  {Bouma}, {Brammer}, {Bray}, {Breytenbach}, {Buddelmeijer}, {Burke},
  {Calderone}, {Cano Rodr{\'\i}guez}, {Cara}, {Cardoso}, {Cheedella}, {Copin},
  {Corrales}, {Crichton}, {D'Avella}, {Deil}, {Depagne}, {Dietrich}, {Donath},
  {Droettboom}, {Earl}, {Erben}, {Fabbro}, {Ferreira}, {Finethy}, {Fox},
  {Garrison}, {Gibbons}, {Goldstein}, {Gommers}, {Greco}, {Greenfield},
  {Groener}, {Grollier}, {Hagen}, {Hirst}, {Homeier}, {Horton}, {Hosseinzadeh},
  {Hu}, {Hunkeler}, {Ivezi{\'c}}, {Jain}, {Jenness}, {Kanarek}, {Kendrew},
  {Kern}, {Kerzendorf}, {Khvalko}, {King}, {Kirkby}, {Kulkarni}, {Kumar},
  {Lee}, {Lenz}, {Littlefair}, {Ma}, {Macleod}, {Mastropietro}, {McCully},
  {Montagnac}, {Morris}, {Mueller}, {Mumford}, {Muna}, {Murphy}, {Nelson},
  {Nguyen}, {Ninan}, {N{\"o}the}, {Ogaz}, {Oh}, {Parejko}, {Parley}, {Pascual},
  {Patil}, {Patil}, {Plunkett}, {Prochaska}, {Rastogi}, {Reddy Janga},
  {Sabater}, {Sakurikar}, {Seifert}, {Sherbert}, {Sherwood-Taylor}, {Shih},
  {Sick}, {Silbiger}, {Singanamalla}, {Singer}, {Sladen}, {Sooley},
  {Sornarajah}, {Streicher}, {Teuben}, {Thomas}, {Tremblay}, {Turner},
  {Terr{\'o}n}, {van Kerkwijk}, {de la Vega}, {Watkins}, {Weaver}, {Whitmore},
  {Woillez}, {Zabalza}, \& {Astropy Contributors}}]{astropy_2018}
{Astropy Collaboration}, {Price-Whelan}, A.~M., {Sip{\H{o}}cz}, B.~M., {et~al.}
  2018, \aj, 156, 123, \dodoi{10.3847/1538-3881/aabc4f}

\bibitem[{{Astropy Collaboration} {et~al.}(2022){Astropy Collaboration},
  {Price-Whelan}, {Lim}, {Earl}, {Starkman}, {Bradley}, {Shupe}, {Patil},
  {Corrales}, {Brasseur}, {N{"o}the}, {Donath}, {Tollerud}, {Morris},
  {Ginsburg}, {Vaher}, {Weaver}, {Tocknell}, {Jamieson}, {van Kerkwijk},
  {Robitaille}, {Merry}, {Bachetti}, {G{"u}nther}, {Aldcroft},
  {Alvarado-Montes}, {Archibald}, {B{'o}di}, {Bapat}, {Barentsen}, {Baz{'a}n},
  {Biswas}, {Boquien}, {Burke}, {Cara}, {Cara}, {Conroy}, {Conseil}, {Craig},
  {Cross}, {Cruz}, {D'Eugenio}, {Dencheva}, {Devillepoix}, {Dietrich},
  {Eigenbrot}, {Erben}, {Ferreira}, {Foreman-Mackey}, {Fox}, {Freij}, {Garg},
  {Geda}, {Glattly}, {Gondhalekar}, {Gordon}, {Grant}, {Greenfield}, {Groener},
  {Guest}, {Gurovich}, {Handberg}, {Hart}, {Hatfield-Dodds}, {Homeier},
  {Hosseinzadeh}, {Jenness}, {Jones}, {Joseph}, {Kalmbach}, {Karamehmetoglu},
  {Ka{l}uszy{'n}ski}, {Kelley}, {Kern}, {Kerzendorf}, {Koch}, {Kulumani},
  {Lee}, {Ly}, {Ma}, {MacBride}, {Maljaars}, {Muna}, {Murphy}, {Norman},
  {O'Steen}, {Oman}, {Pacifici}, {Pascual}, {Pascual-Granado}, {Patil},
  {Perren}, {Pickering}, {Rastogi}, {Roulston}, {Ryan}, {Rykoff}, {Sabater},
  {Sakurikar}, {Salgado}, {Sanghi}, {Saunders}, {Savchenko}, {Schwardt},
  {Seifert-Eckert}, {Shih}, {Jain}, {Shukla}, {Sick}, {Simpson},
  {Singanamalla}, {Singer}, {Singhal}, {Sinha}, {Sip{H{o}}cz}, {Spitler},
  {Stansby}, {Streicher}, {{{S}}umak}, {Swinbank}, {Taranu}, {Tewary},
  {Tremblay}, {Val-Borro}, {Van Kooten}, {Vasovi{'c}}, {Verma}, {de Miranda
  Cardoso}, {Williams}, {Wilson}, {Winkel}, {Wood-Vasey}, {Xue}, {Yoachim},
  {Zhang}, {Zonca}, \& {Astropy Project Contributors}}]{astropy_2022}
{Astropy Collaboration}, {Price-Whelan}, A.~M., {Lim}, P.~L., {et~al.} 2022,
  apj, 935, 167, \dodoi{10.3847/1538-4357/ac7c74}

\bibitem[{{Bensby} {et~al.}(2003){Bensby}, {Feltzing}, \&
  {Lundstr{\"o}m}}]{Bensby_2003}
{Bensby}, T., {Feltzing}, S., \& {Lundstr{\"o}m}, I. 2003, \aap, 410, 527,
  \dodoi{10.1051/0004-6361:20031213}

\bibitem[{{Bergemann} {et~al.}(2014){Bergemann}, {Ruchti}, {Serenelli},
  {Feltzing}, {Alves-Brito}, {Asplund}, {Bensby}, {Gruyters}, {Heiter},
  {Hourihane}, {Korn}, {Lind}, {Marino}, {Jofre}, {Nordlander}, {Ryde},
  {Worley}, {Gilmore}, {Randich}, {Ferguson}, {Jeffries}, {Micela},
  {Negueruela}, {Prusti}, {Rix}, {Vallenari}, {Alfaro}, {Allende Prieto},
  {Bragaglia}, {Koposov}, {Lanzafame}, {Pancino}, {Recio-Blanco}, {Smiljanic},
  {Walton}, {Costado}, {Franciosini}, {Hill}, {Lardo}, {de Laverny}, {Magrini},
  {Maiorca}, {Masseron}, {Morbidelli}, {Sacco}, {Kordopatis}, \&
  {Tautvai{\v{s}}ien{\.{e}}}}]{Bergemann_2014}
{Bergemann}, M., {Ruchti}, G.~R., {Serenelli}, A., {et~al.} 2014, \aap, 565,
  A89, \dodoi{10.1051/0004-6361/201423456}

\bibitem[{{Blanton} {et~al.}(2017){Blanton}, {Bershady}, {Abolfathi},
  {Albareti}, {Allende Prieto}, {Almeida}, {Alonso-Garc{\'\i}a}, {Anders},
  {Anderson}, {Andrews}, {Aquino-Ort{\'\i}z}, {Arag{\'o}n-Salamanca},
  {Argudo-Fern{\'a}ndez}, {Armengaud}, {Aubourg}, {Avila-Reese}, {Badenes},
  {Bailey}, {Barger}, {Barrera-Ballesteros}, {Bartosz}, {Bates}, {Baumgarten},
  {Bautista}, {Beaton}, {Beers}, {Belfiore}, {Bender}, {Berlind}, {Bernardi},
  {Beutler}, {Bird}, {Bizyaev}, {Blanc}, {Blomqvist}, {Bolton}, {Boquien},
  {Borissova}, {van den Bosch}, {Bovy}, {Brandt}, {Brinkmann}, {Brownstein},
  {Bundy}, {Burgasser}, {Burtin}, {Busca}, {Cappellari}, {Delgado Carigi},
  {Carlberg}, {Carnero Rosell}, {Carrera}, {Chanover}, {Cherinka}, {Cheung},
  {G{\'o}mez Maqueo Chew}, {Chiappini}, {Choi}, {Chojnowski}, {Chuang},
  {Chung}, {Cirolini}, {Clerc}, {Cohen}, {Comparat}, {da Costa}, {Cousinou},
  {Covey}, {Crane}, {Croft}, {Cruz-Gonzalez}, {Garrido Cuadra}, {Cunha},
  {Damke}, {Darling}, {Davies}, {Dawson}, {de la Macorra}, {Dell'Agli}, {De
  Lee}, {Delubac}, {Di Mille}, {Diamond-Stanic}, {Cano-D{\'\i}az}, {Donor},
  {Downes}, {Drory}, {du Mas des Bourboux}, {Duckworth}, {Dwelly}, {Dyer},
  {Ebelke}, {Eigenbrot}, {Eisenstein}, {Emsellem}, {Eracleous}, {Escoffier},
  {Evans}, {Fan}, {Fern{\'a}ndez-Alvar}, {Fernandez-Trincado}, {Feuillet},
  {Finoguenov}, {Fleming}, {Font-Ribera}, {Fredrickson}, {Freischlad},
  {Frinchaboy}, {Fuentes}, {Galbany}, {Garcia-Dias},
  {Garc{\'\i}a-Hern{\'a}ndez}, {Gaulme}, {Geisler}, {Gelfand},
  {Gil-Mar{\'\i}n}, {Gillespie}, {Goddard}, {Gonzalez-Perez}, {Grabowski},
  {Green}, {Grier}, {Gunn}, {Guo}, {Guy}, {Hagen}, {Hahn}, {Hall}, {Harding},
  {Hasselquist}, {Hawley}, {Hearty}, {Gonzalez Hern{\'a}ndez}, {Ho}, {Hogg},
  {Holley-Bockelmann}, {Holtzman}, {Holzer}, {Huehnerhoff}, {Hutchinson},
  {Hwang}, {Ibarra-Medel}, {da Silva Ilha}, {Ivans}, {Ivory}, {Jackson},
  {Jensen}, {Johnson}, {Jones}, {J{\"o}nsson}, {Jullo}, {Kamble}, {Kinemuchi},
  {Kirkby}, {Kitaura}, {Klaene}, {Knapp}, {Kneib}, {Kollmeier}, {Lacerna},
  {Lane}, {Lang}, {Law}, {Lazarz}, {Lee}, {Le Goff}, {Liang}, {Li}, {Li},
  {Lian}, {Lima}, {Lin}, {Lin}, {Bertran de Lis}, {Liu}, {de Icaza Lizaola},
  {Long}, {Lucatello}, {Lundgren}, {MacDonald}, {Deconto Machado}, {MacLeod},
  {Mahadevan}, {Geimba Maia}, {Maiolino}, {Majewski}, {Malanushenko},
  {Malanushenko}, {Manchado}, {Mao}, {Maraston}, {Marques-Chaves}, {Masseron},
  {Masters}, {McBride}, {McDermid}, {McGrath}, {McGreer}, {Medina Pe{\~n}a},
  {Melendez}, {Merloni}, {Merrifield}, {Meszaros}, {Meza}, {Minchev},
  {Minniti}, {Miyaji}, {More}, {Mulchaey}, {M{\"u}ller-S{\'a}nchez}, {Muna},
  {Munoz}, {Myers}, {Nair}, {Nandra}, {Correa do Nascimento}, {Negrete},
  {Ness}, {Newman}, {Nichol}, {Nidever}, {Nitschelm}, {Ntelis}, {O'Connell},
  {Oelkers}, {Oravetz}, {Oravetz}, {Pace}, {Padilla}, {Palanque-Delabrouille},
  {Alonso Palicio}, {Pan}, {Parejko}, {Parikh}, {P{\^a}ris}, {Park}, {Patten},
  {Peirani}, {Pellejero-Ibanez}, {Penny}, {Percival}, {Perez-Fournon},
  {Petitjean}, {Pieri}, {Pinsonneault}, {Pisani}, {Poleski}, {Prada},
  {Prakash}, {Queiroz}, {Raddick}, {Raichoor}, {Barboza Rembold}, {Richstein},
  {Riffel}, {Riffel}, {Rix}, {Robin}, {Rockosi}, {Rodr{\'\i}guez-Torres},
  {Roman-Lopes}, {Rom{\'a}n-Z{\'u}{\~n}iga}, {Rosado}, {Ross}, {Rossi}, {Ruan},
  {Ruggeri}, {Rykoff}, {Salazar-Albornoz}, {Salvato}, {S{\'a}nchez}, {Aguado},
  {S{\'a}nchez-Gallego}, {Santana}, {Santiago}, {Sayres}, {Schiavon}, {da Silva
  Schimoia}, {Schlafly}, {Schlegel}, {Schneider}, {Schultheis}, {Schuster},
  {Schwope}, {Seo}, {Shao}, {Shen}, {Shetrone}, {Shull}, {Simon}, {Skinner},
  {Skrutskie}, {Slosar}, {Smith}, {Sobeck}, {Sobreira}, {Somers}, {Souto},
  {Stark}, {Stassun}, {Stauffer}, {Steinmetz}, {Storchi-Bergmann},
  {Streblyanska}, {Stringfellow}, {Su{\'a}rez}, {Sun}, {Suzuki}, {Szigeti},
  {Taghizadeh-Popp}, {Tang}, {Tao}, {Tayar}, {Tembe}, {Teske}, {Thakar},
  {Thomas}, {Thompson}, {Tinker}, {Tissera}, {Tojeiro}, {Hernandez Toledo}, {de
  la Torre}, {Tremonti}, {Troup}, {Valenzuela}, {Martinez Valpuesta},
  {Vargas-Gonz{\'a}lez}, {Vargas-Maga{\~n}a}, {Vazquez}, {Villanova}, {Vivek},
  {Vogt}, {Wake}, {Walterbos}, {Wang}, {Weaver}, {Weijmans}, {Weinberg},
  {Westfall}, {Whelan}, {Wild}, {Wilson}, {Wood-Vasey}, {Wylezalek}, {Xiao},
  {Yan}, {Yang}, {Ybarra}, {Y{\`e}che}, {Zakamska}, {Zamora}, {Zarrouk},
  {Zasowski}, {Zhang}, {Zhao}, {Zheng}, {Zheng}, {Zhou}, {Zhou}, {Zhu},
  {Zoccali}, \& {Zou}}]{Blanton_2017}
{Blanton}, M.~R., {Bershady}, M.~A., {Abolfathi}, B., {et~al.} 2017, \aj, 154,
  28, \dodoi{10.3847/1538-3881/aa7567}

\bibitem[{{Bovy}(2016)}]{bovy_2016}
{Bovy}, J. 2016, \apj, 817, 49, \dodoi{10.3847/0004-637X/817/1/49}

\bibitem[{{Bovy} {et~al.}(2014){Bovy}, {Nidever}, {Rix}, {Girardi}, {Zasowski},
  {Chojnowski}, {Holtzman}, {Epstein}, {Frinchaboy}, {Hayden}, {Rodrigues},
  {Majewski}, {Johnson}, {Pinsonneault}, {Stello}, {Allende Prieto}, {Andrews},
  {Basu}, {Beers}, {Bizyaev}, {Burton}, {Chaplin}, {Cunha}, {Elsworth},
  {Garc{\'\i}a}, {Garc{\'\i}a-Her{\'n}andez}, {Garc{\'\i}a P{\'e}rez},
  {Hearty}, {Hekker}, {Kallinger}, {Kinemuchi}, {Koesterke},
  {M{\'e}sz{\'a}ros}, {Mosser}, {O'Connell}, {Oravetz}, {Pan}, {Robin},
  {Schiavon}, {Schneider}, {Schultheis}, {Serenelli}, {Shetrone}, {Silva
  Aguirre}, {Simmons}, {Skrutskie}, {Smith}, {Stassun}, {Weinberg}, {Wilson},
  \& {Zamora}}]{Bovy_2014}
{Bovy}, J., {Nidever}, D.~L., {Rix}, H.-W., {et~al.} 2014, \apj, 790, 127,
  \dodoi{10.1088/0004-637X/790/2/127}

\bibitem[{{Brand} \& {Blitz}(1993)}]{Brand_1993}
{Brand}, J., \& {Blitz}, L. 1993, \aap, 275, 67

\bibitem[{{Buder} {et~al.}(2019){Buder}, {Lind}, {Ness}, {Asplund}, {Duong},
  {Lin}, {Kos}, {Casagrande}, {Casey}, {Bland-Hawthorn}, {de Silva}, {D'Orazi},
  {Freeman}, {Martell}, {Schlesinger}, {Sharma}, {Simpson}, {Zucker},
  {Zwitter}, {{\v{C}}otar}, {Dotter}, {Hayden}, {Hyde}, {Kafle}, {Lewis},
  {Nataf}, {Nordlander}, {Reid}, {Rix}, {Sk{\'u}lad{\'o}ttir}, {Stello},
  {Ting}, {Traven}, {Wyse}, \& {GALAH Collaboration}}]{Buder_2019}
{Buder}, S., {Lind}, K., {Ness}, M.~K., {et~al.} 2019, \aap, 624, A19,
  \dodoi{10.1051/0004-6361/201833218}

\bibitem[{{Buder} {et~al.}(2022){Buder}, {Lind}, {Ness}, {Feuillet}, {Horta},
  {Monty}, {Buck}, {Nordlander}, {Bland-Hawthorn}, {Casey}, {de Silva},
  {D'Orazi}, {Freeman}, {Hayden}, {Kos}, {Martell}, {Lewis}, {Lin},
  {Schlesinger}, {Sharma}, {Simpson}, {Stello}, {Zucker}, {Zwitter},
  {Ciuc{\u{a}}}, {Horner}, {Kobayashi}, {Ting}, {Wyse}, \& {Wyse}}]{Buder_2022}
---. 2022, \mnras, 510, 2407, \dodoi{10.1093/mnras/stab3504}

\bibitem[{{Clough} {et~al.}(2005){Clough}, {Shephard}, {Mlawer}, {Delamere},
  {Iacono}, {Cady-Pereira}, {Boukabara}, \& {Brown}}]{Clough_2005}
{Clough}, S.~A., {Shephard}, M.~W., {Mlawer}, E.~J., {et~al.} 2005, \jqsrt, 91,
  233, \dodoi{10.1016/j.jqsrt.2004.05.058}

\bibitem[{{Conroy} {et~al.}(2019){Conroy}, {Bonaca}, {Cargile}, {Johnson},
  {Caldwell}, {Naidu}, {Zaritsky}, {Fabricant}, {Moran}, {Rhee},
  {Szentgyorgyi}, {Berlind}, {Calkins}, {Kattner}, \& {Ly}}]{Conroy_2019a}
{Conroy}, C., {Bonaca}, A., {Cargile}, P., {et~al.} 2019, \apj, 883, 107,
  \dodoi{10.3847/1538-4357/ab38b8}

\bibitem[{{Cox} {et~al.}(2014){Cox}, {Cami}, {Kaper}, {Ehrenfreund}, {Foing},
  {Ochsendorf}, {van Hooff}, \& {Salama}}]{Cox_2014}
{Cox}, N.~L.~J., {Cami}, J., {Kaper}, L., {et~al.} 2014, \aap, 569, A117,
  \dodoi{10.1051/0004-6361/201323061}

\bibitem[{{De Silva} {et~al.}(2015){De Silva}, {Freeman}, {Bland-Hawthorn},
  {Martell}, {de Boer}, {Asplund}, {Keller}, {Sharma}, {Zucker}, {Zwitter},
  {Anguiano}, {Bacigalupo}, {Bayliss}, {Beavis}, {Bergemann}, {Campbell},
  {Cannon}, {Carollo}, {Casagrande}, {Casey}, {Da Costa}, {D'Orazi}, {Dotter},
  {Duong}, {Heger}, {Ireland}, {Kafle}, {Kos}, {Lattanzio}, {Lewis}, {Lin},
  {Lind}, {Munari}, {Nataf}, {O'Toole}, {Parker}, {Reid}, {Schlesinger},
  {Sheinis}, {Simpson}, {Stello}, {Ting}, {Traven}, {Watson}, {Wittenmyer},
  {Yong}, \& {{\v{Z}}erjal}}]{DeSilva_2015}
{De Silva}, G.~M., {Freeman}, K.~C., {Bland-Hawthorn}, J., {et~al.} 2015,
  \mnras, 449, 2604, \dodoi{10.1093/mnras/stv327}

\bibitem[{{Ebenbichler} {et~al.}(2022){Ebenbichler}, {Postel}, {Przybilla},
  {Seifahrt}, {We{\ss}mayer}, {Kausch}, {Firnstein}, {Butler}, {Kaufer}, \&
  {Linnartz}}]{Ebenbichler_2022}
{Ebenbichler}, A., {Postel}, A., {Przybilla}, N., {et~al.} 2022, \aap, 662,
  A81, \dodoi{10.1051/0004-6361/202142990}

\bibitem[{{Edvardsson} {et~al.}(1993){Edvardsson}, {Andersen}, {Gustafsson},
  {Lambert}, {Nissen}, \& {Tomkin}}]{Edvardsson_1993}
{Edvardsson}, B., {Andersen}, J., {Gustafsson}, B., {et~al.} 1993, \aap, 275,
  101

\bibitem[{{Eilers} {et~al.}(2022){Eilers}, {Hogg}, {Rix}, {Ness},
  {Price-Whelan}, {M{\'e}sz{\'a}ros}, \& {Nitschelm}}]{Eilers_2022}
{Eilers}, A.-C., {Hogg}, D.~W., {Rix}, H.-W., {et~al.} 2022, \apj, 928, 23,
  \dodoi{10.3847/1538-4357/ac54ad}

\bibitem[{{Eisenstein} {et~al.}(2011){Eisenstein}, {Weinberg}, {Agol},
  {Aihara}, {Allende Prieto}, {Anderson}, {Arns}, {Aubourg}, {Bailey},
  {Balbinot}, {Barkhouser}, {Beers}, {Berlind}, {Bickerton}, {Bizyaev},
  {Blanton}, {Bochanski}, {Bolton}, {Bosman}, {Bovy}, {Brandt}, {Breslauer},
  {Brewington}, {Brinkmann}, {Brown}, {Brownstein}, {Burger}, {Busca},
  {Campbell}, {Cargile}, {Carithers}, {Carlberg}, {Carr}, {Chang}, {Chen},
  {Chiappini}, {Comparat}, {Connolly}, {Cortes}, {Croft}, {Cunha}, {da Costa},
  {Davenport}, {Dawson}, {De Lee}, {Porto de Mello}, {de Simoni}, {Dean},
  {Dhital}, {Ealet}, {Ebelke}, {Edmondson}, {Eiting}, {Escoffier}, {Esposito},
  {Evans}, {Fan}, {Femen{\'\i}a Castell{\'a}}, {Dutra Ferreira}, {Fitzgerald},
  {Fleming}, {Font-Ribera}, {Ford}, {Frinchaboy}, {Garc{\'\i}a P{\'e}rez},
  {Gaudi}, {Ge}, {Ghezzi}, {Gillespie}, {Gilmore}, {Girardi}, {Gott}, {Gould},
  {Grebel}, {Gunn}, {Hamilton}, {Harding}, {Harris}, {Hawley}, {Hearty},
  {Hennawi}, {Gonz{\'a}lez Hern{\'a}ndez}, {Ho}, {Hogg}, {Holtzman},
  {Honscheid}, {Inada}, {Ivans}, {Jiang}, {Jiang}, {Johnson}, {Jordan},
  {Jordan}, {Kauffmann}, {Kazin}, {Kirkby}, {Klaene}, {Knapp}, {Kneib},
  {Kochanek}, {Koesterke}, {Kollmeier}, {Kron}, {Lampeitl}, {Lang}, {Lawler},
  {Le Goff}, {Lee}, {Lee}, {Leisenring}, {Lin}, {Liu}, {Long}, {Loomis},
  {Lucatello}, {Lundgren}, {Lupton}, {Ma}, {Ma}, {MacDonald}, {Mack},
  {Mahadevan}, {Maia}, {Majewski}, {Makler}, {Malanushenko}, {Malanushenko},
  {Mandelbaum}, {Maraston}, {Margala}, {Maseman}, {Masters}, {McBride},
  {McDonald}, {McGreer}, {McMahon}, {Mena Requejo}, {M{\'e}nard},
  {Miralda-Escud{\'e}}, {Morrison}, {Mullally}, {Muna}, {Murayama}, {Myers},
  {Naugle}, {Neto}, {Nguyen}, {Nichol}, {Nidever}, {O'Connell}, {Ogando},
  {Olmstead}, {Oravetz}, {Padmanabhan}, {Paegert}, {Palanque-Delabrouille},
  {Pan}, {Pandey}, {Parejko}, {P{\^a}ris}, {Pellegrini}, {Pepper}, {Percival},
  {Petitjean}, {Pfaffenberger}, {Pforr}, {Phleps}, {Pichon}, {Pieri}, {Prada},
  {Price-Whelan}, {Raddick}, {Ramos}, {Reid}, {Reyle}, {Rich}, {Richards},
  {Rieke}, {Rieke}, {Rix}, {Robin}, {Rocha-Pinto}, {Rockosi}, {Roe},
  {Rollinde}, {Ross}, {Ross}, {Rossetto}, {S{\'a}nchez}, {Santiago}, {Sayres},
  {Schiavon}, {Schlegel}, {Schlesinger}, {Schmidt}, {Schneider}, {Sellgren},
  {Shelden}, {Sheldon}, {Shetrone}, {Shu}, {Silverman}, {Simmerer}, {Simmons},
  {Sivarani}, {Skrutskie}, {Slosar}, {Smee}, {Smith}, {Snedden}, {Stassun},
  {Steele}, {Steinmetz}, {Stockett}, {Stollberg}, {Strauss}, {Szalay},
  {Tanaka}, {Thakar}, {Thomas}, {Tinker}, {Tofflemire}, {Tojeiro}, {Tremonti},
  {Vargas Maga{\~n}a}, {Verde}, {Vogt}, {Wake}, {Wan}, {Wang}, {Weaver},
  {White}, {White}, {Wilson}, {Wisniewski}, {Wood-Vasey}, {Yanny}, {Yasuda},
  {Y{\`e}che}, {York}, {Young}, {Zasowski}, {Zehavi}, \&
  {Zhao}}]{Eisenstein_2011}
{Eisenstein}, D.~J., {Weinberg}, D.~H., {Agol}, E., {et~al.} 2011, \aj, 142,
  72, \dodoi{10.1088/0004-6256/142/3/72}

\bibitem[{{Elyajouri} {et~al.}(2017){Elyajouri}, {Lallement}, {Monreal-Ibero},
  {Capitanio}, \& {Cox}}]{Elyajouri_2017}
{Elyajouri}, M., {Lallement}, R., {Monreal-Ibero}, A., {Capitanio}, L., \&
  {Cox}, N.~L.~J. 2017, \aap, 600, A129, \dodoi{10.1051/0004-6361/201630088}

\bibitem[{{Elyajouri} {et~al.}(2016){Elyajouri}, {Monreal-Ibero}, {Remy}, \&
  {Lallement}}]{Elyajouri_2016}
{Elyajouri}, M., {Monreal-Ibero}, A., {Remy}, Q., \& {Lallement}, R. 2016,
  \apjs, 225, 19, \dodoi{10.3847/0067-0049/225/2/19}

\bibitem[{{Fan} {et~al.}(2019){Fan}, {Hobbs}, {Dahlstrom}, {Welty}, {York},
  {Rachford}, {Snow}, {Sonnentrucker}, {Baskes}, \& {Zhao}}]{Fan_2019}
{Fan}, H., {Hobbs}, L.~M., {Dahlstrom}, J.~A., {et~al.} 2019, \apj, 878, 151,
  \dodoi{10.3847/1538-4357/ab1b74}

\bibitem[{{Feeney} {et~al.}(2021){Feeney}, {Wandelt}, \& {Ness}}]{Feeney_2021}
{Feeney}, S.~M., {Wandelt}, B.~D., \& {Ness}, M.~K. 2021, \mnras, 501, 3258,
  \dodoi{10.1093/mnras/staa3586}

\bibitem[{{Feltzing} \& {Gustafsson}(1998)}]{Feltzing_1998}
{Feltzing}, S., \& {Gustafsson}, B. 1998, \aaps, 129, 237,
  \dodoi{10.1051/aas:1998400}

\bibitem[{Foreman-Mackey(2016)}]{corner_citation}
Foreman-Mackey, D. 2016, The Journal of Open Source Software, 1, 24,
  \dodoi{10.21105/joss.00024}

\bibitem[{{Foreman-Mackey} {et~al.}(2013){Foreman-Mackey}, {Hogg}, {Lang}, \&
  {Goodman}}]{emcee_citation}
{Foreman-Mackey}, D., {Hogg}, D.~W., {Lang}, D., \& {Goodman}, J. 2013, \pasp,
  125, 306, \dodoi{10.1086/670067}

\bibitem[{{Garc{\'\i}a P{\'e}rez} {et~al.}(2016){Garc{\'\i}a P{\'e}rez},
  {Allende Prieto}, {Holtzman}, {Shetrone}, {M{\'e}sz{\'a}ros}, {Bizyaev},
  {Carrera}, {Cunha}, {Garc{\'\i}a-Hern{\'a}ndez}, {Johnson}, {Majewski},
  {Nidever}, {Schiavon}, {Shane}, {Smith}, {Sobeck}, {Troup}, {Zamora},
  {Weinberg}, {Bovy}, {Eisenstein}, {Feuillet}, {Frinchaboy}, {Hayden},
  {Hearty}, {Nguyen}, {O'Connell}, {Pinsonneault}, {Wilson}, \&
  {Zasowski}}]{Garcia-Perez_2016}
{Garc{\'\i}a P{\'e}rez}, A.~E., {Allende Prieto}, C., {Holtzman}, J.~A.,
  {et~al.} 2016, \aj, 151, 144, \dodoi{10.3847/0004-6256/151/6/144}

\bibitem[{{Geballe} {et~al.}(2011){Geballe}, {Najarro}, {Figer},
  {Schlegelmilch}, \& {de La Fuente}}]{Geballe_2011}
{Geballe}, T.~R., {Najarro}, F., {Figer}, D.~F., {Schlegelmilch}, B.~W., \& {de
  La Fuente}, D. 2011, \nat, 479, 200, \dodoi{10.1038/nature10527}

\bibitem[{{Gilmore} {et~al.}(2012){Gilmore}, {Randich}, {Asplund}, {Binney},
  {Bonifacio}, {Drew}, {Feltzing}, {Ferguson}, {Jeffries}, {Micela},
  {Negueruela}, {Prusti}, {Rix}, {Vallenari}, {Alfaro}, {Allende-Prieto},
  {Babusiaux}, {Bensby}, {Blomme}, {Bragaglia}, {Flaccomio}, {Fran{\c{c}}ois},
  {Irwin}, {Koposov}, {Korn}, {Lanzafame}, {Pancino}, {Paunzen},
  {Recio-Blanco}, {Sacco}, {Smiljanic}, {Van Eck}, {Walton}, {Aden}, {Aerts},
  {Affer}, {Alcala}, {Altavilla}, {Alves}, {Antoja}, {Arenou}, {Argiroffi},
  {Asensio Ramos}, {Bailer-Jones}, {Balaguer-Nunez}, {Bayo}, {Barbuy},
  {Barisevicius}, {Barrado y Navascues}, {Battistini}, {Bellas Velidis},
  {Bellazzini}, {Belokurov}, {Bergemann}, {Bertelli}, {Biazzo}, {Bienayme},
  {Bland-Hawthorn}, {Boeche}, {Bonito}, {Boudreault}, {Bouvier}, {Brandao},
  {Brown}, {de Bruijne}, {Burleigh}, {Caballero}, {Caffau}, {Calura},
  {Capuzzo-Dolcetta}, {Caramazza}, {Carraro}, {Casagrande}, {Casewell},
  {Chapman}, {Chiappini}, {Chorniy}, {Christlieb}, {Cignoni}, {Cocozza},
  {Colless}, {Collet}, {Collins}, {Correnti}, {Covino}, {Crnojevic}, {Cropper},
  {Cunha}, {Damiani}, {David}, {Delgado}, {Duffau}, {Edvardsson}, {Eldridge},
  {Enke}, {Eriksson}, {Evans}, {Eyer}, {Famaey}, {Fellhauer}, {Ferreras},
  {Figueras}, {Fiorentino}, {Flynn}, {Folha}, {Franciosini}, {Frasca},
  {Freeman}, {Fremat}, {Friel}, {Gaensicke}, {Gameiro}, {Garzon}, {Geier},
  {Geisler}, {Gerhard}, {Gibson}, {Gomboc}, {Gomez}, {Gonzalez-Fernandez},
  {Gonzalez Hernandez}, {Gosset}, {Grebel}, {Greimel}, {Groenewegen},
  {Grundahl}, {Guarcello}, {Gustafsson}, {Hadrava}, {Hatzidimitriou}, {Hambly},
  {Hammersley}, {Hansen}, {Haywood}, {Heber}, {Heiter}, {Held}, {Helmi},
  {Hensler}, {Herrero}, {Hill}, {Hodgkin}, {Huelamo}, {Huxor}, {Ibata},
  {Jackson}, {de Jong}, {Jonker}, {Jordan}, {Jordi}, {Jorissen}, {Katz},
  {Kawata}, {Keller}, {Kharchenko}, {Klement}, {Klutsch}, {Knude}, {Koch},
  {Kochukhov}, {Kontizas}, {Koubsky}, {Lallement}, {de Laverny}, {van Leeuwen},
  {Lemasle}, {Lewis}, {Lind}, {Lindstrom}, {Lobel}, {Lopez Santiago}, {Lucas},
  {Ludwig}, {Lueftinger}, {Magrini}, {Maiz Apellaniz}, {Maldonado}, {Marconi},
  {Marino}, {Martayan}, {Martinez-Valpuesta}, {Matijevic}, {McMahon},
  {Messina}, {Meyer}, {Miglio}, {Mikolaitis}, {Minchev}, {Minniti}, {Moitinho},
  {Momany}, {Monaco}, {Montalto}, {Monteiro}, {Monier}, {Montes}, {Mora},
  {Moraux}, {Morel}, {Mowlavi}, {Mucciarelli}, {Munari}, {Napiwotzki},
  {Nardetto}, {Naylor}, {Naze}, {Nelemans}, {Okamoto}, {Ortolani}, {Pace},
  {Palla}, {Palous}, {Parker}, {Penarrubia}, {Pillitteri}, {Piotto}, {Posbic},
  {Prisinzano}, {Puzeras}, {Quirrenbach}, {Ragaini}, {Read}, {Read}, {Reyle},
  {De Ridder}, {Robichon}, {Robin}, {Roeser}, {Romano}, {Royer}, {Ruchti},
  {Ruzicka}, {Ryan}, {Ryde}, {Santos}, {Sanz Forcada}, {Sarro Baro},
  {Sbordone}, {Schilbach}, {Schmeja}, {Schnurr}, {Schoenrich}, {Scholz},
  {Seabroke}, {Sharma}, {De Silva}, {Smith}, {Solano}, {Sordo}, {Soubiran},
  {Sousa}, {Spagna}, {Steffen}, {Steinmetz}, {Stelzer}, {Stempels},
  {Tabernero}, {Tautvaisiene}, {Thevenin}, {Torra}, {Tosi}, {Tolstoy}, {Turon},
  {Walker}, {Wambsganss}, {Worley}, {Venn}, {Vink}, {Wyse}, {Zaggia},
  {Zeilinger}, {Zoccali}, {Zorec}, {Zucker}, {Zwitter}, \& {Gaia-ESO Survey
  Team}}]{Gilmore_2012}
{Gilmore}, G., {Randich}, S., {Asplund}, M., {et~al.} 2012, The Messenger, 147,
  25

\bibitem[{{Griffith} {et~al.}(2022){Griffith}, {Weinberg}, {Buder}, {Johnson},
  {Johnson}, \& {Vincenzo}}]{Griffith_2022}
{Griffith}, E.~J., {Weinberg}, D.~H., {Buder}, S., {et~al.} 2022, \apj, 931,
  23, \dodoi{10.3847/1538-4357/ac5826}

\bibitem[{Harris {et~al.}(2020)Harris, Millman, van~der Walt, Gommers,
  Virtanen, Cournapeau, Wieser, Taylor, Berg, Smith, Kern, Picus, Hoyer, van
  Kerkwijk, Brett, Haldane, del R{\'{i}}o, Wiebe, Peterson,
  G{\'{e}}rard-Marchant, Sheppard, Reddy, Weckesser, Abbasi, Gohlke, \&
  Oliphant}]{numpy_citation}
Harris, C.~R., Millman, K.~J., van~der Walt, S.~J., {et~al.} 2020, Nature, 585,
  357, \dodoi{10.1038/s41586-020-2649-2}

\bibitem[{{Hayden} {et~al.}(2015){Hayden}, {Bovy}, {Holtzman}, {Nidever},
  {Bird}, {Weinberg}, {Andrews}, {Majewski}, {Allende Prieto}, {Anders},
  {Beers}, {Bizyaev}, {Chiappini}, {Cunha}, {Frinchaboy},
  {Garc{\'\i}a-Her{\'n}andez}, {Garc{\'\i}a P{\'e}rez}, {Girardi}, {Harding},
  {Hearty}, {Johnson}, {M{\'e}sz{\'a}ros}, {Minchev}, {O'Connell}, {Pan},
  {Robin}, {Schiavon}, {Schneider}, {Schultheis}, {Shetrone}, {Skrutskie},
  {Steinmetz}, {Smith}, {Wilson}, {Zamora}, \& {Zasowski}}]{Hayden_2015}
{Hayden}, M.~R., {Bovy}, J., {Holtzman}, J.~A., {et~al.} 2015, \apj, 808, 132,
  \dodoi{10.1088/0004-637X/808/2/132}

\bibitem[{{Herbig}(1995)}]{Herbig_1995}
{Herbig}, G.~H. 1995, \araa, 33, 19,
  \dodoi{10.1146/annurev.aa.33.090195.000315}

\bibitem[{{Holtzman} {et~al.}(2015){Holtzman}, {Shetrone}, {Johnson}, {Allende
  Prieto}, {Anders}, {Andrews}, {Beers}, {Bizyaev}, {Blanton}, {Bovy},
  {Carrera}, {Chojnowski}, {Cunha}, {Eisenstein}, {Feuillet}, {Frinchaboy},
  {Galbraith-Frew}, {Garc{\'\i}a P{\'e}rez}, {Garc{\'\i}a-Hern{\'a}ndez},
  {Hasselquist}, {Hayden}, {Hearty}, {Ivans}, {Majewski}, {Martell},
  {Meszaros}, {Muna}, {Nidever}, {Nguyen}, {O'Connell}, {Pan}, {Pinsonneault},
  {Robin}, {Schiavon}, {Shane}, {Sobeck}, {Smith}, {Troup}, {Weinberg},
  {Wilson}, {Wood-Vasey}, {Zamora}, \& {Zasowski}}]{Holtzman_2015}
{Holtzman}, J.~A., {Shetrone}, M., {Johnson}, J.~A., {et~al.} 2015, \aj, 150,
  148, \dodoi{10.1088/0004-6256/150/5/148}

\bibitem[{{Holtzman} {et~al.}(2018){Holtzman}, {Hasselquist}, {Shetrone},
  {Cunha}, {Allende Prieto}, {Anguiano}, {Bizyaev}, {Bovy}, {Casey},
  {Edvardsson}, {Johnson}, {J{\"o}nsson}, {Meszaros}, {Smith}, {Sobeck},
  {Zamora}, {Chojnowski}, {Fernandez-Trincado}, {Garcia-Hernandez}, {Majewski},
  {Pinsonneault}, {Souto}, {Stringfellow}, {Tayar}, {Troup}, \&
  {Zasowski}}]{Holtzman_2018}
{Holtzman}, J.~A., {Hasselquist}, S., {Shetrone}, M., {et~al.} 2018, \aj, 156,
  125, \dodoi{10.3847/1538-3881/aad4f9}

\bibitem[{Hunter(2007)}]{matplotlib_citation}
Hunter, J.~D. 2007, Computing in Science \& Engineering, 9, 90,
  \dodoi{10.1109/MCSE.2007.55}

\bibitem[{{Joblin} {et~al.}(1990){Joblin}, {Maillard}, {D'Hendecourt}, \&
  {L{\'e}ger}}]{Joblin_1990}
{Joblin}, C., {Maillard}, J.~P., {D'Hendecourt}, L., \& {L{\'e}ger}, A. 1990,
  \nat, 346, 729, \dodoi{10.1038/346729a0}

\bibitem[{{J{\"o}nsson} {et~al.}(2020){J{\"o}nsson}, {Holtzman}, {Allende
  Prieto}, {Cunha}, {Garc{\'\i}a-Hern{\'a}ndez}, {Hasselquist}, {Masseron},
  {Osorio}, {Shetrone}, {Smith}, {Stringfellow}, {Bizyaev}, {Edvardsson},
  {Majewski}, {M{\'e}sz{\'a}ros}, {Souto}, {Zamora}, {Beaton}, {Bovy}, {Donor},
  {Pinsonneault}, {Poovelil}, \& {Sobeck}}]{Jonsson_2020}
{J{\"o}nsson}, H., {Holtzman}, J.~A., {Allende Prieto}, C., {et~al.} 2020, \aj,
  160, 120, \dodoi{10.3847/1538-3881/aba592}

\bibitem[{{Khanna} {et~al.}(2019){Khanna}, {Sharma}, {Bland-Hawthorn},
  {Hayden}, {Nataf}, {Ting}, {Kos}, {Martell}, {Zwitter}, {De Silva},
  {Asplund}, {Buder}, {Duong}, {Lin}, {Simpson}, {Anguiano}, {Horner}, {Kafle},
  {Lewis}, {Nordlander}, {Wyse}, {Wittenmyer}, \& {Zucker}}]{Khanna_2019}
{Khanna}, S., {Sharma}, S., {Bland-Hawthorn}, J., {et~al.} 2019, \mnras, 482,
  4215, \dodoi{10.1093/mnras/sty2924}

\bibitem[{Kluyver {et~al.}(2016)Kluyver, Ragan-Kelley, P{\'e}rez, Granger,
  Bussonnier, Frederic, Kelley, Hamrick, Grout, Corlay, Ivanov, Avila, Abdalla,
  \& Willing}]{jupyter_citation}
Kluyver, T., Ragan-Kelley, B., P{\'e}rez, F., {et~al.} 2016, in Positioning and
  Power in Academic Publishing: Players, Agents and Agendas, ed. F.~Loizides \&
  B.~Schmidt, IOS Press, 87 -- 90

\bibitem[{{Kollmeier} {et~al.}(2017){Kollmeier}, {Zasowski}, {Rix}, {Johns},
  {Anderson}, {Drory}, {Johnson}, {Pogge}, {Bird}, {Blanc}, {Brownstein},
  {Crane}, {De Lee}, {Klaene}, {Kreckel}, {MacDonald}, {Merloni}, {Ness},
  {O'Brien}, {Sanchez-Gallego}, {Sayres}, {Shen}, {Thakar}, {Tkachenko},
  {Aerts}, {Blanton}, {Eisenstein}, {Holtzman}, {Maoz}, {Nandra}, {Rockosi},
  {Weinberg}, {Bovy}, {Casey}, {Chaname}, {Clerc}, {Conroy}, {Eracleous},
  {G{\"a}nsicke}, {Hekker}, {Horne}, {Kauffmann}, {McQuinn}, {Pellegrini},
  {Schinnerer}, {Schlafly}, {Schwope}, {Seibert}, {Teske}, \& {van
  Saders}}]{Kollmeier_2017}
{Kollmeier}, J.~A., {Zasowski}, G., {Rix}, H.-W., {et~al.} 2017, arXiv
  e-prints, arXiv:1711.03234.
\newblock \doarXiv{1711.03234}

\bibitem[{{Leung} \& {Bovy}(2019)}]{Leung_2019}
{Leung}, H.~W., \& {Bovy}, J. 2019, \mnras, 483, 3255,
  \dodoi{10.1093/mnras/sty3217}

\bibitem[{{Majewski} {et~al.}(2016){Majewski}, {APOGEE Team}, \& {APOGEE-2
  Team}}]{Majewski_2016}
{Majewski}, S.~R., {APOGEE Team}, \& {APOGEE-2 Team}. 2016, Astronomische
  Nachrichten, 337, 863, \dodoi{10.1002/asna.201612387}

\bibitem[{{Majewski} {et~al.}(2017){Majewski}, {Schiavon}, {Frinchaboy},
  {Allende Prieto}, {Barkhouser}, {Bizyaev}, {Blank}, {Brunner}, {Burton},
  {Carrera}, {Chojnowski}, {Cunha}, {Epstein}, {Fitzgerald}, {Garc{\'\i}a
  P{\'e}rez}, {Hearty}, {Henderson}, {Holtzman}, {Johnson}, {Lam}, {Lawler},
  {Maseman}, {M{\'e}sz{\'a}ros}, {Nelson}, {Nguyen}, {Nidever}, {Pinsonneault},
  {Shetrone}, {Smee}, {Smith}, {Stolberg}, {Skrutskie}, {Walker}, {Wilson},
  {Zasowski}, {Anders}, {Basu}, {Beland}, {Blanton}, {Bovy}, {Brownstein},
  {Carlberg}, {Chaplin}, {Chiappini}, {Eisenstein}, {Elsworth}, {Feuillet},
  {Fleming}, {Galbraith-Frew}, {Garc{\'\i}a}, {Garc{\'\i}a-Hern{\'a}ndez},
  {Gillespie}, {Girardi}, {Gunn}, {Hasselquist}, {Hayden}, {Hekker}, {Ivans},
  {Kinemuchi}, {Klaene}, {Mahadevan}, {Mathur}, {Mosser}, {Muna}, {Munn},
  {Nichol}, {O'Connell}, {Parejko}, {Robin}, {Rocha-Pinto}, {Schultheis},
  {Serenelli}, {Shane}, {Silva Aguirre}, {Sobeck}, {Thompson}, {Troup},
  {Weinberg}, \& {Zamora}}]{Majewski_2017}
{Majewski}, S.~R., {Schiavon}, R.~P., {Frinchaboy}, P.~M., {et~al.} 2017, \aj,
  154, 94, \dodoi{10.3847/1538-3881/aa784d}

\bibitem[{{Martell} {et~al.}(2017){Martell}, {Sharma}, {Buder}, {Duong},
  {Schlesinger}, {Simpson}, {Lind}, {Ness}, {Marshall}, {Asplund},
  {Bland-Hawthorn}, {Casey}, {De Silva}, {Freeman}, {Kos}, {Lin}, {Zucker},
  {Zwitter}, {Anguiano}, {Bacigalupo}, {Carollo}, {Casagrande}, {Da Costa},
  {Horner}, {Huber}, {Hyde}, {Kafle}, {Lewis}, {Nataf}, {Navin}, {Stello},
  {Tinney}, {Watson}, \& {Wittenmyer}}]{Martell_2017}
{Martell}, S.~L., {Sharma}, S., {Buder}, S., {et~al.} 2017, \mnras, 465, 3203,
  \dodoi{10.1093/mnras/stw2835}

\bibitem[{{Martig} {et~al.}(2015){Martig}, {Rix}, {Silva Aguirre}, {Hekker},
  {Mosser}, {Elsworth}, {Bovy}, {Stello}, {Anders}, {Garc{\'\i}a}, {Tayar},
  {Rodrigues}, {Basu}, {Carrera}, {Ceillier}, {Chaplin}, {Chiappini},
  {Frinchaboy}, {Garc{\'\i}a-Hern{\'a}ndez}, {Hearty}, {Holtzman}, {Johnson},
  {Majewski}, {Mathur}, {M{\'e}sz{\'a}ros}, {Miglio}, {Nidever}, {Pan},
  {Pinsonneault}, {Schiavon}, {Schneider}, {Serenelli}, {Shetrone}, \&
  {Zamora}}]{Martig_2015}
{Martig}, M., {Rix}, H.-W., {Silva Aguirre}, V., {et~al.} 2015, \mnras, 451,
  2230, \dodoi{10.1093/mnras/stv1071}

\bibitem[{{Mitschang} {et~al.}(2014){Mitschang}, {De Silva}, {Zucker},
  {Anguiano}, {Bensby}, \& {Feltzing}}]{Mitschang_2014}
{Mitschang}, A.~W., {De Silva}, G., {Zucker}, D.~B., {et~al.} 2014, \mnras,
  438, 2753, \dodoi{10.1093/mnras/stt2320}

\bibitem[{{Ness} {et~al.}(2015){Ness}, {Hogg}, {Rix}, {Ho}, \&
  {Zasowski}}]{Ness_2015}
{Ness}, M., {Hogg}, D.~W., {Rix}, H.~W., {Ho}, A. Y.~Q., \& {Zasowski}, G.
  2015, \apj, 808, 16, \dodoi{10.1088/0004-637X/808/1/16}

\bibitem[{{Ness} {et~al.}(2016){Ness}, {Hogg}, {Rix}, {Martig}, {Pinsonneault},
  \& {Ho}}]{Ness_2016}
{Ness}, M., {Hogg}, D.~W., {Rix}, H.~W., {et~al.} 2016, \apj, 823, 114,
  \dodoi{10.3847/0004-637X/823/2/114}

\bibitem[{{Ness} {et~al.}(2019){Ness}, {Johnston}, {Blancato}, {Rix}, {Beane},
  {Bird}, \& {Hawkins}}]{Ness_2019}
{Ness}, M.~K., {Johnston}, K.~V., {Blancato}, K., {et~al.} 2019, \apj, 883,
  177, \dodoi{10.3847/1538-4357/ab3e3c}

\bibitem[{{Ness} {et~al.}(2022){Ness}, {Wheeler}, {McKinnon}, {Horta}, {Casey},
  {Cunningham}, \& {Price-Whelan}}]{Ness_2022}
{Ness}, M.~K., {Wheeler}, A.~J., {McKinnon}, K., {et~al.} 2022, \apj, 926, 144,
  \dodoi{10.3847/1538-4357/ac4754}

\bibitem[{{Nidever} {et~al.}(2014){Nidever}, {Bovy}, {Bird}, {Andrews},
  {Hayden}, {Holtzman}, {Majewski}, {Smith}, {Robin}, {Garc{\'\i}a P{\'e}rez},
  {Cunha}, {Allende Prieto}, {Zasowski}, {Schiavon}, {Johnson}, {Weinberg},
  {Feuillet}, {Schneider}, {Shetrone}, {Sobeck}, {Garc{\'\i}a-Hern{\'a}ndez},
  {Zamora}, {Rix}, {Beers}, {Wilson}, {O'Connell}, {Minchev}, {Chiappini},
  {Anders}, {Bizyaev}, {Brewington}, {Ebelke}, {Frinchaboy}, {Ge}, {Kinemuchi},
  {Malanushenko}, {Malanushenko}, {Marchante}, {M{\'e}sz{\'a}ros}, {Oravetz},
  {Pan}, {Simmons}, \& {Skrutskie}}]{Nidever_2014}
{Nidever}, D.~L., {Bovy}, J., {Bird}, J.~C., {et~al.} 2014, \apj, 796, 38,
  \dodoi{10.1088/0004-637X/796/1/38}

\bibitem[{{Nidever} {et~al.}(2015){Nidever}, {Holtzman}, {Allende Prieto},
  {Beland}, {Bender}, {Bizyaev}, {Burton}, {Desphande}, {Fleming}, {Garc{\'\i}a
  P{\'e}rez}, {Hearty}, {Majewski}, {M{\'e}sz{\'a}ros}, {Muna}, {Nguyen},
  {Schiavon}, {Shetrone}, {Skrutskie}, {Sobeck}, \& {Wilson}}]{Nidever_2015}
{Nidever}, D.~L., {Holtzman}, J.~A., {Allende Prieto}, C., {et~al.} 2015, \aj,
  150, 173, \dodoi{10.1088/0004-6256/150/6/173}

\bibitem[{P\'erez \& Granger(2007)}]{ipython_citation}
P\'erez, F., \& Granger, B.~E. 2007, Computing in Science and Engineering, 9,
  21, \dodoi{10.1109/MCSE.2007.53}

\bibitem[{{Price-Jones} \& {Bovy}(2018)}]{Price-Jones_2018}
{Price-Jones}, N., \& {Bovy}, J. 2018, \mnras, 475, 1410,
  \dodoi{10.1093/mnras/stx3198}

\bibitem[{{Prochaska} {et~al.}(2000){Prochaska}, {Naumov}, {Carney},
  {McWilliam}, \& {Wolfe}}]{Prochaska_2000}
{Prochaska}, J.~X., {Naumov}, S.~O., {Carney}, B.~W., {McWilliam}, A., \&
  {Wolfe}, A.~M. 2000, \aj, 120, 2513, \dodoi{10.1086/316818}

\bibitem[{{Rockosi} {et~al.}(2022){Rockosi}, {Lee}, {Morrison}, {Yanny},
  {Johnson}, {Lucatello}, {Sobeck}, {Beers}, {Allende Prieto}, {An}, {Bizyaev},
  {Blanton}, {Casagrande}, {Eisenstein}, {Gould}, {Gunn}, {Harding}, {Ivans},
  {Jacobson}, {Janesh}, {Knapp}, {Kollmeier}, {L{\'e}pine},
  {L{\'o}pez-Corredoira}, {Ma}, {Newberg}, {Pan}, {Prchlik}, {Sayers},
  {Schlesinger}, {Simmerer}, \& {Weinberg}}]{Rockosi_2022}
{Rockosi}, C.~M., {Lee}, Y.~S., {Morrison}, H.~L., {et~al.} 2022, \apjs, 259,
  60, \dodoi{10.3847/1538-4365/ac5323}

\bibitem[{{Rojas-Arriagada} {et~al.}(2014){Rojas-Arriagada}, {Recio-Blanco},
  {Hill}, {de Laverny}, {Schultheis}, {Babusiaux}, {Zoccali}, {Minniti},
  {Gonzalez}, {Feltzing}, {Gilmore}, {Randich}, {Vallenari}, {Alfaro},
  {Bensby}, {Bragaglia}, {Flaccomio}, {Lanzafame}, {Pancino}, {Smiljanic},
  {Bergemann}, {Costado}, {Damiani}, {Hourihane}, {Jofr{\'e}}, {Lardo},
  {Magrini}, {Maiorca}, {Morbidelli}, {Sbordone}, {Worley}, {Zaggia}, \&
  {Wyse}}]{RojasArriagada_2014}
{Rojas-Arriagada}, A., {Recio-Blanco}, A., {Hill}, V., {et~al.} 2014, \aap,
  569, A103, \dodoi{10.1051/0004-6361/201424121}

\bibitem[{{Saydjari} {et~al.}(2023){Saydjari}, {Uzsoy}, {Zucker}, {Peek}, \&
  {Finkbeiner}}]{Saydjari_2023}
{Saydjari}, A.~K., {Uzsoy}, A. S.~M., {Zucker}, C., {Peek}, J.~E.~G., \&
  {Finkbeiner}, D.~P. 2023, \apj, 954, 141, \dodoi{10.3847/1538-4357/acd454}

\bibitem[{{Schlegel} {et~al.}(1998){Schlegel}, {Finkbeiner}, \&
  {Davis}}]{Schlegel_1998}
{Schlegel}, D.~J., {Finkbeiner}, D.~P., \& {Davis}, M. 1998, \apj, 500, 525,
  \dodoi{10.1086/305772}

\bibitem[{{Sit} \& {Ness}(2020)}]{Sit_2020}
{Sit}, T., \& {Ness}, M.~K. 2020, \apj, 900, 4,
  \dodoi{10.3847/1538-4357/ab9ff6}

\bibitem[{{Smoker} {et~al.}(2023){Smoker}, {M{\"u}ller}, {Monreal Ibero},
  {Elyajouri}, {Evans}, {Najarro}, {Farhang}, {Cox}, {Minniti}, {Smith},
  {Pritchard}, {Lallement}, {Smette}, {Boffin}, {Cordiner}, \&
  {Cami}}]{Smoker_2023}
{Smoker}, J.~V., {M{\"u}ller}, A., {Monreal Ibero}, A., {et~al.} 2023, \aap,
  672, A181, \dodoi{10.1051/0004-6361/202142267}

\bibitem[{{Steinmetz} {et~al.}(2006){Steinmetz}, {Zwitter}, {Siebert},
  {Watson}, {Freeman}, {Munari}, {Campbell}, {Williams}, {Seabroke}, {Wyse},
  {Parker}, {Bienaym{\'e}}, {Roeser}, {Gibson}, {Gilmore}, {Grebel}, {Helmi},
  {Navarro}, {Burton}, {Cass}, {Dawe}, {Fiegert}, {Hartley}, {Russell},
  {Saunders}, {Enke}, {Bailin}, {Binney}, {Bland-Hawthorn}, {Boeche}, {Dehnen},
  {Eisenstein}, {Evans}, {Fiorucci}, {Fulbright}, {Gerhard}, {Jauregi}, {Kelz},
  {Mijovi{\'c}}, {Minchev}, {Parmentier}, {Pe{\~n}arrubia}, {Quillen}, {Read},
  {Ruchti}, {Scholz}, {Siviero}, {Smith}, {Sordo}, {Veltz}, {Vidrih}, {von
  Berlepsch}, {Boyle}, \& {Schilbach}}]{Steinmetz_2006}
{Steinmetz}, M., {Zwitter}, T., {Siebert}, A., {et~al.} 2006, \aj, 132, 1645,
  \dodoi{10.1086/506564}

\bibitem[{{Tchernyshyov} \& {Peek}(2017)}]{Tchernyshyov_2017}
{Tchernyshyov}, K., \& {Peek}, J.~E.~G. 2017, \aj, 153, 8,
  \dodoi{10.3847/1538-3881/153/1/8}

\bibitem[{{Tchernyshyov} {et~al.}(2018){Tchernyshyov}, {Peek}, \&
  {Zasowski}}]{Tchernyshyov_2018}
{Tchernyshyov}, K., {Peek}, J.~E.~G., \& {Zasowski}, G. 2018, \aj, 156, 248,
  \dodoi{10.3847/1538-3881/aae68d}

\bibitem[{{Ting} {et~al.}(2015){Ting}, {Conroy}, \& {Goodman}}]{Ting_2015}
{Ting}, Y.-S., {Conroy}, C., \& {Goodman}, A. 2015, \apj, 807, 104,
  \dodoi{10.1088/0004-637X/807/1/104}

\bibitem[{{Ting} {et~al.}(2019){Ting}, {Conroy}, {Rix}, \&
  {Cargile}}]{Ting_2019}
{Ting}, Y.-S., {Conroy}, C., {Rix}, H.-W., \& {Cargile}, P. 2019, \apj, 879,
  69, \dodoi{10.3847/1538-4357/ab2331}

\bibitem[{Virtanen {et~al.}(2020)Virtanen, Gommers, Oliphant, Haberland, Reddy,
  Cournapeau, Burovski, Peterson, Weckesser, Bright, {van der Walt}, Brett,
  Wilson, Millman, Mayorov, Nelson, Jones, Kern, Larson, Carey, Polat, Feng,
  Moore, {VanderPlas}, Laxalde, Perktold, Cimrman, Henriksen, Quintero, Harris,
  Archibald, Ribeiro, Pedregosa, {van Mulbregt}, \& {SciPy 1.0
  Contributors}}]{scipy_citation}
Virtanen, P., Gommers, R., Oliphant, T.~E., {et~al.} 2020, Nature Methods, 17,
  261, \dodoi{10.1038/s41592-019-0686-2}

\bibitem[{{Weinberg} {et~al.}(2022){Weinberg}, {Holtzman}, {Johnson}, {Hayes},
  {Hasselquist}, {Shetrone}, {Ting}, {Beaton}, {Beers}, {Bird}, {Bizyaev},
  {Blanton}, {Cunha}, {Fern{\'a}ndez-Trincado}, {Frinchaboy},
  {Garc{\'\i}a-Hern{\'a}ndez}, {Griffith}, {Johnson}, {J{\"o}nsson}, {Lane},
  {Leung}, {Mackereth}, {Majewski}, {M{\'e}sz{\'a}ros}, {Nitschelm}, {Pan},
  {Schiavon}, {Schneider}, {Schultheis}, {Smith}, {Sobeck}, {Stassun},
  {Stringfellow}, {Vincenzo}, {Wilson}, \& {Zasowski}}]{Weinberg_2022}
{Weinberg}, D.~H., {Holtzman}, J.~A., {Johnson}, J.~A., {et~al.} 2022, \apjs,
  260, 32, \dodoi{10.3847/1538-4365/ac6028}

\bibitem[{{Wilson} {et~al.}(2012){Wilson}, {Hearty}, {Skrutskie}, {Majewski},
  {Schiavon}, {Eisenstein}, {Gunn}, {Holtzman}, {Nidever}, {Gillespie},
  {Weinberg}, {Blank}, {Henderson}, {Smee}, {Barkhouser}, {Harding}, {Hope},
  {Fitzgerald}, {Stolberg}, {Arns}, {Nelson}, {Brunner}, {Burton}, {Walker},
  {Lam}, {Maseman}, {Barr}, {Leger}, {Carey}, {MacDonald}, {Ebelke}, {Beland},
  {Horne}, {Young}, {Rieke}, {Rieke}, {O'Brien}, {Crane}, {Carr}, {Harrison},
  {Stoll}, {Vernieri}, {Shetrone}, {Allende-Prieto}, {Johnson}, {Frinchaboy},
  {Zasowski}, {Garcia Perez}, {Bizyaev}, {Cunha}, {Smith}, {Meszaros}, {Zhao},
  {Hayden}, {Chojnowski}, {Andrews}, {Loomis}, {Owen}, {Klaene}, {Brinkmann},
  {Stauffer}, {Long}, {Jordan}, {Holder}, {Cope}, {Naugle}, {Pfaffenberger},
  {Schlegel}, {Blanton}, {Muna}, {Weaver}, {Snedden}, {Pan}, {Brewington},
  {Malanushenko}, {Malanushenko}, {Simmons}, {Oravetz}, {Mahadevan}, \&
  {Halverson}}]{Wilson_2012}
{Wilson}, J.~C., {Hearty}, F., {Skrutskie}, M.~F., {et~al.} 2012, in Society of
  Photo-Optical Instrumentation Engineers (SPIE) Conference Series, Vol. 8446,
  Ground-based and Airborne Instrumentation for Astronomy IV, ed. I.~S.
  {McLean}, S.~K. {Ramsay}, \& H.~{Takami}, 84460H, \dodoi{10.1117/12.927140}

\bibitem[{{Wylie} {et~al.}(2021){Wylie}, {Gerhard}, {Ness}, {Clarke},
  {Freeman}, \& {Bland-Hawthorn}}]{Wylie_2021}
{Wylie}, S.~M., {Gerhard}, O.~E., {Ness}, M.~K., {et~al.} 2021, \aap, 653,
  A143, \dodoi{10.1051/0004-6361/202140990}

\bibitem[{{Yanny} {et~al.}(2009){Yanny}, {Rockosi}, {Newberg}, {Knapp},
  {Adelman-McCarthy}, {Alcorn}, {Allam}, {Allende Prieto}, {An}, {Anderson},
  {Anderson}, {Bailer-Jones}, {Bastian}, {Beers}, {Bell}, {Belokurov},
  {Bizyaev}, {Blythe}, {Bochanski}, {Boroski}, {Brinchmann}, {Brinkmann},
  {Brewington}, {Carey}, {Cudworth}, {Evans}, {Evans}, {Gates}, {G{\"a}nsicke},
  {Gillespie}, {Gilmore}, {Nebot Gomez-Moran}, {Grebel}, {Greenwell}, {Gunn},
  {Jordan}, {Jordan}, {Harding}, {Harris}, {Hendry}, {Holder}, {Ivans},
  {Ivezi{\v{c}}}, {Jester}, {Johnson}, {Kent}, {Kleinman}, {Kniazev},
  {Krzesinski}, {Kron}, {Kuropatkin}, {Lebedeva}, {Lee}, {French Leger},
  {L{\'e}pine}, {Levine}, {Lin}, {Long}, {Loomis}, {Lupton}, {Malanushenko},
  {Malanushenko}, {Margon}, {Martinez-Delgado}, {McGehee}, {Monet}, {Morrison},
  {Munn}, {Neilsen}, {Nitta}, {Norris}, {Oravetz}, {Owen}, {Padmanabhan},
  {Pan}, {Peterson}, {Pier}, {Platson}, {Re Fiorentin}, {Richards}, {Rix},
  {Schlegel}, {Schneider}, {Schreiber}, {Schwope}, {Sibley}, {Simmons},
  {Snedden}, {Allyn Smith}, {Stark}, {Stauffer}, {Steinmetz}, {Stoughton},
  {SubbaRao}, {Szalay}, {Szkody}, {Thakar}, {Sivarani}, {Tucker}, {Uomoto},
  {Vanden Berk}, {Vidrih}, {Wadadekar}, {Watters}, {Wilhelm}, {Wyse}, {Yarger},
  \& {Zucker}}]{Yanny_2009}
{Yanny}, B., {Rockosi}, C., {Newberg}, H.~J., {et~al.} 2009, \aj, 137, 4377,
  \dodoi{10.1088/0004-6256/137/5/4377}

\bibitem[{{York} {et~al.}(2000){York}, {Adelman}, {Anderson}, {Anderson},
  {Annis}, {Bahcall}, {Bakken}, {Barkhouser}, {Bastian}, {Berman}, {Boroski},
  {Bracker}, {Briegel}, {Briggs}, {Brinkmann}, {Brunner}, {Burles}, {Carey},
  {Carr}, {Castander}, {Chen}, {Colestock}, {Connolly}, {Crocker}, {Csabai},
  {Czarapata}, {Davis}, {Doi}, {Dombeck}, {Eisenstein}, {Ellman}, {Elms},
  {Evans}, {Fan}, {Federwitz}, {Fiscelli}, {Friedman}, {Frieman}, {Fukugita},
  {Gillespie}, {Gunn}, {Gurbani}, {de Haas}, {Haldeman}, {Harris}, {Hayes},
  {Heckman}, {Hennessy}, {Hindsley}, {Holm}, {Holmgren}, {Huang}, {Hull},
  {Husby}, {Ichikawa}, {Ichikawa}, {Ivezi{\'c}}, {Kent}, {Kim}, {Kinney},
  {Klaene}, {Kleinman}, {Kleinman}, {Knapp}, {Korienek}, {Kron}, {Kunszt},
  {Lamb}, {Lee}, {Leger}, {Limmongkol}, {Lindenmeyer}, {Long}, {Loomis},
  {Loveday}, {Lucinio}, {Lupton}, {MacKinnon}, {Mannery}, {Mantsch}, {Margon},
  {McGehee}, {McKay}, {Meiksin}, {Merelli}, {Monet}, {Munn}, {Narayanan},
  {Nash}, {Neilsen}, {Neswold}, {Newberg}, {Nichol}, {Nicinski}, {Nonino},
  {Okada}, {Okamura}, {Ostriker}, {Owen}, {Pauls}, {Peoples}, {Peterson},
  {Petravick}, {Pier}, {Pope}, {Pordes}, {Prosapio}, {Rechenmacher}, {Quinn},
  {Richards}, {Richmond}, {Rivetta}, {Rockosi}, {Ruthmansdorfer}, {Sandford},
  {Schlegel}, {Schneider}, {Sekiguchi}, {Sergey}, {Shimasaku}, {Siegmund},
  {Smee}, {Smith}, {Snedden}, {Stone}, {Stoughton}, {Strauss}, {Stubbs},
  {SubbaRao}, {Szalay}, {Szapudi}, {Szokoly}, {Thakar}, {Tremonti}, {Tucker},
  {Uomoto}, {Vanden Berk}, {Vogeley}, {Waddell}, {Wang}, {Watanabe},
  {Weinberg}, {Yanny}, {Yasuda}, \& {SDSS Collaboration}}]{York_2000}
{York}, D.~G., {Adelman}, J., {Anderson}, John~E., J., {et~al.} 2000, \aj, 120,
  1579, \dodoi{10.1086/301513}

\bibitem[{{Zasowski} {et~al.}(2015){Zasowski}, {M{\'e}nard}, {Bizyaev},
  {Garc{\'\i}a-Hern{\'a}ndez}, {Garc{\'\i}a P{\'e}rez}, {Hayden}, {Holtzman},
  {Johnson}, {Kinemuchi}, {Majewski}, {Nidever}, {Shetrone}, \&
  {Wilson}}]{Zasowski_2015}
{Zasowski}, G., {M{\'e}nard}, B., {Bizyaev}, D., {et~al.} 2015, \apj, 798, 35,
  \dodoi{10.1088/0004-637X/798/1/35}

\bibitem[{{Zasowski} {et~al.}(2019){Zasowski}, {Schultheis}, {Hasselquist},
  {Cunha}, {Sobeck}, {Johnson}, {Rojas-Arriagada}, {Majewski}, {Andrews},
  {J{\"o}nsson}, {Beers}, {Chojnowski}, {Frinchaboy}, {Holtzman}, {Minniti},
  {Nidever}, \& {Nitschelm}}]{Zasowski_2019}
{Zasowski}, G., {Schultheis}, M., {Hasselquist}, S., {et~al.} 2019, \apj, 870,
  138, \dodoi{10.3847/1538-4357/aaeff4}

\bibitem[{{Zhao} {et~al.}(2012){Zhao}, {Zhao}, {Chu}, {Jing}, \&
  {Deng}}]{Zhao_2012}
{Zhao}, G., {Zhao}, Y.-H., {Chu}, Y.-Q., {Jing}, Y.-P., \& {Deng}, L.-C. 2012,
  Research in Astronomy and Astrophysics, 12, 723,
  \dodoi{10.1088/1674-4527/12/7/002}

\end{thebibliography}
\bibliographystyle{aasjournal}

\begin{longrotatetable}
\begin{deluxetable}{cccccccccccc}
\tabletypesize{\footnotesize}
\tablecaption{Summary of the \nfinaldibs\@ DIB features (\nfinaldibsABSORPTION\@ in absorption, \nfinaldibsEMISSION\@ in emission) where the best sorting parameter is either $A_K$ or 15272~\AA\@ DIB-strength. Only features where $\mathrm{Pr}\left( \sigma_{\mathrm{SORT}} > \sigma_{\mathrm{RANDOM}}\right) > 0.5$ are included.
\label{tab:new_DIB_summary}}
\tablehead{
\colhead{$\lambda_{0}$\tablenotemark{a}} & \colhead{Hydrogen} & \colhead{Feature} & \colhead{Wavelength Range\tablenotemark{c}} & \multicolumn{2}{c}{$\mathrm{Pr}\left( \sigma_{\mathrm{SORT}} > \sigma_{\mathrm{RANDOM}}\right)$} & \colhead{Best Sort} & \multicolumn{4}{c}{EW\tablenotemark{d} (m\AA) in bin} \\ \cline{5-6} \cline{8-11}
\colhead{(\AA)} & \colhead{Transition\tablenotemark{b}} & \colhead{Type} & \colhead{(\AA)} & \colhead{DIB} & \colhead{$A_K$} & \colhead{Parameter} & \colhead{min DIB} & \colhead{max DIB} & \colhead{min $A_K$} & \colhead{max $A_K$}
} 
\startdata
15172.08 &   & Absorption & $15169.6 - 15175.4$ & 0.9607 & 0.8572 &  DIB  & $-0.8 \pm 0.9$ & $5.6 \pm 1.5$ & $-0.8 \pm 0.8$ & $4.6 \pm 1.5$ \\
15178.79 &   & Absorption & $15176.1 - 15181.3$ & 0.9565 & 0.6423 &  DIB  & $-1.4 \pm 0.8$ & $4.3 \pm 1.4$ & $0.3 \pm 0.7$ & $2.1 \pm 1.3$ \\
15189.07 &   & Emission & $15186.8 - 15192.2$ & 0.8240 & 0.6193 &  DIB  & $-3.4 \pm 0.8$ & $-10.0 \pm 1.6$ & $-3.6 \pm 0.8$ & $-6.1 \pm 1.4$ \\
15191.80 &   & Absorption & $15189.5 - 15194.5$ & 0.2387 & 0.6785 & $A_K$ & $2.3 \pm 0.7$ & $3.7 \pm 1.3$ & $1.4 \pm 0.7$ & $0.8 \pm 1.2$ \\
15206.50 &   & Absorption & $15203.1 - 15210.1$ & 0.8433 & 0.8584 & $A_K$ & $3.0 \pm 0.6$ & $4.9 \pm 1.3$ & $4.4 \pm 0.7$ & $1.8 \pm 1.3$ \\
15215.54 &   & Absorption & $15211.3 - 15220.8$ & 0.9289 & 0.9387 &  DIB  & $-1.8 \pm 0.9$ & $7.8 \pm 1.8$ & $-0.9 \pm 1.0$ & $4.3 \pm 1.8$ \\
15224.37\tablenotemark{*} &   & Absorption & $15221.8 - 15226.7$ & 0.6093 & 0.9189 & $A_K$ & $-0.0 \pm 0.5$ & $1.2 \pm 1.0$ & $-1.5 \pm 0.5$ & $2.1 \pm 0.9$ \\
15237.62 &   & Absorption & $15233.2 - 15245.0$ & 0.8508 & 0.9808 & $A_K$ & $16.0 \pm 1.0$ & $23.0 \pm 2.4$ & $13.2 \pm 1.3$ & $10.2 \pm 2.2$ \\
15258.27 &   & Absorption & $15255.3 - 15264.0$ & 0.9202 & 0.8982 &  DIB  & $2.4 \pm 0.8$ & $6.8 \pm 1.5$ & $6.0 \pm 0.8$ & $0.5 \pm 1.3$ \\
15272.40\tablenotemark{*} &   & Absorption & $15264.6 - 15281.0$ & 0.9999 & 0.9999 &  DIB  & $59.4 \pm 1.5$ & $303.0 \pm 2.8$ & $75.3 \pm 1.5$ & $215.5 \pm 2.6$ \\
15292.03 &   & Absorption & $15286.1 - 15296.3$ & 0.9883 & 0.9240 &  DIB  & $5.8 \pm 1.1$ & $15.7 \pm 2.2$ & $3.5 \pm 1.1$ & $12.3 \pm 2.0$ \\
15319.31 &   & Emission & $15315.7 - 15325.2$ & 0.9906 & 0.9707 &  DIB  & $-0.2 \pm 1.0$ & $-14.7 \pm 1.8$ & $2.7 \pm 1.0$ & $-8.5 \pm 1.8$ \\
15347.49 & Br18 & Absorption & $15344.5 - 15351.5$ & 0.5131 & 0.8855 & $A_K$ & $-0.2 \pm 0.6$ & $3.7 \pm 1.1$ & $-0.4 \pm 0.6$ & $5.1 \pm 1.2$ \\
15370.61 &   & Absorption & $15362.8 - 15375.5$ & 0.9578 & 0.8730 &  DIB  & $3.4 \pm 1.1$ & $14.9 \pm 2.3$ & $5.0 \pm 1.1$ & $9.1 \pm 1.9$ \\
15384.21 &   & Absorption & $15376.1 - 15391.9$ & 0.8128 & 0.8068 &  DIB  & $5.3 \pm 1.4$ & $9.8 \pm 2.8$ & $1.8 \pm 1.4$ & $2.7 \pm 2.7$ \\
15394.84 &   & Absorption & $15392.1 - 15397.8$ & 0.4916 & 0.9186 & $A_K$ & $2.6 \pm 0.7$ & $5.5 \pm 1.3$ & $4.3 \pm 0.7$ & $3.2 \pm 1.3$ \\
15407.18 &   & Absorption & $15404.6 - 15410.6$ & 0.7304 & 0.5966 &  DIB  & $1.3 \pm 0.6$ & $4.0 \pm 1.1$ & $1.2 \pm 0.6$ & $2.5 \pm 1.0$ \\
15441.28 & Br17 & Absorption & $15437.6 - 15450.2$ & 0.8960 & 0.8995 &  DIB  & $-1.3 \pm 1.0$ & $10.0 \pm 2.0$ & $3.4 \pm 1.0$ & $5.3 \pm 1.8$ \\
15477.59 &   & Emission & $15469.9 - 15481.2$ & 0.8836 & 0.8983 & $A_K$ & $-5.9 \pm 1.0$ & $-15.9 \pm 2.0$ & $-7.5 \pm 1.0$ & $-12.2 \pm 1.7$ \\
15481.22 &   & Absorption & $15477.8 - 15485.7$ & 0.9155 & 0.9664 & $A_K$ & $0.2 \pm 0.7$ & $8.3 \pm 1.3$ & $6.1 \pm 0.8$ & $3.4 \pm 1.4$ \\
15537.79 &   & Absorption & $15532.2 - 15541.2$ & 0.3082 & 0.5685 & $A_K$ & $0.1 \pm 1.5$ & $3.1 \pm 2.7$ & $0.8 \pm 1.4$ & $2.3 \pm 2.8$ \\
15549.17 &   & Absorption & $15545.1 - 15554.5$ & 0.9757 & 0.9101 &  DIB  & $3.4 \pm 1.0$ & $16.6 \pm 1.8$ & $6.0 \pm 1.0$ & $7.4 \pm 1.8$ \\
15560.13 & Br16 & Absorption & $15556.9 - 15564.4$ & 0.9910 & 0.8949 &  DIB  & $-2.4 \pm 0.7$ & $8.2 \pm 1.2$ & $0.3 \pm 0.7$ & $-0.2 \pm 1.1$ \\
15601.89 &   & Absorption & $15599.5 - 15604.9$ & 0.4632 & 0.8414 & $A_K$ & $-0.6 \pm 0.6$ & $2.5 \pm 1.2$ & $1.1 \pm 0.6$ & $-0.0 \pm 1.0$ \\
15615.91\tablenotemark{*} &   & Absorption & $15611.2 - 15624.3$ & 0.9999 & 0.9999 &  DIB  & $10.0 \pm 1.0$ & $44.0 \pm 2.2$ & $16.1 \pm 1.1$ & $42.4 \pm 1.7$ \\
15633.18 &   & Absorption & $15627.4 - 15637.7$ & 0.8891 & 0.9183 & $A_K$ & $3.1 \pm 0.9$ & $8.1 \pm 1.6$ & $2.4 \pm 0.9$ & $5.3 \pm 1.6$ \\
15643.55 &   & Absorption & $15641.0 - 15646.4$ & 0.8174 & 0.7844 &  DIB  & $0.5 \pm 0.5$ & $3.9 \pm 0.8$ & $-0.1 \pm 0.5$ & $1.4 \pm 0.8$ \\
15652.63\tablenotemark{*} &   & Absorption & $15646.1 - 15662.8$ & 0.9999 & 0.9999 &  DIB  & $14.2 \pm 1.6$ & $76.6 \pm 2.8$ & $10.9 \pm 1.5$ & $62.0 \pm 2.5$ \\
15671.89\tablenotemark{*} &   & Absorption & $15663.7 - 15679.2$ & 0.9999 & 0.9999 &  DIB  & $21.7 \pm 1.2$ & $76.0 \pm 2.4$ & $26.6 \pm 1.3$ & $59.3 \pm 2.3$ \\
15706.13 & Br15 & Absorption & $15701.8 - 15711.1$ & 0.9953 & 0.9764 &  DIB  & $2.5 \pm 1.2$ & $20.7 \pm 2.1$ & $3.2 \pm 1.0$ & $14.0 \pm 2.0$ \\
15722.41 &   & Absorption & $15718.9 - 15726.8$ & 0.9921 & 0.9520 &  DIB  & $-1.7 \pm 0.9$ & $7.6 \pm 1.5$ & $-0.8 \pm 0.9$ & $5.7 \pm 1.4$ \\
15733.93 &   & Absorption & $15728.1 - 15738.9$ & 0.8863 & 0.9278 & $A_K$ & $1.6 \pm 1.2$ & $10.6 \pm 2.3$ & $0.9 \pm 1.4$ & $12.3 \pm 2.1$ \\
15751.55 &   & Emission & $15746.8 - 15760.3$ & 0.7631 & 0.8419 & $A_K$ & $-1.9 \pm 1.2$ & $-10.4 \pm 2.5$ & $-6.7 \pm 1.3$ & $-8.9 \pm 2.3$ \\
15769.62 &   & Absorption & $15767.0 - 15772.5$ & 0.8378 & 0.7406 &  DIB  & $1.5 \pm 0.7$ & $3.2 \pm 1.2$ & $2.3 \pm 0.8$ & $4.1 \pm 1.2$ \\
15878.05 &   & Absorption & $15873.5 - 15882.2$ & 0.9611 & 0.6855 &  DIB  & $1.3 \pm 0.9$ & $13.3 \pm 1.6$ & $3.5 \pm 0.9$ & $9.4 \pm 1.7$ \\
15889.24 &   & Absorption & $15886.6 - 15894.3$ & 0.6431 & 0.8079 & $A_K$ & $2.6 \pm 0.8$ & $5.6 \pm 1.5$ & $5.6 \pm 0.9$ & $3.5 \pm 1.4$ \\
15919.35 &   & Absorption & $15913.9 - 15922.6$ & 0.8737 & 0.5808 &  DIB  & $-1.4 \pm 0.8$ & $5.1 \pm 1.4$ & $1.0 \pm 0.7$ & $0.5 \pm 1.4$ \\
15927.93 &   & Absorption & $15925.1 - 15930.8$ & 0.8308 & 0.2518 &  DIB  & $0.1 \pm 0.5$ & $3.0 \pm 0.9$ & $0.3 \pm 0.5$ & $2.0 \pm 0.9$ \\
15937.83 &   & Absorption & $15931.9 - 15944.0$ & 0.9847 & 0.7884 &  DIB  & $4.1 \pm 1.0$ & $16.5 \pm 1.8$ & $6.4 \pm 0.9$ & $10.3 \pm 1.7$ \\
15949.95 &   & Absorption & $15944.7 - 15954.4$ & 0.8917 & 0.9338 & $A_K$ & $1.6 \pm 0.8$ & $8.5 \pm 1.4$ & $5.7 \pm 0.8$ & $6.4 \pm 1.2$ \\
15977.29 &   & Absorption & $15974.4 - 15981.5$ & 0.6406 & 0.5648 &  DIB  & $0.3 \pm 0.7$ & $4.0 \pm 1.4$ & $1.3 \pm 0.7$ & $2.5 \pm 1.2$ \\
15989.22\tablenotemark{*} &   & Absorption & $15983.7 - 15993.9$ & 0.9771 & 0.9274 &  DIB  & $2.8 \pm 0.9$ & $14.3 \pm 1.6$ & $7.7 \pm 1.0$ & $8.2 \pm 1.4$ \\
15999.38 &   & Absorption & $15995.0 - 16004.0$ & 0.7028 & 0.4841 &  DIB  & $0.3 \pm 0.8$ & $5.6 \pm 1.2$ & $3.4 \pm 0.7$ & $4.5 \pm 1.3$ \\
16029.47 &   & Absorption & $16025.7 - 16043.2$ & 0.7538 & 0.8039 & $A_K$ & $-3.9 \pm 2.1$ & $12.0 \pm 3.9$ & $10.5 \pm 2.0$ & $2.5 \pm 3.7$ \\
16046.53 &   & Absorption & $16044.3 - 16049.2$ & 0.4056 & 0.7245 & $A_K$ & $-0.0 \pm 0.8$ & $3.7 \pm 1.8$ & $1.2 \pm 0.7$ & $0.8 \pm 1.6$ \\
16059.40 &   & Absorption & $16054.7 - 16062.5$ & 0.9423 & 0.6806 &  DIB  & $1.5 \pm 0.8$ & $4.7 \pm 1.6$ & $1.1 \pm 0.8$ & $5.8 \pm 1.5$ \\
16113.62 & Br13 & Absorption & $16101.8 - 16119.4$ & 0.8491 & 0.8396 &  DIB  & $3.3 \pm 1.8$ & $17.1 \pm 2.9$ & $6.3 \pm 1.7$ & $15.4 \pm 2.9$ \\
16132.78 &   & Absorption & $16130.5 - 16135.5$ & 0.5925 & 0.3200 &  DIB  & $0.5 \pm 0.6$ & $3.2 \pm 1.1$ & $1.0 \pm 0.6$ & $2.7 \pm 1.1$ \\
16141.92 &   & Absorption & $16138.4 - 16145.7$ & 0.8523 & 0.2659 &  DIB  & $2.0 \pm 0.6$ & $3.4 \pm 1.2$ & $1.5 \pm 0.6$ & $1.1 \pm 1.0$ \\
16147.94 &   & Emission & $16142.1 - 16157.3$ & 0.6719 & 0.2360 &  DIB  & $-4.0 \pm 1.3$ & $-12.2 \pm 2.3$ & $-5.4 \pm 1.2$ & $-8.3 \pm 2.0$ \\
16157.09 &   & Absorption & $16152.9 - 16166.5$ & 0.9850 & 0.9613 &  DIB  & $-2.3 \pm 1.2$ & $14.0 \pm 2.1$ & $-0.2 \pm 1.2$ & $6.8 \pm 2.2$ \\
16219.72 &   & Absorption & $16216.4 - 16224.9$ & 0.9462 & 0.9046 &  DIB  & $0.9 \pm 0.7$ & $8.0 \pm 1.5$ & $1.1 \pm 0.7$ & $6.5 \pm 1.4$ \\
16233.39 &   & Absorption & $16224.9 - 16237.6$ & 0.9373 & 0.9335 &  DIB  & $4.3 \pm 1.4$ & $18.2 \pm 3.2$ & $7.8 \pm 1.3$ & $14.9 \pm 2.5$ \\
16265.94 &   & Absorption & $16262.1 - 16270.0$ & 0.9133 & 0.8788 &  DIB  & $2.4 \pm 0.7$ & $7.6 \pm 1.3$ & $3.3 \pm 0.7$ & $4.7 \pm 1.2$ \\
16280.11 &   & Absorption & $16277.0 - 16286.6$ & 0.8855 & 0.8796 &  DIB  & $1.4 \pm 0.8$ & $8.2 \pm 1.5$ & $-0.3 \pm 0.8$ & $4.4 \pm 1.4$ \\
16290.68 &   & Absorption & $16288.0 - 16294.7$ & 0.7890 & 0.9301 & $A_K$ & $1.3 \pm 0.6$ & $4.6 \pm 1.2$ & $1.9 \pm 0.6$ & $4.4 \pm 1.1$ \\
16371.00 &   & Absorption & $16368.5 - 16375.8$ & 0.9926 & 0.7010 &  DIB  & $2.4 \pm 0.7$ & $9.1 \pm 1.4$ & $1.3 \pm 0.7$ & $4.7 \pm 1.3$ \\
16382.09 &   & Absorption & $16378.9 - 16385.0$ & 0.3357 & 0.9270 & $A_K$ & $0.6 \pm 0.7$ & $3.7 \pm 1.3$ & $0.0 \pm 0.6$ & $2.7 \pm 1.2$ \\
16411.31 & Br12 & Absorption & $16406.3 - 16414.5$ & 0.6866 & 0.9621 & $A_K$ & $0.0 \pm 1.0$ & $8.6 \pm 2.2$ & $-5.1 \pm 1.0$ & $-0.3 \pm 1.9$ \\
16518.68 &   & Absorption & $16515.5 - 16522.3$ & 0.9870 & 0.8953 &  DIB  & $-0.6 \pm 0.7$ & $8.1 \pm 1.3$ & $-0.2 \pm 0.7$ & $4.1 \pm 1.1$ \\
16525.52 &   & Absorption & $16522.8 - 16528.5$ & 0.9339 & 0.6967 &  DIB  & $1.0 \pm 0.6$ & $4.2 \pm 1.1$ & $0.4 \pm 0.6$ & $1.6 \pm 1.0$ \\
16532.60 &   & Absorption & $16529.9 - 16536.3$ & 0.8580 & 0.8020 &  DIB  & $1.5 \pm 0.7$ & $3.5 \pm 1.3$ & $2.4 \pm 0.7$ & $-2.0 \pm 1.1$ \\
16536.49 &   & Emission & $16532.4 - 16550.2$ & 0.9392 & 0.5342 &  DIB  & $-3.3 \pm 1.6$ & $-19.3 \pm 3.1$ & $-8.1 \pm 1.7$ & $-2.9 \pm 3.2$ \\
16549.51 &   & Absorption & $16541.7 - 16554.5$ & 0.8462 & 0.8512 & $A_K$ & $-0.3 \pm 1.5$ & $5.7 \pm 2.7$ & $3.2 \pm 1.5$ & $8.5 \pm 2.5$ \\
16573.31\tablenotemark{*} &   & Absorption & $16567.8 - 16579.5$ & 0.9998 & 0.9919 &  DIB  & $5.9 \pm 1.1$ & $30.3 \pm 2.1$ & $7.1 \pm 1.0$ & $22.9 \pm 1.9$ \\
16584.99 &   & Absorption & $16580.4 - 16589.8$ & 0.9906 & 0.9814 &  DIB  & $8.7 \pm 0.9$ & $18.3 \pm 1.8$ & $3.5 \pm 0.9$ & $14.7 \pm 1.6$ \\
16594.62 &   & Absorption & $16590.5 - 16598.1$ & 0.9346 & 0.6475 &  DIB  & $2.9 \pm 0.7$ & $9.0 \pm 1.5$ & $1.3 \pm 0.7$ & $4.5 \pm 1.3$ \\
16601.73 &   & Absorption & $16598.7 - 16605.2$ & 0.8588 & 0.7641 &  DIB  & $0.3 \pm 0.6$ & $5.3 \pm 1.2$ & $0.1 \pm 0.6$ & $2.6 \pm 1.2$ \\
16608.38 &   & Absorption & $16605.8 - 16611.4$ & 0.3847 & 0.5903 & $A_K$ & $1.4 \pm 0.6$ & $2.2 \pm 1.4$ & $1.6 \pm 0.6$ & $4.2 \pm 1.1$ \\
16621.92 &   & Absorption & $16617.3 - 16627.2$ & 0.9017 & 0.6553 &  DIB  & $-1.4 \pm 0.9$ & $7.8 \pm 2.0$ & $0.9 \pm 0.9$ & $-0.3 \pm 1.8$ \\
16655.25 &   & Absorption & $16645.8 - 16661.2$ & 0.6496 & 0.8407 & $A_K$ & $1.2 \pm 1.6$ & $7.2 \pm 3.5$ & $-2.4 \pm 1.7$ & $11.8 \pm 2.6$ \\
16711.03 &   & Absorption & $16708.3 - 16715.2$ & 0.7723 & 0.9000 & $A_K$ & $1.0 \pm 1.0$ & $6.9 \pm 2.0$ & $1.5 \pm 1.0$ & $1.3 \pm 2.0$ \\
16718.42 &   & Absorption & $16716.1 - 16722.8$ & 0.7223 & 0.3377 &  DIB  & $0.2 \pm 0.8$ & $3.9 \pm 1.7$ & $1.0 \pm 0.8$ & $1.0 \pm 1.4$ \\
16750.09 &   & Absorption & $16739.7 - 16753.3$ & 0.6424 & 0.8165 & $A_K$ & $0.9 \pm 1.5$ & $12.5 \pm 3.3$ & $3.7 \pm 1.5$ & $11.0 \pm 2.9$ \\
16757.04 &   & Absorption & $16753.8 - 16762.1$ & 0.8341 & 0.9163 & $A_K$ & $0.2 \pm 1.2$ & $9.1 \pm 2.2$ & $2.9 \pm 1.0$ & $0.8 \pm 2.0$ \\
16770.47 &   & Absorption & $16764.7 - 16777.7$ & 0.3466 & 0.7885 & $A_K$ & $5.1 \pm 1.4$ & $10.0 \pm 3.2$ & $5.4 \pm 1.5$ & $4.8 \pm 3.0$ \\
16780.44 &   & Absorption & $16778.1 - 16783.0$ & 0.6389 & 0.2577 &  DIB  & $0.9 \pm 0.6$ & $2.2 \pm 1.2$ & $0.7 \pm 0.6$ & $0.4 \pm 1.1$ \\
16826.17 &   & Absorption & $16823.4 - 16828.7$ & 0.8639 & 0.5821 &  DIB  & $-0.5 \pm 0.6$ & $2.5 \pm 1.5$ & $0.0 \pm 0.6$ & $0.9 \pm 1.2$ \\
16842.68 &   & Emission & $16835.5 - 16846.4$ & 0.8304 & 0.1570 &  DIB  & $-2.4 \pm 1.3$ & $-8.6 \pm 2.9$ & $-5.7 \pm 1.3$ & $-2.1 \pm 2.7$ \\
16845.94 &   & Absorption & $16843.2 - 16849.0$ & 0.4582 & 0.5016 & $A_K$ & $-0.8 \pm 0.8$ & $5.6 \pm 1.7$ & $2.7 \pm 0.8$ & $1.8 \pm 1.5$ \\
16885.78 &   & Absorption & $16881.1 - 16888.8$ & 0.7304 & 0.5061 &  DIB  & $2.1 \pm 1.0$ & $3.7 \pm 2.4$ & $0.3 \pm 1.0$ & $3.0 \pm 1.8$ \\
16901.42 &   & Absorption & $16894.7 - 16905.4$ & 0.4721 & 0.6623 & $A_K$ & $6.6 \pm 2.0$ & $2.8 \pm 5.5$ & $1.6 \pm 2.0$ & $-0.5 \pm 3.8$ \\
16916.84 &   & Emission & $16911.7 - 16921.8$ & 0.9711 & 0.8659 &  DIB  & $-2.0 \pm 1.3$ & $-14.2 \pm 3.4$ & $-4.8 \pm 1.3$ & $-9.4 \pm 2.9$ \\
16920.58 &   & Absorption & $16917.3 - 16924.3$ & 0.8827 & 0.7339 &  DIB  & $1.8 \pm 0.9$ & $8.2 \pm 2.4$ & $2.5 \pm 0.9$ & $2.6 \pm 2.0$ \\
\enddata
\tablenotetext{a}{Wavelength, in the 15272~\AA\@ DIB rest-frame, of the local extremum found by smoothing the highest DIB-strength spectrum in Figure \ref{fig:coadded_specrta_EW_binned} with a 5~pixel Gaussian kernel.}
\tablenotetext{b}{Hydrogen transition wavelengths that are within 2~\AA\@ of the feature's central wavelength.}
\tablenotetext{c}{Region, in the 15272~\AA\@ DIB rest-frame, that we measure the equivalent width from. The wavelengths used to estimate the local continuum are taken to be the 5 pixels nearest to the Wavelength Range values.}
\tablenotetext{d}{Equivalent width measured when binning by $A_K$ or the 15272~\AA\@ DIB strength, in the minimum and maximum bins.}
\tablenotetext{*}{Nearest detected DIB to the previously-known DIBs in Table \ref{tab:previous_DIBs}.}
\end{deluxetable}
\end{longrotatetable}

\appendix

\section{Spectral Combinations} \label{sec:spectral_coaddition}
Our analysis, like many studies involving spectroscopy, requires the combination of multiple spectral observations. The standard approach would be to use an inverse-variance weighted combination using the fluxes and corresponding uncertainties at each pixel, but this results in a combined uncertainty that is often overly constraining and which becomes smaller for any increase in the number of observations. Instead, we argue that fitting a population-level distribution is generally the better approach when combining spectra, particularly in cases where the uncertainty on the combined measurement is important. This technique returns population-level means and variances that incorporate the uncertainties in each individual measurement as well as the dispersion in those measurements. 

\begin{figure}[ht]
\begin{center}
\includegraphics[width=\linewidth]{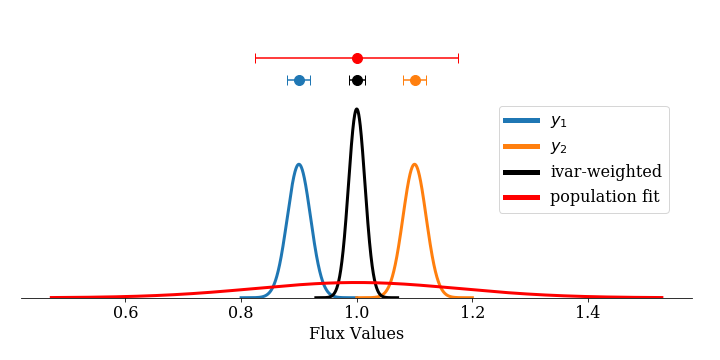}
\caption{Example of the differences between the inverse-variance weighted combination of multiple data measurements and Bayesian population-level distribution fitting. The lines show normal distributions that define each set of measurements, and the points above the distributions show the mean and standard deviation of each distribution. The blue and orange lines correspond to the two data measurements while the other two lines correspond to different methods of combining those data.}
\label{fig:population_flux_fit_example}
\end{center}
\end{figure}

As an illustrative example, two flux measurements of the same star in one pixel might differ from one another by an amount larger than is described by their uncertainties; this case is displayed in Figure \ref{fig:population_flux_fit_example}. Combining these measurements using inverse-variance weighting (black line) produces a mean that is, again, statistically far away from either measurement and has a high confidence (small uncertainty). If we instead fit a population-level distribution (red line), the resulting population width we measure is much larger than the inverse-variance width so that the population distribution can capture the large distance between the data\footnote{While this is a helpful visual example, we should be cautious about performing population fits on a very small number of measurements.}. 

\begin{figure}[ht]
\begin{center}
\includegraphics[width=\linewidth]{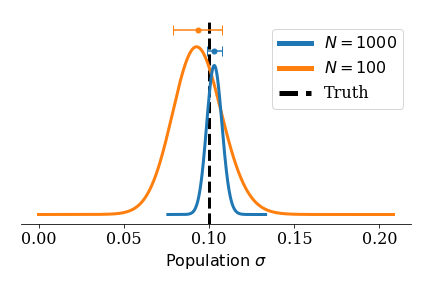}
\caption{Illustration of the impact that the population size has on the distribution of the population width/uncertainty as measured from Bayesian population fitting. The vertical dashed black line shows the input true population width that was used to generate the data, and the points above the distributions show the median and 68\% region for each distribution.}
\label{fig:population_flux_uncertainty_example}
\end{center}
\end{figure}

For another example, consider set of $N$ normalized flux measurements that all have SNR of $10~\mathrm{pixel}^{-1}$\@. The inverse-variance weighted combination will return an uncertainty of $\left( 10/\sqrt{N} \right) ~\mathrm{pixel}^{-1}$, which will clearly decrease in size as we add more measurements, regardless of how similar or disparate those measurements may be to one another. Using population fitting instead, the distribution of the population width becomes narrower and can converge on an underlying true width, assuming one exists. This is illustrated in Figure \ref{fig:population_flux_uncertainty_example}. 

For this paper's analysis of residuals\footnote{NOTE: for this example, we are assuming that each star has one observation for clarity of the math, but the approach is easily expanded to include multiple observations of the same star.}, we consider the measured/observed residual flux at pixel $i$ for star $j$ to be $r_{i,j}$ with corresponding uncertainty $\sigma_{r,i,j}$. Then, we define the following hierarchical statistical model that describes the relationship between the individual measurements, their uncertainties, and the population distribution: 
\begin{equation*}
    \begin{split}
        p\left(\hat \sigma_{r,i} \right) &\propto 1\\
        p\left( \hat r_i | \hat \sigma_{r,i} \right) &\propto 1\\
        \left( r'_{i,j} | \hat r_i, \hat \sigma_{r,i} \right) &\sim \mathcal{N}\left(\hat r_i, \hat \sigma_{r,i}^2 \right)\\
        \left( r_{i,j} | r'_{i,j}, \sigma_{r,i,j} \right) &\sim \mathcal{N}\left(r'_{i,j}, \sigma_{r,i,j}^2 \right)\\
    \end{split}
\end{equation*}
where $\hat r_i$ is the population mean of the residual fluxes in pixel $i$, $\hat \sigma_{r,i}$ is the population width/uncertainty/standard deviation of the residuals in pixel $i$, and $r'_{i,j}$ is the true residual flux for star $j$ in pixel $i$. We have chosen flat priors on the population parameters, though these could be changed to other distributions if there was good reason for it (e.g. Gaussian for $p\left( \hat r_i | \hat \sigma_{r,i} \right)$). The full posterior distribution is then:
\begin{equation}
    \label{eq:posterior_population}
    \begin{split}
        p(\hat r_i,& \hat \sigma_{r,i}^2, r'_{i,1}, \dots, r'_{i,n_*} |  r_{i,1}, \sigma^2_{i,1}, \dots, r_{i,n_*}, \sigma^2_{i,n_*}) \\
        &\propto p\left( \hat r_i,\sigma_{r,i} \right) \cdot \prod_j^{n_*} \mathcal{N}\left(r'_{i,j} | \hat r_i, \hat \sigma_{r,i}^2 \right) \cdot \mathcal{N}\left(r_{i,j} | r'_{i,j}, \sigma_{r,i,j}^2 \right)\\
    \end{split}
\end{equation}

We next see that the posterior full conditional on $r'_{i,j}$ is given by:
\begin{equation}
    \label{eq:posterior_full_conditional_rprime}
    \begin{split}
        p(r'_{i,j} | \hat r_{i}, \dots) &\propto \mathcal{N}\left(r'_{i,j} | \hat r_i, \hat \sigma_{r,i}^2 \right) \cdot \mathcal{N}\left(r'_{i,j} | r_{i,j}, \sigma_{r,i,j}^2 \right)\\
        &= \mathcal{N}\left( r'_{i,j} | \mu_{r',i,j}, \sigma_{r',i,j}^2 \right)\\
    \end{split}
\end{equation}
where $$\sigma_{r',i,j}^2 = \left[\hat \sigma_{r,i}^{-2}+\sigma_{r,i,j}^{-2}\right]^{-1}$$ and $$\mu_{r',i,j} = \sigma_{r',i,j}^2\cdot \left[\hat \sigma_{r,i}^{-2}\cdot \hat r_i +\sigma_{r,i,j}^{-2}\cdot r_{i,j} \right].$$ 

We can then use these results to integrate over $r'_{i,j}$ in the full posterior distribution of Equation \ref{eq:posterior_population} to find the marginal posterior of $\left( \hat r_i | \hat \sigma_{r,i}, \mathrm{data} \right)$:
\begin{equation} \label{eq:population_mean}
    \begin{split}
        p\left(\hat r_i | \hat \sigma_{r,i}, \mathrm{data} \right) = \mathcal{N}\left(\mu_{\hat r,i}, \sigma_{\hat r,i}^2 \right)
    \end{split}
\end{equation}
where $$\sigma_{\hat r,i}^2 = \left[\sum_{j}^{n_*} \left( \hat \sigma_{r,i}^{2}+\sigma_{r,i,j}^{2} \right)^{-1} \right]^{-1}$$ and $$\mu_{\hat r,i} = \sigma_{\hat r,i}^2\cdot \left[\sum_{j}^{n_*} \left( \hat \sigma_{ r,i}^{2}+\sigma_{r,i,j}^{2} \right)^{-1} \cdot r_{i,j} \right].$$ 

Using Bayes' Law, we can find the marginal posterior for $p\left(\hat \sigma_{r,i} | \mathrm{data} \right)$ as:
\begin{equation} \label{eq:population_dispersion}
    \begin{split}
        p(\hat \sigma_{r,i} | \mathrm{data} ) \propto p(\hat \sigma_{r,i}) \cdot \hat \sigma_{r,i}^{1/2} \cdot \prod_{j}^{n_*} \left(\hat \sigma_{r,i}^{2}+\sigma_{r,i,j}^{2} \right)^{-1/2} \cdot 
        \exp\left( -\frac{\left(r_{i,j}-\hat r_i\right)^2}{2 \left( \hat \sigma_{r,i}^{2}+\sigma_{r,i,j}^{2} \right)^2}\right)
    \end{split}
\end{equation}

With the functional forms of the distributions in Equation \ref{eq:population_mean} and \ref{eq:population_dispersion} in hand, we are able to draw samples of $(\hat r_i, \hat \sigma_{r,i} | \mathrm{data})$ fairly quickly. First, we evaluate $p(\hat  \sigma_{r,i} | \mathrm{data} )$ for a reasonable range and number of $\hat \sigma_{r,i}$ values, then we use those probabilities to draw $\hat \sigma_{r,i}$ samples. Next, we use those $\hat \sigma_{r,i}$ samples to draw samples from the $\left(\hat r_i | \hat \sigma_{\hat r,i}, \mathrm{data} \right)$ Gaussian distribution, which is an relatively easy and efficient step. Once we've repeated this process a sufficient number of times, we can take the median of the $\left(\hat r_i, \hat \sigma_{r,i}  | \mathrm{data} \right)$ samples as our best estimate of the underlying population mean and width. Sometimes, like in the main text of this work, we are also interested in the distribution on $\left(\hat r_i, \hat \sigma_{r,i}  | \mathrm{data} \right)$ itself, and this technique allows us to study this distribution.

\end{document}